\documentclass[12pt]{article}

\usepackage{graphicx}
\usepackage{amsmath}
\usepackage{tabularx}
\usepackage{color}

\newcommand{\pr}[1]{\mathcal{P}^{(#1)}}

\begin{document}

\title{{\bf Use of $SU(3)$ flavor projection operators to construct baryon-meson scattering amplitudes in the $1/N_c$ expansion}}

\author{V{\'\i}ctor Miguel Banda Guzm\'an \\
	{\it \normalsize Universidad Polit\'ecnica de San Luis Potos{\'\i}} \\
	{\it \normalsize Urbano Villal\'on 500, Col.\ La Ladrillera, San Luis Potos{\'\i}, 78363, S.L.P., M\'exico}
	\and
	Rub\'en Flores-Mendieta \\
	{\it \normalsize Instituto de F{\'\i}sica, Universidad Aut\'onoma de San Luis Potos{\'\i}} \\
	{\it \normalsize \'Alvaro Obreg\'on 64, Zona Centro, San Luis Potos{\'\i}, 78000, San Luis Potos{\'\i}, M\'exico}
        \and
        Johann Hern\'andez \\
	{\it \normalsize Instituto de F{\'\i}sica, Universidad Aut\'onoma de San Luis Potos{\'\i}} \\
	{\it \normalsize \'Alvaro Obreg\'on 64, Zona Centro, San Luis Potos{\'\i}, 78000, San Luis Potos{\'\i}, M\'exico}
	}

\maketitle

\abstract{
An $SU(3)$ flavor projection operator technique is implemented to construct the baryon-meson scattering amplitude in the framework of the $1/N_c$ expansion of QCD, where $N_c$ is the number of color charges. The operator technique is implemented to evaluate not only the lowest-order scattering amplitude but also effects coming from first-order perturbative $SU(3)$ flavor symmetry breaking and strong isospin breaking. The most general expression is obtained by accounting for explicitly the effects of the decuplet-octet baryon mass difference. At order $\mathcal{O}(1/N_c^2)$, a large number of unknown operator coefficients appear, so there is little additional predictive power unless leading and subleading terms are retained. Although the resultant expression is general enough that it can be applied to any incoming and outgoing baryons and pseudo scalar mesons, provided that the Gell-Mann--Nishijima scheme is respected, results for $N\pi\to N\pi$ scattering processes are explicitly dealt with.
}

\section{Introduction}

Quantum chromodynamics (QCD) is commonly accepted as the theory of the strong interaction, with quarks and gluons as fundamental fields. QCD is a gauge theory with the local symmetry group $SU(N_c)$, acting in the internal space of color degrees of freedom, with $N_c=3$ color charges. The analytical computation of hadron properties from first principles, however, is hampered due to the fact that QCD is strongly coupled at low energies. Two major theories have shed light on the static properties of hadrons. One of them is the large-$N_c$ limit and the other one is chiral perturbation theory (ChPT).

The generalization of QCD from $N_c=3$ to $N_c\to \infty$, commonly referred to as large-$N_c$ QCD, has become a remarkable tool to study the structure and interactions of mesons \cite{tHooft,ven} and baryons \cite{witten} in more generality. Physical quantities evaluated in the large-$N_c$ limit get corrections of relative orders $1/N_c$, $1/N_c^2$, and so on, which originates the $1/N_c$ expansion of QCD.

Baryon-meson scattering is a fundamental nuclear physics process which has been analyzed within the large-$N_c$ limit (and of course ChPT and several other approaches). The earliest analysis of baryon-meson scattering amplitudes in the context of the $1/N_c$ expansion was introduced in the seminal paper by Witten \cite{witten}. On general grounds, it takes $N_c$ quarks (in a totally antisymmetric color state) to make up a baryon, so Witten proposed to split the problem into two parts, to first use graphical methods to study $n$-quark forces in the large-$N_c$ limit, and then to use other methods to analyze the effects of these forces on an $N_c$-body state. From the analysis of large-$N_c$ counting rules for baryon-meson scattering, Witten concluded that the corresponding amplitude at fixed energy must be of order one.

Later on, Gervais and Sakita \cite{sakita} and Dashen and Manohar \cite{dm1} independently proved that large-$N_c$ QCD has a contracted $SU(4)$ symmetry (for two flavors of light quarks) and derived a set of consistency conditions that must be satisfied. The equations obtained from these consistency conditions admit a unique (minimal) solution for the baryon-meson coupling constants, which are identical to those of the Skyrme model or non-relativistic quark model. Dashen, Jenkins, and Manohar applied the approach to show that large-$N_c$ power counting rules for multimeson--baryon-baryon scattering amplitudes lead to important constraints on baryon static properties \cite{djm94,djm95}. In the same context, Flores-Mendieta, Hofmann, and Jenkins \cite{rfm00} studied tree-level amplitudes for baryon-meson scattering and obtained generalized large-$N_c$ consistency conditions valid to all orders in the baryon mass splitting $\Delta\equiv M_T-M_B$, where $M_T$ and $M_B$ represent the baryon decuplet and baryon octet masses, respectively. Cohen, Lebed and collaborators implemented a systematic method for deriving linear relations among meson-baryon scattering amplitudes combining the $1/N_c$ expansion of QCD with the Wigner-Eckart theorem applied to both angular momentum and isospin \cite{c1,kwee,c2}. In this framework, the scattering amplitudes are expressed via partial wave expansions, where the mesons carry fixed orbital angular momentum and the baryons possess definite spin and isospin, with baryon recoil effects neglected. The scattering matrix elements are further specified by the total spin and total isospin of the meson-baryon system. This approach, together with the $1/N_c$ corrections to the $t$-channel isospin and angular momentum exchange quantum numbers, $I_t = J_t $, enables the derivation of multiple linear relations among partial-wave amplitudes for meson-baryon scattering.

In the context of baryon chiral perturbation theory (BChPT), important advancements have been made on baryon-meson scattering over the past three decades. A detailed account of phenomenological models and/or different approaches proposed prior 2016 is presented in Ref.~\cite{yao}. Apart from the heavy baryon approach (HBChPT) \cite{bernard,fettes}, some fully relativistic methods are noteworthy, namely, the infrared regularization of covariant BChPT \cite{becher} and the extended-on-mass-shell scheme for BChPT \cite{geg1,geg2}, to name but a few. Further improvements in HBChPT to orders $\mathcal{O}(p^3)$ and $\mathcal{O}(p^4)$ have been performed recently \cite{huangA,huangB}.

Despite the important progress achieved in the understanding of baryon-meson scattering processes in both the phenomenological and experimental bent \cite{part}, various challenges still remain unsolved. In view of this, lattice QCD has become an essential non-perturbative tool to tackle some issues with first-principles QCD calculations which can not be dealt with otherwise. A comprehensive description of the state-of-the art computation of scattering amplitude for the baryon-meson system within lattice QCD can be found in Ref.~\cite{bulava}.

It is evident that the baryon-meson scattering problem is a mature area of research which has been tackled from a number of different perspectives. Nonetheless, the aim of the present work is to analytically compute baryon-meson scattering amplitudes at leading and subleading orders in the framework of the $1/N_c$ expansion, using widely the projection operator technique developed in Ref.~\cite{banda}. This approach introduces new and unique elements into the theory of baryon-meson scattering, expanding existing concepts and insights. At the first stage in the analysis, the primary objective will be to perform a calculation in the exact $SU(3)$ symmetry limit. At the second stage, the effects of first-order perturbative $SU(3)$ flavor symmetry breaking (SB) {\it and} strong isospin symmetry breaking (IB) will be separately incorporated. Thus, flavor projection operators will be useful to fully classify all flavor representations involved in the structure of the scattering amplitude. From this perspective, the present analysis is fundamentally different from previous works \cite{c1,kwee,c2}. Loop graphs contributing to the scattering amplitude can be consistently analyzed in a combined formalism between chiral and $1/N_c$ corrections, the so-called large-$N_c$ chiral perturbation theory based on the chiral Lagrangian introduced in Ref.~\cite{jen96}. This, however, requires a non-negligible effort which will be deferred to later work.

The paper is organized as follows. In Sec.~\ref{sec:tree} some elementary material about scattering processes is presented, along with a brief review of large-$N_c$ QCD to introduce notation and conventions. The $1/N_c$ expansion of the baryon operator whose matrix elements between baryon states yields the scattering amplitude in the limit of exact $SU(3)$ limit is constructed. The most complete form of this amplitude is obtained by accounting for the decuplet-octet baryon mass difference explicitly. In Sec.~\ref{sec:nucleonpion} results are particularized to the $N\pi$ system; some isospin relations are checked to be respected by the expressions obtained. In Sec.~\ref{sec:str} the analysis is applied to two process including strangeness only as case studies. In Sec.~\ref{sec:sb} the effects of first-order SB are evaluated; for this purpose, flavor projection operators are constructed and extensively used to rigorously identify components from different $SU(3)$ flavor representations participating in the breaking. First-order IB effects to the scattering amplitude are also evaluated. Violations to the isospin relations discussed in Sec.~\ref{sec:nucleonpion} are straightforwardly found. A comparison of nucleon-pion scattering amplitudes within this formalism and HBChPT are outlined in Sec.~\ref{sec:compa}. In Sec.~\ref{sec:sa} applications to scattering lengths are sketched. Some concluding remarks are given in Sec.~\ref{sec:cr}. In Appendix \ref{sec:appa} the baryon operator basis used in the scattering amplitude is listed. The paper is complemented by some supplementary material, loosely referred to as the Online Resource, which contains: 1) the reduction of the different baryon structures in terms of an operator basis of linearly independent operators; 2) the full list of the pertinent coefficients that accompany the baryon operators of Appendix \ref{sec:appa}; and 3) the operator basis used to evaluate SB effects along with their respective matrix elements listed in tables.

\section{\label{sec:tree}Baryon-meson scattering amplitude at leading and subleading orders}

\begin{figure}[h]
\centering
\includegraphics{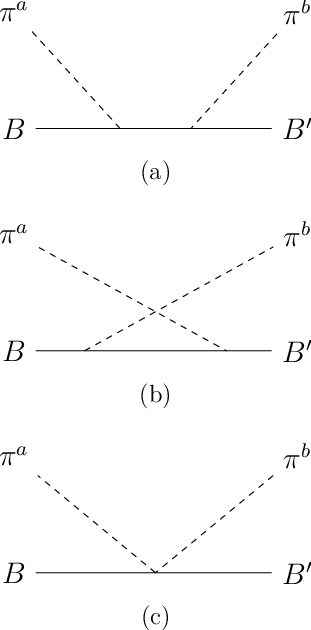}
\caption{\label{fig:sp}Leading-order diagrams for the scattering $B + \pi \to B^\prime + \pi$.}
\end{figure}

In this section, the analytical computation of the amplitude of baryon-meson scattering presented in Ref.~\cite{rfm00} is explicitly carried out, specialized to the process
\begin{equation}
B(p) + \pi^a(k) \to B^\prime(p^\prime) + \pi^b(k^\prime). \label{eq:scatt}
\end{equation}
The amplitude for baryon-meson scattering at fixed meson energy is dominated in the large-$N_c$ limit by the diagrams displayed in Fig.~\ref{fig:sp}. In Eq.~(\ref{eq:scatt}), $\pi$ denotes one of the nine pseudo scalar mesons $\pi$, $K$, $\eta$ and $\eta^\prime$ of momenta $k=(k^0,k^1,k^2,k^3)$ and $k^\prime=({k^\prime}^0,{k^\prime}^1,{k^\prime}^2,{k^\prime}^3)$ and flavors $a$ and $b$ for the incoming and outgoing mesons, respectively, and $B$ and $B^\prime$ denote the incoming and outgoing baryons of momenta $p$ and $p^\prime$, respectively. Soft mesons with energies of order unity are considered in the process. The goal is to explicitly evaluate the corresponding scattering amplitude at leading and subleading orders, incorporating the effects of the baryon mass splitting $\Delta$ defined in the previous section. Before tackling the problem, it is convenient to introduce some key concepts on large-$N_c$ QCD to set notation and conventions. Further details on the formalism can be found in Refs.~\cite{djm94,djm95}.

In the large-$N_c$ limit, the baryon sector has a contracted $SU(2N_f)$ spin-flavor symmetry, where $N_f$ is the number of light quark flavors. For $N_f=3$ the lowest lying baryon states fall into a representation of the spin-flavor group $SU(6)$. When $N_c=3$, this corresponds to the $\mathbf{56}$ dimensional representation of $SU(6)$.

The $1/N_c$ expansion of a QCD operator can be written in terms of $1/N_c$-suppressed operators with well-defined spin-flavor transformation properties. A complete set of operators can be constructed using the 0-body operator ${\mathcal{I}}$ and the 1-body operators
\begin{subequations}
\begin{eqnarray}
J^k & = & q^\dagger \left[\dfrac{\sigma^k}{2} \otimes {\mathcal{I}} \right]q, \qquad \qquad (1,1) \\
T^c & = & q^\dagger \left[{\mathcal{I}} \otimes \dfrac{\lambda^c}{2} \right]q, \qquad \qquad (0,8) \\
G^{kc} & = & q^\dagger \left[\dfrac{\sigma}{2} \otimes \dfrac{\lambda^c}{2} \right]q, \qquad \qquad (1,8)
\end{eqnarray}
\end{subequations}
where $J^k$, $T^c$ and $G^{kc}$ are the baryon spin, baryon flavor and baryon spin-flavor generators, respectively, which transform under $SU(2) \times SU(3)$ as $(j,\mathrm{dim})$, where $j$ is the spin and $\mathrm{dim}$ is the dimension of the $SU(3)$ flavor representation. The $SU(2N_f)$ spin-flavor generators satisfy well-known commutation relations \cite{djm95}.

The Feynman diagrams displayed in Fig.~\ref{fig:sp} will be analyzed separately as they contribute differently to the scattering process.

\subsection{\label{sec:figab} Scattering amplitude from Fig.~\ref{fig:sp}(a,b)}

The amplitude for the scattering process (\ref{eq:scatt}) represented in Fig.~\ref{fig:sp}(a,b), \textit{in the rest frame of the initial baryon}, can be represented by the baryon operator \cite{rfm00}
\begin{equation}
\mathcal{A}_\mathrm{LO}^{ab} = -\frac{1}{f^2} k^i {k^\prime}^j \left[ \frac{1}{k^0} \sum_{n=0}^\infty \frac{1}{{k^0}^n} [A^{jb},\underbrace{[\mathcal{M},[\mathcal{M},\ldots [\mathcal{M}}_{n \,\, \mathrm{insertions}},A^{ia}\underbrace{] \ldots ]]}] \right], \label{eq:atree}
\end{equation}
where $f\approx 93\, \mathrm{MeV}$ is the pion decay constant, $A^{ia}$ is the baryon axial vector current and $\mathcal{M}$ is the baryon mass operator. Explicitly, the $1/N_c$ expansion of $A^{ia}$, at $N_c=3$, is given by \cite{djm95}
\begin{equation}
A^{ia} = a_1 G^{ia} + \frac{1}{N_c} b_2 \mathcal{D}_2^{ia} + \frac{1}{N_c^2} b_3 \mathcal{D}_3^{ia} + \frac{1}{N_c^2} c_3 \mathcal{O}_3^{ia},
\end{equation}
where $a_1$, $b_2$, $b_3$, and $c_3$ are unknown coefficients of order one, and the $2$- and $3$-body operators $\mathcal{D}_2^{ia}$ and $\mathcal{D}_3^{ia}$ and $\mathcal{O}_3^{ia}$ read,
\begin{subequations}
\begin{eqnarray}
\mathcal{D}_2^{ia} & = & J^iT^a, \\
\mathcal{D}_3^{ia} & = & \{J^i,\{J^r,G^{ra}\}\}, \\
\mathcal{O}_3^{ia} & = & \{J^2,G^{ia}\} - \frac12 \{J^i,\{J^r,G^{ra}\}\}. \\
\end{eqnarray}
\end{subequations}

The baryon mass operator is expressed as \cite{djm95}
\begin{eqnarray}
\mathcal{M} = m_0 N_c {\mathcal{I}} + \sum_{n=2,4}^{N_c-1} m_{n} \frac{1}{N_c^{n-1}} J^n, \label{eq:mop}
\end{eqnarray}
where $m_n$ are unknown coefficients. While the first term on the right-hand side is the overall spin-independent mass of the baryon multiplet, the remaining terms are spin-dependent and make up $\mathcal{M}_{\text{hyperfine}}$. At $N_c=3$, $\mathcal{M}_{\text{hyperfine}}$ is simply
\begin{eqnarray}
\mathcal{M}_{\text{hyperfine}} = \frac{m_2}{N_c} J^2, \label{eq:smop}
\end{eqnarray}
where $m_2$ can be set to $\Delta$. Numerically, the average value is $\Delta=0.237$ GeV \cite{rfm24}.

The series (\ref{eq:atree}) with the first three summands reads,
\begin{equation}
\mathcal{A}_\mathrm{LO}^{ab} = - \frac{1}{f^2} k^i {k^\prime}^j \left[ \frac{1}{k^0}[A^{jb},A^{ia}] + \frac{1}{{k^0}^2} [A^{jb},[\mathcal{M},A^{ia}]] + \frac{1}{{k^0}^3} [A^{jb},[\mathcal{M},[\mathcal{M},A^{ia}]]] + \ldots \right]. \label{eq:atree3}
\end{equation}

The constraint that $\mathcal{A}_\mathrm{LO}^{ab}$ be at most $\mathcal{O}(1)$ in the large-$N_c$ limit sets the consistency conditions \cite{dm1,rfm00}
\begin{subequations}
\label{eq:cc}
\begin{align}
[A^{jb},A^{ia}] & \leq \mathcal{O}(N_c), \\
[A^{jb},[\mathcal{M},A^{ia}]] & \leq \mathcal{O}(N_c), \\
[A^{jb},[\mathcal{M},[\mathcal{M},A^{ia}]]] & \leq \mathcal{O}(N_c), \\
& \vdots \nonumber
\end{align}
\end{subequations}
where $k^0$, $f$, and $\Delta$ are orders $\mathcal{O}(1)$, $\mathcal{O}(\sqrt{N_c})$, and $\mathcal{O}(N_c^{-1})$ in that limit, respectively. Explicit analytical computations of the first three operator structures in (\ref{eq:cc}) is the aim of this work; results will be discussed in the following sections.

\subsubsection{Spin-flavor transformation properties of $\mathcal{A}_\mathrm{LO}^{ab}$}

The baryon operator $\mathcal{A}_\mathrm{LO}^{ab}$ is a spin-zero object and contains two adjoint (octet) indices. The tensor product of two adjoint representations $8 \otimes 8$ can be split into the symmetric product $(8 \otimes 8)_S$ and the antisymmetric product $(8 \otimes 8)_A$ \cite{djm95}, which in turn can be decomposed in terms of $SU(3)$ multiplets as
\begin{subequations}
\label{eq:su3deco}
\begin{eqnarray}
& & (8 \otimes 8)_S = 1 \oplus 8 \oplus 27, \\
& & (8 \otimes 8)_A = 8 \oplus 10 \oplus \overline{10}.
\end{eqnarray}
\end{subequations}

In order to exploit the transformation properties of $\mathcal{A}_\mathrm{LO}^{ab}$ under the $SU(2) \times SU(3)$ spin-flavor symmetry, the spin and flavor projectors introduced in Ref.~\cite{banda} become handy. In a few words, this technique exploits the decomposition of the tensor space formed by the product of the adjoint space with itself $n$ times, $\prod_{i=1}^n adj \otimes$, into subspaces which can be labeled by a specific eigenvalue of the quadratic Casimir operator $C$ of the Lie algebra of $SU(N)$. In particular, for the product of two $SU(3)$ adjoints, the flavor projectors $[\pr{\mathrm{dim}}]^{abcd}$ for the irreducible representation of dimension $\mathrm{dim}$ contained in (\ref{eq:su3deco}) are given by \cite{banda}
\begin{equation}
[\pr{1}]^{abcd} = \frac{1}{N_f^2-1} \delta^{ab} \delta^{cd},
\end{equation}
\begin{equation}
[\pr{8}]^{abcd} = \frac{N_f}{N_f^2-4} d^{abe} d^{cde},
\end{equation}
\begin{equation}
[\pr{27}]^{abcd} = \frac12 (\delta^{ac} \delta^{bd} + \delta^{bc} \delta^{ad}) - \frac{1}{N_f^2-1} \delta^{ab} \delta^{cd} - \frac{N_f}{N_f^2-4} d^{abe} d^{cde},
\end{equation}
\begin{equation}
[\pr{8_A}]^{abcd} = \frac{1}{N_f} f^{abe} f^{cde},
\end{equation}
and
\begin{equation}
[\pr{10+\overline{10}}]^{abcd} = \frac12 (\delta^{ac} \delta^{bd} - \delta^{bc} \delta^{ad}) - \frac{1}{N_f} f^{abe} f^{cde},
\end{equation}
which fulfill the completeness relation
\begin{equation}
[\pr{1} + \pr{8} + \pr{27} + \pr{8_A} + \pr{10+\overline{10}}]^{abcd} = \delta^{ac} \delta^{bd}.
\end{equation}

Therefore, $[\pr{\mathrm{dim}} \mathcal{A}_\mathrm{LO}]^{ab}$ effectively projects out the piece of $\mathcal{A}_\mathrm{LO}^{ab}$ that transforms under the flavor representation of dimension $\mathrm{dim}$ according to decompositions (\ref{eq:su3deco}). However, for computational purposes, it is more convenient to group the operators $[\pr{\mathrm{dim}} \mathcal{A}_\mathrm{LO}]^{ab}$ based on their symmetry transformation properties under the interchange of $a$ and $b$. Accordingly, $[\pr{1} + \pr{8} + \pr{27}]^{abcd}$ and $[\pr{8_A} + \pr{10+\overline{10}}]^{abcd}$, acting respectively on the symmetric and antisymmetric [antisymmetric and symmetric] pieces of $\mathcal{A}_\mathrm{LO}^{cd}$ will provide the symmetric [antisymmetric] piece of $\mathcal{A}_\mathrm{LO}^{ab}$ under the interchange of $a$ and $b$.

\subsubsection{Explicit form of the scattering amplitude}

A more specialized and detailed calculation beyond the qualitative analyses of baryon-meson scattering presented in previous works \cite{dm1,rfm00} can be performed by explicitly evaluating the first summands displayed in Eq.~(\ref{eq:atree3}); succinctly, all baryon operators allowed at $N_c=3$ ({\it i.e.}, up to relative order $1/N_c^2$ not discussed so far in the literature) are accounted for in the terms kept in the series (\ref{eq:atree3}).

As a starting point, it should be recalled that the commutator of an $n$-body operator and an $m$-body operator is an $(n+m-1)$-body operator, {\it i.e.},
\begin{equation}
[O_m,O_n] = O_{m+n-1}. \label{eq:comon}
\end{equation}
A close inspection of Eq.~(\ref{eq:atree3}) reveals that $[A^{jb},A^{ia}]$, for $N_c=3$, yields at most the operator structure $[\mathcal{O}_3^{jb},\mathcal{O}_3^{ia}]$ and, according to relation (\ref{eq:comon}), retains up to $5$-body operators. Sequential insertions of one and two $J^2$ operators add up $6$- and $7$-body operators, respectively. Thus, the first three summands displayed in Eq.~(\ref{eq:atree3}) will be explicitly evaluated and they will suffice to draw some conclusions. Clearly, it would be desirable to perform a calculation including 8-body operators and higher, but this is beyond the scope of this work due to the considerable amount of group theory involved.

The task now is to rewrite the baryon operators involved in $[\pr{\mathrm{dim}} \mathcal{A}_\mathrm{LO}]^{ab}$ in terms of a set of linearly independent operators, up to $7$-body operators. A convenient operator basis $S_m^{(ij)(ab)}$, for $m=1,\ldots,139$, is listed in Appendix \ref{sec:appa}. It is a straightforward albeit long and tedious exercise to compute those reductions. However, due to the length and unilluminating nature of the full expressions, only the symmetric and antisymmetric pieces rather than the individual results for each representation are listed in the Online Resource. In passing, it is straightforward to verify that the consistency conditions (\ref{eq:cc}) are fulfilled by all these reduced structures.

The matrix elements of $\mathcal{A}_\mathrm{LO}^{ab}$ given in Eq.~(\ref{eq:atree}) between $SU(6)$ baryon states, where mesons are labelled with flavors $a$ and $b$, yield the corresponding scattering amplitude, namely,
\begin{equation}
\mathcal{A}_\mathrm{LO}(B + \pi^a \to B^\prime + \pi^b) \equiv \langle \pi^b B^\prime|\mathcal{A}_\mathrm{LO}^{ab}|\pi^a B\rangle. \label{eq:mtxe}
\end{equation}
The flavors associated to mesons are conventionally given by $\Big\{\frac{1\mp i2}{\sqrt{2}},3,\frac{4\mp i5}{\sqrt{2}},
\frac{6-i7}{\sqrt{2}},\frac{6+i7}{\sqrt{2}},8\Big\}$ for $\{\pi^\pm,\pi^0,K^\pm,K^0,\overline{K}^0,\eta\}$, respectively.\footnote{For simplicity only the octet of mesons is considered. Extending the analysis to include the $\eta^\prime$ is straightforward by using the baryon axial vector current $A^i\equiv A^{i9}$, which is written in terms of the 1-body operators $G^{i9} = \frac{1}{\sqrt{6}} J^i$ and $T^9 = \frac{1}{\sqrt{6}} N_c{\mathcal{I}}$ \cite{jen96}.} For instance, an expressions such as $\mathcal{A}_\mathrm{LO}(p + \pi^- \to n + \pi^0)$ should be understood as $\langle \pi^0 n|\mathcal{A}_\mathrm{LO}^{13} + i \mathcal{A}_\mathrm{LO}^{23}|\pi^- p\rangle/\sqrt{2}$.

Thus, with the operator reductions listed in the Online Resource, the scattering amplitude for process (\ref{eq:scatt}), arising from Fig.~\ref{fig:sp}(a,b), can be organized as
\begin{equation}
\mathcal{A}_\mathrm{LO}(B + \pi^a \to B^\prime + \pi^b) = -\frac{1}{f^2 k^0} \sum_{m=1}^{139} (c^{(\mathrm{s})}_m + c^{(\mathrm{a})}_m) k^i{k^\prime}^j\langle \pi^b B^\prime|S_m^{(ij)(ab)}|\pi^a B\rangle, \label{eq:sam1}
\end{equation}
where $S_m^{(ij)(ab)}$ constitute a basis of linearly independent spin-2 baryon operators with two adjoint indices and $c^{(\mathrm{s})}_m$ and $c^{(\mathrm{a})}_m$ are well-defined coefficients which come along with the symmetric and antisymmetric pieces of $\mathcal{A}_\mathrm{LO}^{ab}$; these coefficients are listed in the Online Resource too. Notice that in Eq.~(\ref{eq:sam1}) the sum over spin indices is implicit.

\subsection{\label{sec:figc} Scattering amplitude from Fig.~\ref{fig:sp}(c)}

The 2-meson-baryon-baryon contact interaction represented in Fig.~\ref{fig:sp}(c) contributes to the baryon-meson scattering amplitude with a term \cite{rfm00}
\begin{equation}
\mathcal{A}_\mathrm{vertex}^{ab} = -\frac{1}{2f^2}(2k^0+M-M^\prime)if^{abc}T^c, \label{eq:vtx}
\end{equation}
where $M$ and $M^\prime$ denote the masses of the initial and final baryons, respectively. Since $\mathcal{A}_\mathrm{vertex}^{ab}$ is already antisymmetric under the interchange of $a$ and $b$, the only term that remains once the projection operators are applied is $[\pr{8_A} A_\mathrm{vertex}]^{ab}$, so this term only contributes to the octet piece. $\mathcal{A}_\mathrm{vertex}^{ab}$ is also order $\mathcal{O}(1)$, so together with $\mathcal{A}_\mathrm{LO}^{ab}$ Eq.~(\ref{eq:atree}), they both yield the leading order $\mathcal{O}(1)$ scattering amplitude for baryons with spin $J\sim \mathcal{O}(1)$.

\section{\label{sec:nucleonpion}Application: $N\pi \to N\pi$ scattering amplitude}

The formalism presented so far can be implemented to study scattering processes of the form $B+\pi^a\to B^\prime+\pi^b$ provided that
reactions in which these particles are produced have equal total strangeness on each side, according to the Gell-Mann--Nishijima scheme. Since $B$ and $B^\prime$ can be either octet or decuplet baryons, from the theoretical point of view the possibilities are numerous; examples are $\Lambda + K^+ \to p + \pi^0$, ${\Xi^*}^- + K^+ \to {\Sigma^*}^0 + \pi^0$, ${\Xi^*}^- + K^0 \to \Sigma^- + \pi^0$, and so on. For definiteness, the $N\pi\to N\pi$ scattering processes will be analyzed to exemplify the approach.

A pion, $I=1$, and a nucleon, $I=\frac12$, can be combined in a $I=\frac32$ or a $I=\frac12$ state following the usual addition rules of angular momenta \cite{add}. The allowed states for the $N\pi$ system are listed in Table \ref{t:isos2}.
\begin{table}[h]
\caption{Allowed states for $N\pi$ system}\label{t:isos2}
\begin{center}
\begin{tabular}{lcc}
\hline
& $I = \frac32$ & $I = \frac12$ \\
\hline
$I_3 = +\frac32$ & $|\pi^+p \rangle$ & \\
$I_3 = +\frac12$ & $\sqrt{\frac13}|\pi^+n \rangle + \sqrt{\frac23}|\pi^0p \rangle$ & $\sqrt{\frac23}|\pi^+n \rangle -\sqrt{\frac13}|\pi^0p\rangle$ \\
$I_3 = -\frac12$ & $\sqrt{\frac23}|\pi^0n \rangle + \sqrt{\frac13}|\pi^-p \rangle$ & $\sqrt{\frac13}|\pi^0n \rangle -\sqrt{\frac23}|\pi^-p\rangle$ \\
$I_3 = -\frac32$ & $|\pi^-p \rangle$ & \\
\hline
\end{tabular}
\end{center}
\end{table}

The elastic scattering amplitude for the process (\ref{eq:scatt}) can be decomposed, by the usual Clebsch-Gordan technique, into two non-interfering amplitudes, $\mathcal{A}^{(T)}$, with $I=\frac32$ and $I=\frac12$. Thus, starting from the $s$-channel isospin eigenstates
\begin{eqnarray}
& &|\pi^+p \rangle = \left|\frac32,+\frac32 \right\rangle, \\
& &|\pi^+n \rangle = \sqrt{\frac13} \left|\frac32,+\frac12 \right\rangle +\sqrt{\frac23} \left|\frac12,+\frac12 \right\rangle, \\
& &|\pi^0p \rangle = \sqrt{\frac23} \left|\frac32,+\frac12 \right\rangle -\sqrt{\frac13} \left|\frac12,+\frac12 \right\rangle, \\
& &|\pi^0n \rangle = \sqrt{\frac23} \left|\frac32,-\frac12 \right\rangle +\sqrt{\frac13} \left|\frac12,-\frac12 \right\rangle, \\
& &|\pi^-p \rangle = \sqrt{\frac13} \left|\frac32,-\frac12 \right\rangle -\sqrt{\frac23} \left|\frac12,-\frac12 \right\rangle, \\
& &|\pi^-n \rangle = \left|\frac32,-\frac32 \right\rangle,
\end{eqnarray}
it is straightforward to obtain \cite{ditsche}
\begin{eqnarray}
& & \mathcal{A}_\mathrm{LO}(p + \pi^+ \to p + \pi^+) = \mathcal{A}_\mathrm{LO}(n + \pi^- \to n + \pi^-) = \mathcal{A}^{(3/2)}, \nonumber \\
& & \mathcal{A}_\mathrm{LO}(p + \pi^- \to p + \pi^-) = \mathcal{A}_\mathrm{LO}(n + \pi^+ \to n + \pi^+) = \frac13 \mathcal{A}^{(3/2)} + \frac23 \mathcal{A}^{(1/2)}, \nonumber \\
& & \mathcal{A}_\mathrm{LO}(p + \pi^0 \to p + \pi^0) = \mathcal{A}_\mathrm{LO}(n + \pi^0 \to n + \pi^0) = \frac23 \mathcal{A}^{(3/2)} + \frac13 \mathcal{A}^{(1/2)}, \nonumber \\
& &\sqrt{2} \mathcal{A}_\mathrm{LO}(p + \pi^- \to n + \pi^0) =\sqrt{2} \mathcal{A}_\mathrm{LO}(n + \pi^+ \to p + \pi^0) = \frac23 \mathcal{A}^{(3/2)} - \frac23 \mathcal{A}^{(1/2)}. \nonumber \\
\end{eqnarray}

Furthermore, an alternative set of invariant amplitudes, $\mathcal{A}^{(+)}$ and $\mathcal{A}^{(-)}$, can be introduced for the $N\pi$ system, which are defined as \cite{moo}
\begin{equation}
\mathcal{A}^{(+)} = \frac23 \mathcal{A}^{(3/2)} + \frac13 \mathcal{A}^{(1/2)}, \qquad \mathcal{A}^{(-)} = - \frac13 \mathcal{A}^{(3/2)} + \frac13 \mathcal{A}^{(1/2)},
\end{equation}
so
\begin{equation}
\mathcal{A}^{(3/2)} = \mathcal{A}^{(+)} - \mathcal{A}^{(-)}, \qquad \mathcal{A}^{(1/2)} = \mathcal{A}^{(+)} + 2 \mathcal{A}^{(-)}.
\end{equation}

The non-trivial matrix elements $k^i{k^\prime}^j \langle \pi^b B^\prime|S_r^{(ij)(ab)}|\pi^a B\rangle$ are displayed in Tables \ref{t:psca} and \ref{t:nsca} for proton-pion and neutron-pion processes ($N\pi\to N\pi$ processes for short), respectively.\footnote{Here, non-trivial matrix elements are those which are either zero or obtained as anticommutators with $J^2$. For instance, for the $N\pi$ system, $k^i{k^\prime}^j\langle \pi^b B^\prime|S_{17}^{(ij)(ab)}|\pi^a B\rangle$ vanishes whereas $k^i{k^\prime}^j\langle \pi^b B^\prime|S_{37}^{(ij)(ab)}|\pi^a B\rangle = \frac32 k^i{k^\prime}^j\langle \pi^b B^\prime|S_{15}^{(ij)(ab)}|\pi^a B\rangle$, so they are not listed.} It can be easily verified that the symmetric and antisymmetric pieces of $\mathcal{A}_\mathrm{LO}(B+\pi^a\to B^\prime+\pi^b)$ are respectively proportional to $\mathbf{k} \cdot \mathbf{k}^\prime$ and the third component of $i(\mathbf{k} \times \mathbf{k}^\prime)$, which will be denoted hereafter by $i(\mathbf{k} \times \mathbf{k}^\prime)_3$. The latter can also be rewritten as $i \epsilon^{ij3}k^i{k^\prime}^j=i(k^1{k^\prime}^2-k^2{k^\prime}^1)$.

\begin{table}[h]
\caption{Non-trivial matrix elements of operators involved in proton-pion scattering processes}\label{t:psca}
\begin{center}
{\small
\begin{tabular}{lcccc}
\hline
Operator & $p + \pi^+ \to p + \pi^+$ & $p + \pi^- \to p + \pi^-$ & $p + \pi^0 \to p + \pi^0$ & $p + \pi^- \to n + \pi^0$ \\
\hline
$k^i{k^\prime}^j \langle S_{1}^{(ij)(ab)} \rangle$ & $-\mathbf{k} \cdot \mathbf{k}^\prime$ & $\mathbf{k} \cdot \mathbf{k}^\prime$ & $0$ & $-\mathbf{k} \cdot \mathbf{k}^\prime$ \\
$k^i{k^\prime}^j \langle S_{2}^{(ij)(ab)} \rangle$ & $ i(\mathbf{k} \times \mathbf{k}^\prime)_3$ & $ i(\mathbf{k} \times \mathbf{k}^\prime)_3$ & $\frac12 i(\mathbf{k} \times \mathbf{k}^\prime)_3$ & $0$ \\
$k^i{k^\prime}^j \langle S_{3}^{(ij)(ab)} \rangle$ & $\frac16 i(\mathbf{k} \times \mathbf{k}^\prime)_3$ & $\frac16 i(\mathbf{k} \times \mathbf{k}^\prime)_3$ & $\frac{1}{12} i(\mathbf{k} \times \mathbf{k}^\prime)_3$ & $0$ \\
$k^i{k^\prime}^j \langle S_{4}^{(ij)(ab)} \rangle$ & $\mathbf{k} \cdot \mathbf{k}^\prime$ & $\mathbf{k} \cdot \mathbf{k}^\prime$ & $\frac12 \mathbf{k} \cdot \mathbf{k}^\prime$ & $0$ \\
$k^i{k^\prime}^j \langle S_{5}^{(ij)(ab)} \rangle$ & $\frac32 \mathbf{k} \cdot \mathbf{k}^\prime$ & $\frac32 \mathbf{k} \cdot \mathbf{k}^\prime$ & $\frac34 \mathbf{k} \cdot \mathbf{k}^\prime$ & $0$ \\
$k^i{k^\prime}^j \langle S_{6}^{(ij)(ab)} \rangle$ & $\frac{19}{12}\mathbf{k} \cdot \mathbf{k}^\prime-\frac{11}{12}i(\mathbf{k} \times \mathbf{k}^\prime)_3$ & $\frac{19}{12}\mathbf{k} \cdot \mathbf{k}^\prime+\frac{11}{12}i(\mathbf{k} \times \mathbf{k}^\prime)_3$ & $\frac{19}{24}\mathbf{k} \cdot \mathbf{k}^\prime$ & $-\frac{11}{12}i(\mathbf{k} \times \mathbf{k}^\prime)_3$ \\
$k^i{k^\prime}^j \langle S_{7}^{(ij)(ab)} \rangle$ & $\frac{19}{12}\mathbf{k} \cdot \mathbf{k}^\prime+\frac{11}{12}i(\mathbf{k} \times \mathbf{k}^\prime)_3$ & $\frac{19}{12}\mathbf{k} \cdot \mathbf{k}^\prime-\frac{11}{12}i(\mathbf{k} \times \mathbf{k}^\prime)_3$ & $\frac{19}{24}\mathbf{k} \cdot \mathbf{k}^\prime$ & $\frac{11}{12}i(\mathbf{k} \times \mathbf{k}^\prime)_3$ \\
$k^i{k^\prime}^j \langle S_{8}^{(ij)(ab)} \rangle$ & $\frac{19}{4}\mathbf{k} \cdot \mathbf{k}^\prime$ & $\frac{19}{4}\mathbf{k} \cdot \mathbf{k}^\prime$ & $\frac{19}{8}\mathbf{k} \cdot \mathbf{k}^\prime$ & $0$ \\
$k^i{k^\prime}^j \langle S_{9}^{(ij)(ab)} \rangle$ & $\frac56 i(\mathbf{k} \times \mathbf{k}^\prime)_3$ & $\frac56 i(\mathbf{k} \times \mathbf{k}^\prime)_3$ & $\frac{5}{12} i(\mathbf{k} \times \mathbf{k}^\prime)_3$ & $0$ \\
$k^i{k^\prime}^j \langle S_{10}^{(ij)(ab)} \rangle$ & $\frac56 i(\mathbf{k} \times \mathbf{k}^\prime)_3$ & $\frac56 i(\mathbf{k} \times \mathbf{k}^\prime)_3$ & $\frac{5}{12} i(\mathbf{k} \times \mathbf{k}^\prime)_3$ & $0$ \\
$k^i{k^\prime}^j \langle S_{11}^{(ij)(ab)} \rangle$ & $\frac16 \mathbf{k} \cdot \mathbf{k}^\prime$ & $\frac16 \mathbf{k} \cdot \mathbf{k}^\prime$ & $\frac{1}{12}\mathbf{k} \cdot \mathbf{k}^\prime$ & $0$ \\
$k^i{k^\prime}^j \langle S_{12}^{(ij)(ab)} \rangle$ & $-\frac56 \mathbf{k} \cdot \mathbf{k}^\prime$ & $\frac56 \mathbf{k} \cdot \mathbf{k}^\prime$ & $0$ & $-\frac56 \mathbf{k} \cdot \mathbf{k}^\prime$ \\
$k^i{k^\prime}^j \langle S_{13}^{(ij)(ab)} \rangle$ & $-\frac56 \mathbf{k} \cdot \mathbf{k}^\prime$ & $\frac56 \mathbf{k} \cdot \mathbf{k}^\prime$ & $0$ & $-\frac56 \mathbf{k} \cdot \mathbf{k}^\prime$ \\
$k^i{k^\prime}^j \langle S_{14}^{(ij)(ab)} \rangle$ & $\frac12 \mathbf{k} \cdot \mathbf{k}^\prime$ & $\frac12 \mathbf{k} \cdot \mathbf{k}^\prime$ & $\frac14 \mathbf{k} \cdot \mathbf{k}^\prime$ & $0$ \\
$k^i{k^\prime}^j \langle S_{15}^{(ij)(ab)} \rangle$ & $\frac12 i(\mathbf{k} \times \mathbf{k}^\prime)_3$ & $-\frac12 i(\mathbf{k} \times \mathbf{k}^\prime)_3$ & $0$ & $\frac12 i(\mathbf{k} \times \mathbf{k}^\prime)_3$ \\
$k^i{k^\prime}^j \langle S_{16}^{(ij)(ab)} \rangle$ & $\frac12 i(\mathbf{k} \times \mathbf{k}^\prime)_3$ & $\frac12 i(\mathbf{k} \times \mathbf{k}^\prime)_3$ & $\frac14 i(\mathbf{k} \times \mathbf{k}^\prime)_3$ & $0$ \\
$k^i{k^\prime}^j \langle S_{18}^{(ij)(ab)} \rangle$ & $ i(\mathbf{k} \times \mathbf{k}^\prime)_3$ & $ i(\mathbf{k} \times \mathbf{k}^\prime)_3$ & $\frac12 i(\mathbf{k} \times \mathbf{k}^\prime)_3$ & $0$ \\
$k^i{k^\prime}^j \langle S_{19}^{(ij)(ab)} \rangle$ & $\frac{19}{4} i(\mathbf{k} \times \mathbf{k}^\prime)_3$ & $\frac{19}{4} i(\mathbf{k} \times \mathbf{k}^\prime)_3$ & $\frac{19}{8} i(\mathbf{k} \times \mathbf{k}^\prime)_3$ & $0$ \\
$k^i{k^\prime}^j \langle S_{20}^{(ij)(ab)} \rangle$ & $-\mathbf{k} \cdot \mathbf{k}^\prime$ & $\mathbf{k} \cdot \mathbf{k}^\prime$ & $0$ & $-\mathbf{k} \cdot \mathbf{k}^\prime$ \\
$k^i{k^\prime}^j \langle S_{23}^{(ij)(ab)} \rangle$ & $-\frac43 \mathbf{k} \cdot \mathbf{k}^\prime-\frac43 i(\mathbf{k} \times \mathbf{k}^\prime)_3$ & $\frac43 \mathbf{k} \cdot \mathbf{k}^\prime-\frac43 i(\mathbf{k} \times \mathbf{k}^\prime)_3$ & $-\frac23 i(\mathbf{k} \times \mathbf{k}^\prime)_3$ & $-\frac43 \mathbf{k} \cdot \mathbf{k}^\prime$ \\
$k^i{k^\prime}^j \langle S_{24}^{(ij)(ab)} \rangle$ & $-\frac43 \mathbf{k} \cdot \mathbf{k}^\prime+\frac43 i(\mathbf{k} \times \mathbf{k}^\prime)_3$ & $\frac43 \mathbf{k} \cdot \mathbf{k}^\prime+\frac43 i(\mathbf{k} \times \mathbf{k}^\prime)_3$ & $\frac23 i(\mathbf{k} \times \mathbf{k}^\prime)_3$ & $-\frac43 \mathbf{k} \cdot \mathbf{k}^\prime$ \\
$k^i{k^\prime}^j \langle S_{25}^{(ij)(ab)} \rangle$ & $\frac43 \mathbf{k} \cdot \mathbf{k}^\prime-\frac43 i(\mathbf{k} \times \mathbf{k}^\prime)_3$ & $-\frac43 \mathbf{k} \cdot \mathbf{k}^\prime-\frac43 i(\mathbf{k} \times \mathbf{k}^\prime)_3$ & $-\frac23 i(\mathbf{k} \times \mathbf{k}^\prime)_3$ & $\frac43 \mathbf{k} \cdot \mathbf{k}^\prime$ \\
$k^i{k^\prime}^j \langle S_{26}^{(ij)(ab)} \rangle$ & $\frac43 \mathbf{k} \cdot \mathbf{k}^\prime+\frac43 i(\mathbf{k} \times \mathbf{k}^\prime)_3$ & $-\frac43 \mathbf{k} \cdot \mathbf{k}^\prime+\frac43 i(\mathbf{k} \times \mathbf{k}^\prime)_3$ & $\frac23 i(\mathbf{k} \times \mathbf{k}^\prime)_3$ & $\frac43 \mathbf{k} \cdot \mathbf{k}^\prime$ \\
$k^i{k^\prime}^j \langle S_{27}^{(ij)(ab)} \rangle$ & $\frac{25}{12} i(\mathbf{k} \times \mathbf{k}^\prime)_3$ & $\frac{25}{12} i(\mathbf{k} \times \mathbf{k}^\prime)_3$ & $\frac{25}{24}i(\mathbf{k} \times \mathbf{k}^\prime)_3$ & $0$ \\
$k^i{k^\prime}^j \langle S_{28}^{(ij)(ab)} \rangle$ & $\frac{25}{12} i(\mathbf{k} \times \mathbf{k}^\prime)_3$ & $\frac{25}{12} i(\mathbf{k} \times \mathbf{k}^\prime)_3$ & $\frac{25}{24}i(\mathbf{k} \times \mathbf{k}^\prime)_3$ & $0$ \\
$k^i{k^\prime}^j \langle S_{29}^{(ij)(ab)} \rangle$ & $-\frac54 \mathbf{k} \cdot \mathbf{k}^\prime$ & $\frac54 \mathbf{k} \cdot \mathbf{k}^\prime$ & $0$ & $-\frac54 \mathbf{k} \cdot \mathbf{k}^\prime$ \\
$k^i{k^\prime}^j \langle S_{30}^{(ij)(ab)} \rangle$ & $\frac54 \mathbf{k} \cdot \mathbf{k}^\prime$ & $-\frac54 \mathbf{k} \cdot \mathbf{k}^\prime$ & $0$ & $\frac54 \mathbf{k} \cdot \mathbf{k}^\prime$ \\
$k^i{k^\prime}^j \langle S_{38}^{(ij)(ab)} \rangle$ & $-\frac52 \mathbf{k} \cdot \mathbf{k}^\prime$ & $\frac52 \mathbf{k} \cdot \mathbf{k}^\prime$ & $0$ & $-\frac52 \mathbf{k} \cdot \mathbf{k}^\prime$ \\
$k^i{k^\prime}^j \langle S_{39}^{(ij)(ab)} \rangle$ & $\frac54 i(\mathbf{k} \times \mathbf{k}^\prime)_3$ & $\frac54 i(\mathbf{k} \times \mathbf{k}^\prime)_3$ & $\frac58 i(\mathbf{k} \times \mathbf{k}^\prime)_3$ & $0$ \\
$k^i{k^\prime}^j \langle S_{40}^{(ij)(ab)} \rangle$ & $\frac54 i(\mathbf{k} \times \mathbf{k}^\prime)_3$ & $\frac54 i(\mathbf{k} \times \mathbf{k}^\prime)_3$ & $\frac58 i(\mathbf{k} \times \mathbf{k}^\prime)_3$ & $0$ \\
$k^i{k^\prime}^j \langle S_{47}^{(ij)(ab)} \rangle$ & $\frac12 \mathbf{k} \cdot \mathbf{k}^\prime$ & $\frac12 \mathbf{k} \cdot \mathbf{k}^\prime$ & $\frac14 \mathbf{k} \cdot \mathbf{k}^\prime$ & $0$ \\
$k^i{k^\prime}^j \langle S_{49}^{(ij)(ab)} \rangle$ & $4 i(\mathbf{k} \times \mathbf{k}^\prime)_3$ & $-4 i(\mathbf{k} \times \mathbf{k}^\prime)_3$ & $0$ & $4 i(\mathbf{k} \times \mathbf{k}^\prime)_3$ \\
$k^i{k^\prime}^j \langle S_{54}^{(ij)(ab)} \rangle$ & $4\mathbf{k} \cdot \mathbf{k}^\prime-i(\mathbf{k} \times \mathbf{k}^\prime)_3$ & $4\mathbf{k} \cdot \mathbf{k}^\prime+i(\mathbf{k} \times \mathbf{k}^\prime)_3$ & $2\mathbf{k} \cdot \mathbf{k}^\prime$ & $-i(\mathbf{k} \times \mathbf{k}^\prime)_3$ \\
$k^i{k^\prime}^j \langle S_{56}^{(ij)(ab)} \rangle$ & $\frac{25}{8} i(\mathbf{k} \times \mathbf{k}^\prime)_3$ & $-\frac{25}{8} i(\mathbf{k} \times \mathbf{k}^\prime)_3$ & $0$ & $\frac{25}{8} i(\mathbf{k} \times \mathbf{k}^\prime)_3$ \\
$k^i{k^\prime}^j \langle S_{60}^{(ij)(ab)} \rangle$ & $-8\mathbf{k} \cdot \mathbf{k}^\prime+2 i(\mathbf{k} \times \mathbf{k}^\prime)_3$ & $-8\mathbf{k} \cdot \mathbf{k}^\prime-2 i(\mathbf{k} \times \mathbf{k}^\prime)_3$ & $-4\mathbf{k} \cdot \mathbf{k}^\prime$ & $2 i(\mathbf{k} \times \mathbf{k}^\prime)_3$ \\
$k^i{k^\prime}^j \langle S_{67}^{(ij)(ab)} \rangle$ & $\frac54 \mathbf{k} \cdot \mathbf{k}^\prime$ & $\frac54 \mathbf{k} \cdot \mathbf{k}^\prime$ & $\frac58\mathbf{k} \cdot \mathbf{k}^\prime$ & $0$ \\
$k^i{k^\prime}^j \langle S_{68}^{(ij)(ab)} \rangle$ & $\frac54 \mathbf{k} \cdot \mathbf{k}^\prime$ & $\frac54 \mathbf{k} \cdot \mathbf{k}^\prime$ & $\frac58\mathbf{k} \cdot \mathbf{k}^\prime$ & $0$ \\
$k^i{k^\prime}^j \langle S_{83}^{(ij)(ab)} \rangle$ & $\frac{25}{4}i(\mathbf{k} \times \mathbf{k}^\prime)_3$ & $\frac{25}{4}i(\mathbf{k} \times \mathbf{k}^\prime)_3$ & $\frac{25}{8} i(\mathbf{k} \times \mathbf{k}^\prime)_3$ & $0$ \\
$k^i{k^\prime}^j \langle S_{84}^{(ij)(ab)} \rangle$ & $\frac52 \mathbf{k} \cdot \mathbf{k}^\prime$ & $\frac52 \mathbf{k} \cdot \mathbf{k}^\prime$ & $\frac54 \mathbf{k} \cdot \mathbf{k}^\prime$ & $0$ \\
$k^i{k^\prime}^j \langle S_{85}^{(ij)(ab)} \rangle$ & $\frac52 \mathbf{k} \cdot \mathbf{k}^\prime$ & $\frac52 \mathbf{k} \cdot \mathbf{k}^\prime$ & $\frac54 \mathbf{k} \cdot \mathbf{k}^\prime$ & $0$ \\
$k^i{k^\prime}^j \langle S_{86}^{(ij)(ab)} \rangle$ & $4\mathbf{k} \cdot \mathbf{k}^\prime$ & $-4\mathbf{k} \cdot \mathbf{k}^\prime$ & $0$ & $4\mathbf{k} \cdot \mathbf{k}^\prime$ \\
$k^i{k^\prime}^j \langle S_{91}^{(ij)(ab)} \rangle$ & $6\mathbf{k} \cdot \mathbf{k}^\prime-6 i(\mathbf{k} \times \mathbf{k}^\prime)_3$ & $-6\mathbf{k} \cdot \mathbf{k}^\prime-6 i(\mathbf{k} \times \mathbf{k}^\prime)_3$ & $-3i(\mathbf{k} \times \mathbf{k}^\prime)_3$ & $6\mathbf{k} \cdot \mathbf{k}^\prime$ \\
\hline
\end{tabular}
}
\end{center}
\end{table}

\begin{table}[h]
\caption{\label{t:nsca}Non-trivial matrix elements of operators involved in neutron-pion scattering processes}
\begin{center}
{\small
\begin{tabular}{lcccc}
\hline
Operator & $n + \pi^+ \to n + \pi^+$ & $n + \pi^- \to n + \pi^-$ & $n + \pi^0 \to n + \pi^0$ & $n + \pi^+ \to p + \pi^0$ \\
\hline
$k^i{k^\prime}^j \langle S_{1}^{(ij)(ab)} \rangle$ & $\mathbf{k} \cdot \mathbf{k}^\prime$ & $-\mathbf{k} \cdot \mathbf{k}^\prime$ & $0$ & $-\mathbf{k} \cdot \mathbf{k}^\prime$ \\
$k^i{k^\prime}^j \langle S_{2}^{(ij)(ab)} \rangle$ & $ i(\mathbf{k} \times \mathbf{k}^\prime)_3$ & $ i(\mathbf{k} \times \mathbf{k}^\prime)_3$ & $-\frac12 i(\mathbf{k} \times \mathbf{k}^\prime)_3$ & $0$ \\
$k^i{k^\prime}^j \langle S_{3}^{(ij)(ab)} \rangle$ & $\frac16 i(\mathbf{k} \times \mathbf{k}^\prime)_3$ & $\frac16 i(\mathbf{k} \times \mathbf{k}^\prime)_3$ & $-\frac{1}{12} i(\mathbf{k} \times \mathbf{k}^\prime)_3$ & $0$ \\
$k^i{k^\prime}^j \langle S_{4}^{(ij)(ab)} \rangle$ & $\mathbf{k} \cdot \mathbf{k}^\prime$ & $\mathbf{k} \cdot \mathbf{k}^\prime$ & $\frac12 \mathbf{k} \cdot \mathbf{k}^\prime$ & $0$ \\
$k^i{k^\prime}^j \langle S_{5}^{(ij)(ab)} \rangle$ & $\frac32 \mathbf{k} \cdot \mathbf{k}^\prime$ & $\frac32 \mathbf{k} \cdot \mathbf{k}^\prime$ & $\frac34 \mathbf{k} \cdot \mathbf{k}^\prime$ & $0$ \\
$k^i{k^\prime}^j \langle S_{6}^{(ij)(ab)} \rangle$ & $\frac{19}{12}\mathbf{k} \cdot \mathbf{k}^\prime+\frac{11}{12}i(\mathbf{k} \times \mathbf{k}^\prime)_3$ & $\frac{19}{12}\mathbf{k} \cdot \mathbf{k}^\prime-\frac{11}{12}i(\mathbf{k} \times \mathbf{k}^\prime)_3$ & $\frac{19}{24}\mathbf{k} \cdot \mathbf{k}^\prime$ & $-\frac{11}{12}i(\mathbf{k} \times \mathbf{k}^\prime)_3$ \\
$k^i{k^\prime}^j \langle S_{7}^{(ij)(ab)} \rangle$ & $\frac{19}{12}\mathbf{k} \cdot \mathbf{k}^\prime-\frac{11}{12}i(\mathbf{k} \times \mathbf{k}^\prime)_3$ & $\frac{19}{12}\mathbf{k} \cdot \mathbf{k}^\prime+\frac{11}{12}i(\mathbf{k} \times \mathbf{k}^\prime)_3$ & $\frac{19}{24}\mathbf{k} \cdot \mathbf{k}^\prime$ & $\frac{11}{12}i(\mathbf{k} \times \mathbf{k}^\prime)_3$ \\
$k^i{k^\prime}^j \langle S_{8}^{(ij)(ab)} \rangle$ & $\frac{19}{4}\mathbf{k} \cdot \mathbf{k}^\prime$ & $\frac{19}{4}\mathbf{k} \cdot \mathbf{k}^\prime$ & $-\frac{19}{8}\mathbf{k} \cdot \mathbf{k}^\prime$ & $0$ \\
$k^i{k^\prime}^j \langle S_{9}^{(ij)(ab)} \rangle$ & $\frac56 i(\mathbf{k} \times \mathbf{k}^\prime)_3$ & $\frac56 i(\mathbf{k} \times \mathbf{k}^\prime)_3$ & $\frac{5}{12} i(\mathbf{k} \times \mathbf{k}^\prime)_3$ & $0$ \\
$k^i{k^\prime}^j \langle S_{10}^{(ij)(ab)} \rangle$ & $\frac56 i(\mathbf{k} \times \mathbf{k}^\prime)_3$ & $\frac56 i(\mathbf{k} \times \mathbf{k}^\prime)_3$ & $\frac{5}{12} i(\mathbf{k} \times \mathbf{k}^\prime)_3$ & $0$ \\
$k^i{k^\prime}^j \langle S_{11}^{(ij)(ab)} \rangle$ & $\frac16 \mathbf{k} \cdot \mathbf{k}^\prime$ & $\frac16 \mathbf{k} \cdot \mathbf{k}^\prime$ & $\frac{1}{12}\mathbf{k} \cdot \mathbf{k}^\prime$ & $0$ \\
$k^i{k^\prime}^j \langle S_{12}^{(ij)(ab)} \rangle$ & $\frac56 \mathbf{k} \cdot \mathbf{k}^\prime$ & $-\frac56 \mathbf{k} \cdot \mathbf{k}^\prime$ & $0$ & $-\frac56 \mathbf{k} \cdot \mathbf{k}^\prime$ \\
$k^i{k^\prime}^j \langle S_{13}^{(ij)(ab)} \rangle$ & $\frac56 \mathbf{k} \cdot \mathbf{k}^\prime$ & $-\frac56 \mathbf{k} \cdot \mathbf{k}^\prime$ & $0$ & $-\frac56 \mathbf{k} \cdot \mathbf{k}^\prime$ \\
$k^i{k^\prime}^j \langle S_{14}^{(ij)(ab)} \rangle$ & $\frac12 \mathbf{k} \cdot \mathbf{k}^\prime$ & $\frac12 \mathbf{k} \cdot \mathbf{k}^\prime$ & $\frac14 \mathbf{k} \cdot \mathbf{k}^\prime$ & $0$ \\
$k^i{k^\prime}^j \langle S_{15}^{(ij)(ab)} \rangle$ & $-\frac12 i(\mathbf{k} \times \mathbf{k}^\prime)_3$ & $\frac12 i(\mathbf{k} \times \mathbf{k}^\prime)_3$ & $0$ & $\frac12 i(\mathbf{k} \times \mathbf{k}^\prime)_3$ \\
$k^i{k^\prime}^j \langle S_{16}^{(ij)(ab)} \rangle$ & $\frac12 i(\mathbf{k} \times \mathbf{k}^\prime)_3$ & $\frac12 i(\mathbf{k} \times \mathbf{k}^\prime)_3$ & $\frac14 i(\mathbf{k} \times \mathbf{k}^\prime)_3$ & $0$ \\
$k^i{k^\prime}^j \langle S_{18}^{(ij)(ab)} \rangle$ & $ i(\mathbf{k} \times \mathbf{k}^\prime)_3$ & $ i(\mathbf{k} \times \mathbf{k}^\prime)_3$ & $\frac12 i(\mathbf{k} \times \mathbf{k}^\prime)_3$ & $0$ \\
$k^i{k^\prime}^j \langle S_{19}^{(ij)(ab)} \rangle$ & $\frac{19}{4} i(\mathbf{k} \times \mathbf{k}^\prime)_3$ & $\frac{19}{4} i(\mathbf{k} \times \mathbf{k}^\prime)_3$ & $\frac{19}{8} i(\mathbf{k} \times \mathbf{k}^\prime)_3$ & $0$ \\
$k^i{k^\prime}^j \langle S_{20}^{(ij)(ab)} \rangle$ & $\mathbf{k} \cdot \mathbf{k}^\prime$ & $-\mathbf{k} \cdot \mathbf{k}^\prime$ & $0$ & $-\mathbf{k} \cdot \mathbf{k}^\prime$ \\
$k^i{k^\prime}^j \langle S_{23}^{(ij)(ab)} \rangle$ & $\frac43 \mathbf{k} \cdot \mathbf{k}^\prime-\frac43 i(\mathbf{k} \times \mathbf{k}^\prime)_3$ & $-\frac43 \mathbf{k} \cdot \mathbf{k}^\prime-\frac43i(\mathbf{k} \times \mathbf{k}^\prime)_3$ & $-\frac23 i(\mathbf{k} \times \mathbf{k}^\prime)_3$ & $-\frac43 \mathbf{k} \cdot \mathbf{k}^\prime$ \\
$k^i{k^\prime}^j \langle S_{24}^{(ij)(ab)} \rangle$ & $\frac43 \mathbf{k} \cdot \mathbf{k}^\prime+\frac43 i(\mathbf{k} \times \mathbf{k}^\prime)_3$ & $-\frac43 \mathbf{k} \cdot \mathbf{k}^\prime+\frac43 i(\mathbf{k} \times \mathbf{k}^\prime)_3$ & $\frac23 i(\mathbf{k} \times \mathbf{k}^\prime)_3$ & $-\frac43 \mathbf{k} \cdot \mathbf{k}^\prime$ \\
$k^i{k^\prime}^j \langle S_{25}^{(ij)(ab)} \rangle$ & $-\frac43 \mathbf{k} \cdot \mathbf{k}^\prime-\frac43 i(\mathbf{k} \times \mathbf{k}^\prime)_3$ & $\frac43 \mathbf{k} \cdot \mathbf{k}^\prime-\frac43 i(\mathbf{k} \times \mathbf{k}^\prime)_3$ & $-\frac23 i(\mathbf{k} \times \mathbf{k}^\prime)_3$ & $\frac43 \mathbf{k} \cdot \mathbf{k}^\prime$ \\
$k^i{k^\prime}^j \langle S_{26}^{(ij)(ab)} \rangle$ & $-\frac43 \mathbf{k} \cdot \mathbf{k}^\prime+\frac43 i(\mathbf{k} \times \mathbf{k}^\prime)_3$ & $\frac43 \mathbf{k} \cdot \mathbf{k}^\prime+\frac43 i(\mathbf{k} \times \mathbf{k}^\prime)_3$ & $\frac23 i(\mathbf{k} \times \mathbf{k}^\prime)_3$ & $\frac43 \mathbf{k} \cdot \mathbf{k}^\prime$ \\
$k^i{k^\prime}^j \langle S_{27}^{(ij)(ab)} \rangle$ & $\frac{25}{12} i(\mathbf{k} \times \mathbf{k}^\prime)_3$ & $\frac{25}{12} i(\mathbf{k} \times \mathbf{k}^\prime)_3$ & $\frac{25}{24}i(\mathbf{k} \times \mathbf{k}^\prime)_3$ & $0$ \\
$k^i{k^\prime}^j \langle S_{28}^{(ij)(ab)} \rangle$ & $\frac{25}{12} i(\mathbf{k} \times \mathbf{k}^\prime)_3$ & $\frac{25}{12} i(\mathbf{k} \times \mathbf{k}^\prime)_3$ & $\frac{25}{24}i(\mathbf{k} \times \mathbf{k}^\prime)_3$ & $0$ \\
$k^i{k^\prime}^j \langle S_{29}^{(ij)(ab)} \rangle$ & $\frac54 \mathbf{k} \cdot \mathbf{k}^\prime$ & $-\frac54 \mathbf{k} \cdot \mathbf{k}^\prime$ & $0$ & $-\frac54 \mathbf{k} \cdot \mathbf{k}^\prime$ \\
$k^i{k^\prime}^j \langle S_{30}^{(ij)(ab)} \rangle$ & $-\frac54 \mathbf{k} \cdot \mathbf{k}^\prime$ & $\frac54 \mathbf{k} \cdot \mathbf{k}^\prime$ & $0$ & $\frac54 \mathbf{k} \cdot \mathbf{k}^\prime$ \\
$k^i{k^\prime}^j \langle S_{38}^{(ij)(ab)} \rangle$ & $\frac52 \mathbf{k} \cdot \mathbf{k}^\prime$ & $-\frac52 \mathbf{k} \cdot \mathbf{k}^\prime$ & $0$ & $-\frac52 \mathbf{k} \cdot \mathbf{k}^\prime$ \\
$k^i{k^\prime}^j \langle S_{39}^{(ij)(ab)} \rangle$ & $\frac54 i(\mathbf{k} \times \mathbf{k}^\prime)_3$ & $\frac54 i(\mathbf{k} \times \mathbf{k}^\prime)_3$ & $\frac58 i(\mathbf{k} \times \mathbf{k}^\prime)_3$ & $0$ \\
$k^i{k^\prime}^j \langle S_{40}^{(ij)(ab)} \rangle$ & $\frac54 i(\mathbf{k} \times \mathbf{k}^\prime)_3$ & $\frac54 i(\mathbf{k} \times \mathbf{k}^\prime)_3$ & $\frac58 i(\mathbf{k} \times \mathbf{k}^\prime)_3$ & $0$ \\
$k^i{k^\prime}^j \langle S_{47}^{(ij)(ab)} \rangle$ & $\frac12 \mathbf{k} \cdot \mathbf{k}^\prime$ & $\frac12 \mathbf{k} \cdot \mathbf{k}^\prime$ & $\frac14 \mathbf{k} \cdot \mathbf{k}^\prime$ & $0$ \\
$k^i{k^\prime}^j \langle S_{49}^{(ij)(ab)} \rangle$ & $-4 i(\mathbf{k} \times \mathbf{k}^\prime)_3$ & $4 i(\mathbf{k} \times \mathbf{k}^\prime)_3$ & $0$ & $4 i(\mathbf{k} \times \mathbf{k}^\prime)_3$ \\
$k^i{k^\prime}^j \langle S_{54}^{(ij)(ab)} \rangle$ & $4\mathbf{k} \cdot \mathbf{k}^\prime+i(\mathbf{k} \times \mathbf{k}^\prime)_3$ & $4\mathbf{k} \cdot \mathbf{k}^\prime-i(\mathbf{k} \times \mathbf{k}^\prime)_3$ & $2\mathbf{k} \cdot \mathbf{k}^\prime$ & $-i(\mathbf{k} \times \mathbf{k}^\prime)_3$ \\
$k^i{k^\prime}^j \langle S_{56}^{(ij)(ab)} \rangle$ & $-\frac{25}{8} i(\mathbf{k} \times \mathbf{k}^\prime)_3$ & $\frac{25}{8} i(\mathbf{k} \times \mathbf{k}^\prime)_3$ & $0$ & $\frac{25}{8} i(\mathbf{k} \times \mathbf{k}^\prime)_3$ \\
$k^i{k^\prime}^j \langle S_{60}^{(ij)(ab)} \rangle$ & $-8\mathbf{k} \cdot \mathbf{k}^\prime-2 i(\mathbf{k} \times \mathbf{k}^\prime)_3$ & $-8\mathbf{k} \cdot \mathbf{k}^\prime+2 i(\mathbf{k} \times \mathbf{k}^\prime)_3$ & $-4\mathbf{k} \cdot \mathbf{k}^\prime$ & $2 i(\mathbf{k} \times \mathbf{k}^\prime)_3$ \\
$k^i{k^\prime}^j \langle S_{67}^{(ij)(ab)} \rangle$ & $\frac54 \mathbf{k} \cdot \mathbf{k}^\prime$ & $\frac54 \mathbf{k} \cdot \mathbf{k}^\prime$ & $\frac58\mathbf{k} \cdot \mathbf{k}^\prime$ & $0$ \\
$k^i{k^\prime}^j \langle S_{68}^{(ij)(ab)} \rangle$ & $\frac54 \mathbf{k} \cdot \mathbf{k}^\prime$ & $\frac54 \mathbf{k} \cdot \mathbf{k}^\prime$ & $\frac58\mathbf{k} \cdot \mathbf{k}^\prime$ & $0$ \\
$k^i{k^\prime}^j \langle S_{83}^{(ij)(ab)} \rangle$ & $\frac{25}{4}i(\mathbf{k} \times \mathbf{k}^\prime)_3$ & $\frac{25}{4}i(\mathbf{k} \times \mathbf{k}^\prime)_3$ & $\frac{25}{8} i(\mathbf{k} \times \mathbf{k}^\prime)_3$ & $0$ \\
$k^i{k^\prime}^j \langle S_{84}^{(ij)(ab)} \rangle$ & $\frac52 \mathbf{k} \cdot \mathbf{k}^\prime$ & $\frac52 \mathbf{k} \cdot \mathbf{k}^\prime$ & $\frac54 \mathbf{k} \cdot \mathbf{k}^\prime$ & $0$ \\
$k^i{k^\prime}^j \langle S_{85}^{(ij)(ab)} \rangle$ & $\frac52 \mathbf{k} \cdot \mathbf{k}^\prime$ & $\frac52 \mathbf{k} \cdot \mathbf{k}^\prime$ & $\frac54 \mathbf{k} \cdot \mathbf{k}^\prime$ & $0$ \\
$k^i{k^\prime}^j \langle S_{86}^{(ij)(ab)} \rangle$ & $-4\mathbf{k} \cdot \mathbf{k}^\prime$ & $4\mathbf{k} \cdot \mathbf{k}^\prime$ & $0$ & $4\mathbf{k} \cdot \mathbf{k}^\prime$ \\
$k^i{k^\prime}^j \langle S_{91}^{(ij)(ab)} \rangle$ & $-6\mathbf{k} \cdot \mathbf{k}^\prime-6 i(\mathbf{k} \times \mathbf{k}^\prime)_3$ & $6\mathbf{k} \cdot \mathbf{k}^\prime-6 i(\mathbf{k} \times \mathbf{k}^\prime)_3$ & $-3 i(\mathbf{k} \times \mathbf{k}^\prime)_3$ & $6\mathbf{k} \cdot \mathbf{k}^\prime$ \\
\hline
\end{tabular}
}
\end{center}
\end{table}

\subsection{\label{sec:figabR}Scattering amplitude from Fig.~\ref{fig:sp}(a,b)}

Collecting partial results, from Eq.~(\ref{eq:sam1}) the scattering amplitude for the $N\pi$ system can be cast into
\begin{eqnarray}
& & f^2 k^0 \mathcal{A}_\mathrm{LO} (p + \pi^+ \to p + \pi^+) \nonumber \\
& & \mbox{\hglue0.2truecm} = \left[ -\frac{25}{72} a_1^2 - \frac{5}{36} a_1b_2 - \frac{25}{108} a_1b_3 - \frac{1}{72} b_2^2 - \frac{5}{108} b_2b_3 - \frac{25}{648} b_3^2 \right. \nonumber \\
& & \mbox{\hglue0.6truecm} + \left. \frac29 \left[ 1 - \frac{2\Delta}{k^0} + \frac{\Delta^2}{{k^0}^2} \right] \left[ a_1^2 + a_1c_3 + \frac14 c_3^2 \right] \right] \mathbf{k} \cdot \mathbf{k}^\prime \nonumber \\
& & \mbox{\hglue0.6truecm} + \left[ \frac{25}{72} a_1^2 + \frac{5}{36} a_1b_2 + \frac{25}{108} a_1b_3 + \frac{1}{72} b_2^2 + \frac{5}{108} b_2b_3 + \frac{25}{648} b_3^2 \right. \nonumber \\
& & \mbox{\hglue1.1truecm} - \left. \frac29 \left[ 1 - \frac12 \frac{\Delta}{k^0} + \frac{\Delta^2}{{k^0}^2} \right] \left[ a_1^2 + a_1c_3 + \frac14 c_3^2 \right] \right] i(\mathbf{k} \times \mathbf{k}^\prime)_3 + \mathcal{O}\left[\frac{\Delta^3}{{k^0}^3}\right] \nonumber \\
& & \mbox{\hglue0.2truecm} = f^2 k^0 \mathcal{A}_\mathrm{LO} (n + \pi^- \to n + \pi^-), \label{eq:pppppp}
\end{eqnarray}

\begin{eqnarray}
& & f^2 k^0\mathcal{A}_\mathrm{LO}(p + \pi^- \to p + \pi^-) \nonumber \\
& & \mbox{\hglue0.2truecm} = \left[ \frac{25}{72} a_1^2 + \frac{5}{36} a_1b_2 + \frac{25}{108} a_1b_3 + \frac{1}{72} b_2^2 + \frac{5}{108} b_2b_3 + \frac{25}{648} b_3^2 \right. \nonumber \\
& & \mbox{\hglue0.6truecm} - \left. \frac29 \left[ 1 + \frac{2\Delta}{k^0} + \frac{\Delta^2}{{k^0}^2} \right] \left[ a_1^2 + a_1c_3 + \frac14 c_3^2 \right] \right] \mathbf{k} \cdot \mathbf{k}^\prime \nonumber \\
& & \mbox{\hglue0.6truecm} + \left[ \frac{25}{72} a_1^2 + \frac{5}{36} a_1b_2 + \frac{25}{108} a_1b_3 + \frac{1}{72} b_2^2 + \frac{5}{108} b_2b_3 + \frac{25}{648} b_3^2 \right. \nonumber \\
& & \mbox{\hglue1.1truecm} - \left. \frac29 \left[ 1 + \frac12 \frac{\Delta}{k^0} + \frac{\Delta^2}{{k^0}^2} \right] \left[ a_1^2 + a_1c_3 + \frac14 c_3^2 \right] \right] i(\mathbf{k} \times \mathbf{k}^\prime)_3 + \mathcal{O}\left[\frac{\Delta^3}{{k^0}^3}\right] \nonumber \\
& & \mbox{\hglue0.2truecm} = f^2 k^0\mathcal{A}_\mathrm{LO}(n + \pi^+ \to n + \pi^+),
\end{eqnarray}

\begin{eqnarray}
& & f^2 k^0 \mathcal{A}_\mathrm{LO} (p + \pi^0 \to p + \pi^0) \nonumber \\
& & \mbox{\hglue0.2truecm} = - \frac49 \frac{\Delta}{k^0} \left[ a_1^2 + a_1c_3 + \frac14 c_3^2 \right] \mathbf{k} \cdot \mathbf{k}^\prime \nonumber \\
& & \mbox{\hglue0.6truecm} + \left[ \frac{25}{72} a_1^2 + \frac{5}{36} a_1b_2 + \frac{25}{108} a_1b_3 + \frac{1}{72} b_2^2 + \frac{5}{108} b_2b_3 + \frac{25}{648} b_3^2 \right. \nonumber \\
& & \mbox{\hglue1.1truecm} \left. - \frac29 \left[ 1 + \frac{\Delta^2}{{k^0}^2} \right] \left[ a_1^2 + a_1c_3 + \frac14 c_3^2 \right] \right] i(\mathbf{k} \times \mathbf{k}^\prime)_3 + \mathcal{O}\left[\frac{\Delta^3}{{k^0}^3}\right] \nonumber \\
& & \mbox{\hglue0.2truecm} = f^2 k^0 \mathcal{A}_\mathrm{LO} (n + \pi^0 \to n + \pi^0),
\end{eqnarray}

\begin{eqnarray}
& & \sqrt{2}f^2 k^0 \mathcal{A}_\mathrm{LO}(p + \pi^- \to n + \pi^0) \nonumber \\
& & \mbox{\hglue0.2truecm} = \left[ - \frac{25}{36} a_1^2 - \frac{5}{18} a_1b_2 - \frac{25}{54} a_1b_3 - \frac{1}{36} b_2^2 - \frac{5}{54} b_2b_3 - \frac{25}{324} b_3^2 \right. \nonumber \\
& & \mbox{\hglue0.6truecm} + \left. \frac49 \left[ 1 + \frac{\Delta^2}{{k^0}^2} \right] \left[ a_1^2 + a_1c_3 + \frac14 c_3^2 \right] \right] \mathbf{k} \cdot \mathbf{k}^\prime \nonumber \\
& & \mbox{\hglue0.6truecm} + \frac29 \frac{\Delta}{k^0} \left[ a_1^2 + a_1c_3 + \frac14 c_3^2 \right] i(\mathbf{k} \times \mathbf{k}^\prime)_3 + \mathcal{O}\left[\frac{\Delta^3}{{k^0}^3}\right] \nonumber \\
& & \mbox{\hglue0.2truecm} = \sqrt{2}f^2 k^0 \mathcal{A}_\mathrm{LO}(n + \pi^+ \to p + \pi^0). \label{eq:nppppz}
\end{eqnarray}

The urge to evaluate scattering amplitudes including all operator structures allowed for $N_c=3$ in Eq.~(\ref{eq:atree3}) becomes now manifest because the above results can be rewritten in terms of the $SU(3)$ invariant couplings $D$, $F$, $\mathcal{C}$, and $\mathcal{H}$ introduced in HBChPT \cite{jm255,jm259}. These couplings are related to the $1/N_c$ coefficients $a_1$, $b_2$, $b_3$, and $c_3$ for $N_c=3$ as \cite{jen96}
\begin{subequations}
\label{eq:rel1}
\begin{eqnarray}
& & D = \frac12 a_1 + \frac16 b_3, \\
& & F = \frac13 a_1 + \frac16 b_2 + \frac19 b_3, \\
& & \mathcal{C} = - a_1 - \frac12 c_3, \\
& & \mathcal{H} = -\frac32 a_1 - \frac32 b_2 - \frac52 b_3.
\end{eqnarray}
\end{subequations}

A further simplification can be achieved if the power series expansion in $\Delta$ of the function
\begin{equation}
t_1 \frac{k^0}{k^0 - \Delta} + t_2 \frac{k^0}{k^0 + \Delta} = t_1+t_2 + (t_1-t_2) \frac{\Delta}{k^0} + (t_1+t_2) \frac{\Delta^2}{{k^0}^2} + (t_1-t_2) \frac{\Delta^3}{{k^0}^3} + \mathcal{O}\left[\frac{\Delta^4}{{k^0}^4}\right],
\end{equation}
where $t_k$ are some coefficients, is substituted into Eqs.~(\ref{eq:pppppp})-(\ref{eq:nppppz}) to rewrite the final forms of the scattering amplitudes as
\begin{eqnarray}
& & f^2 k^0 \mathcal{A}_\mathrm{LO} (p + \pi^+ \to p + \pi^+) \nonumber \\
& & \mbox{\hglue0.2truecm} = \left[ - \frac12 (D + F)^2 + \frac19 \left[ - \frac{k^0}{k^0 - \Delta} + 3 \frac{k^0}{k^0 + \Delta} \right] \mathcal{C}^2 \right] \mathbf{k} \cdot \mathbf{k}^\prime \nonumber \\
& & \mbox{\hglue0.6truecm} + \left[ \frac12 (D + F)^2 - \frac{1}{18} \left[ \frac{k^0}{k^0 - \Delta} + 3 \frac{k^0}{k^0 + \Delta} \right] \mathcal{C}^2 \right] i(\mathbf{k} \times \mathbf{k}^\prime)_3 \nonumber \\
& & \mbox{\hglue0.2truecm} = f^2 k^0 \mathcal{A}_\mathrm{LO} (n + \pi^- \to n + \pi^-), \label{eq:ppppppch}
\end{eqnarray}
\begin{eqnarray}
& & f^2 k^0\mathcal{A}_\mathrm{LO}(p + \pi^- \to p + \pi^-) \nonumber \\
& & \mbox{\hglue0.2truecm} = \left[ \frac12 (D + F)^2 - \frac19 \left[ 3 \frac{k^0}{k^0 - \Delta} - \frac{k^0}{k^0 + \Delta} \right] \mathcal{C}^2 \right] \mathbf{k} \cdot \mathbf{k}^\prime \nonumber \\
& & \mbox{\hglue0.6truecm} + \left[ \frac12 (D + F)^2 -\frac{1}{18} \left[ 3 \frac{k^0}{k^0 - \Delta} + \frac{k^0}{k^0 + \Delta} \right] \mathcal{C}^2 \right] i(\mathbf{k} \times \mathbf{k}^\prime)_3 \nonumber \\
& & \mbox{\hglue0.2truecm} = f^2 k^0\mathcal{A}_\mathrm{LO}(n + \pi^+ \to n + \pi^+),
\end{eqnarray}
\begin{eqnarray}
& & f^2 k^0 \mathcal{A}_\mathrm{LO} (p + \pi^0 \to p + \pi^0) \nonumber \\
& & \mbox{\hglue0.2truecm} = - \frac29 \left[ \frac{k^0}{k^0 - \Delta} - \frac{k^0}{k^0 + \Delta} \right] \mathcal{C}^2 \mathbf{k} \cdot \mathbf{k}^\prime \nonumber \\
& & \mbox{\hglue0.6truecm} + \left[ \frac12 (D + F)^2 - \frac19 \left[ \frac{k^0}{k^0 - \Delta} + \frac{k^0}{k^0 + \Delta} \right] \mathcal{C}^2 \right] i(\mathbf{k} \times \mathbf{k}^\prime)_3 \nonumber \\
& & \mbox{\hglue0.2truecm} = f^2 k^0 \mathcal{A}_\mathrm{LO} (n + \pi^0 \to n + \pi^0),
\end{eqnarray}
\begin{eqnarray}
& & \sqrt{2}f^2 k^0 \mathcal{A}_\mathrm{LO}(p + \pi^- \to n + \pi^0) \nonumber \\
& & \mbox{\hglue0.2truecm} = \left[ -(D + F)^2 + \frac29 \left[ \frac{k^0}{k^0 - \Delta} + \frac{k^0}{k^0 + \Delta} \right] \mathcal{C}^2 \right] \mathbf{k} \cdot \mathbf{k}^\prime \nonumber \\
& & \mbox{\hglue0.6truecm} + \frac19 \left[ \frac{k^0}{k^0 - \Delta} - \frac{k^0}{k^0 + \Delta} \right] \mathcal{C}^2 i(\mathbf{k} \times \mathbf{k}^\prime)_3 \nonumber \\
& & \mbox{\hglue0.2truecm} = \sqrt{2}f^2 k^0 \mathcal{A}_\mathrm{LO}(n + \pi^+ \to p + \pi^0), \label{eq:nppppzch}
\end{eqnarray}
which are valid to order $\mathcal{O}(\Delta^3/{k^0}^3)$.

A glance at the above expressions shows that the scattering amplitudes for $N\pi\to N\pi$ processes are written in terms of the $SU(3)$ invariants $F$, $D$, and $\mathcal{C}$ \cite{jm255,jm259}, a totally expected and consistent result because $F$ and $D$ come along $BB\pi$ vertices, where $g_A=D+F$ is the axial coupling for neutron beta decay in the limit of exact $SU(3)$ symmetry, whereas $\mathcal{C}$ comes along $TB\pi$ vertices. Also, in the limit $\Delta\to 0$ the coefficients of the $\mathcal{C}^2$ terms do not vanish.

As for the $\mathcal{A}_\mathrm{LO}^{(3/2)}$ and $\mathcal{A}_\mathrm{LO}^{(1/2)}$ amplitudes, they are found to be
\begin{eqnarray}
& & f^2k^0 \mathcal{A}_\mathrm{LO}^{(3/2)} \nonumber \\
& & \mbox{\hglue0.2truecm} = \left[ -\frac{25}{72} a_1^2 - \frac{5}{36} a_1b_2 - \frac{25}{108} a_1b_3 - \frac{1}{72} b_2^2 - \frac{5}{108} b_2b_3 - \frac{25}{648} b_3^2 \right. \\
& & \mbox{\hglue0.2truecm} + \left. \frac29 \left[ 1 - \frac{2\Delta}{k^0} + \frac{\Delta^2}{{k^0}^2} \right] \left[ a_1^2 + a_1c_3 + \frac14 c_3^2 \right] \right] \mathbf{k} \cdot \mathbf{k}^\prime \nonumber \\
& & \mbox{\hglue0.6truecm} + \left[ \frac{25}{72} a_1^2 + \frac{5}{36} a_1b_2 + \frac{25}{108} a_1b_3 + \frac{1}{72} b_2^2 + \frac{5}{108} b_2b_3 + \frac{25}{648} b_3^2 \right. \nonumber \\
& & \mbox{\hglue1.1truecm} \left. - \frac29 \left[ 1 - \frac12 \frac{\Delta}{k^0} + \frac{\Delta^2}{{k^0}^2} \right] \left[ a_1^2 + a_1c_3 + \frac14 c_3^2 \right] \right] i(\mathbf{k} \times \mathbf{k}^\prime)_3 + \mathcal{O}\left[\frac{\Delta^3}{{k^0}^3}\right],
\end{eqnarray}
and
\begin{eqnarray}
& & f^2 k^0 \mathcal{A}_\mathrm{LO}^{(1/2)} \nonumber \\
& & \mbox{\hglue0.2truecm} = \left[ \frac{25}{36} a_1^2 + \frac{5}{18} a_1b_2 + \frac{25}{54} a_1b_3 + \frac{1}{36} b_2^2 + \frac{5}{54} b_2b_3 + \frac{25}{324} b_3^2 \right. \\
& & \mbox{\hglue0.2truecm} - \left. \frac49 \left[ 1 + \frac{\Delta}{k^0} + \frac{\Delta^2}{{k^0}^2} \right] \left[ a_1^2 + a_1c_3 + \frac14 c_3^2 \right] \right] \mathbf{k} \cdot \mathbf{k}^\prime \nonumber \\
& & \mbox{\hglue0.6truecm} + \left[ \frac{25}{72} a_1^2 + \frac{5}{36} a_1b_2 + \frac{25}{108} a_1b_3 + \frac{1}{72} b_2^2 + \frac{5}{108} b_2b_3 + \frac{25}{648} b_3^2 \right. \nonumber \\
& & \mbox{\hglue1.1truecm} \left. - \frac29 \left[ 1 + \frac{\Delta}{k^0} + \frac{\Delta^2}{{k^0}^2} \right]\left[ a_1^2 + a_1c_3 + \frac14 c_3^2 \right] \right] i(\mathbf{k} \times \mathbf{k}^\prime)_3 + \mathcal{O}\left[\frac{\Delta^3}{{k^0}^3}\right],
\end{eqnarray}
or equivalently,
\begin{eqnarray}
f^2k^0 \mathcal{A}_\mathrm{LO}^{(3/2)} & = & \left[ - \frac12 (D + F)^2 + \frac19 \left[ - \frac{k^0}{k^0 - \Delta} + 3 \frac{k^0}{k^0 + \Delta} \right] \mathcal{C}^2 \right] \mathbf{k} \cdot \mathbf{k}^\prime \nonumber \\
& & \mbox{} + \left[ \frac12 (D + F)^2 - \frac{1}{18} \left[ \frac{k^0}{k^0 - \Delta} + 3 \frac{k^0}{k^0 + \Delta} \right] \mathcal{C}^2 \right] i(\mathbf{k} \times \mathbf{k}^\prime)_3, \label{eq:a32}
\end{eqnarray}
and
\begin{equation}
f^2 k^0\mathcal{A}_\mathrm{LO}^{(1/2)} = \left[ (D + F)^2 - \frac49 \frac{k^0}{k^0 - \Delta}\mathcal{C}^2 \right] \mathbf{k} \cdot \mathbf{k}^\prime + \left[ \frac12 (D + F)^2 - \frac29 \frac{k^0}{k^0 - \Delta} \mathcal{C}^2 \right] i(\mathbf{k} \times \mathbf{k}^\prime)_3, \label{eq:a12}
\end{equation}
which are valid to order $\mathcal{O}(\Delta^3/{k^0}^3)$.

\subsubsection{Isospin relations}

The $N\pi \to N\pi$ scattering amplitudes satisfy the following isospin relations,
\begin{equation}
\mathcal{A}_\mathrm{LO}(p + \pi^- \to p + \pi^-) - \mathcal{A}_\mathrm{LO}(p + \pi^0 \to p + \pi^0) + \frac{1}{\sqrt{2}} \mathcal{A}_\mathrm{LO}(p + \pi^- \to n + \pi^0) = 0, \label{eq:is1}
\end{equation}

\begin{equation}
\mathcal{A}_\mathrm{LO}(p + \pi^+ \to p + \pi^+) - \mathcal{A}_\mathrm{LO}(p + \pi^- \to p + \pi^-) - \sqrt{2} \mathcal{A}_\mathrm{LO}(p + \pi^- \to n + \pi^0) = 0,
\end{equation}

\begin{equation}
\mathcal{A}_\mathrm{LO}(p + \pi^+ \to p + \pi^+) + \mathcal{A}_\mathrm{LO}(p + \pi^- \to p + \pi^-) - 2 \mathcal{A}_\mathrm{LO}(p + \pi^0 \to p + \pi^0) = 0,
\end{equation}

\begin{equation}
\mathcal{A}_\mathrm{LO}(n + \pi^- \to n + \pi^-) - \mathcal{A}_\mathrm{LO}(n + \pi^0 \to n + \pi^0) - \frac{1}{\sqrt{2}} \mathcal{A}_\mathrm{LO}(n + \pi^+ \to p + \pi^0) = 0,
\end{equation}

\begin{equation}
\mathcal{A}_\mathrm{LO}(n + \pi^+ \to n + \pi^+) - \mathcal{A}_\mathrm{LO}(n + \pi^- \to n + \pi^-) + \sqrt{2} \mathcal{A}_\mathrm{LO}(n + \pi^+ \to p + \pi^0) = 0,
\end{equation}

\begin{equation}
\mathcal{A}_\mathrm{LO}(n + \pi^+ \to n + \pi^+) + \mathcal{A}_\mathrm{LO}(n + \pi^- \to n + \pi^-) - 2 \mathcal{A}_\mathrm{LO}(n + \pi^0 \to n + \pi^0) = 0. \label{eq:is6}
\end{equation}

\subsection{Scattering amplitude from Fig.~\ref{fig:sp}(c)}

Following the lines of Eq.~(\ref{eq:mtxe}), for the $N\pi$ system, the scattering amplitudes arising from Fig.~\ref{fig:sp}(c) read
\begin{eqnarray}
\mathcal{A}_\mathrm{vertex} (p + \pi^ + \to p + \pi^+) & = & \frac14 \frac{k^0}{f^2} \nonumber \\
& = & \mathcal{A}_\mathrm{vertex} (n + \pi^- \to n + \pi^-), \label{eq:ppppppvtx}
\end{eqnarray}
\begin{eqnarray}
\mathcal{A}_\mathrm{vertex} (p + \pi^- \to p + \pi^-) & = & - \frac14 \frac{k^0}{f^2} \nonumber \\
& = & \mathcal{A}_\mathrm{vertex} (n + \pi^+ \to n + \pi^+),
\end{eqnarray}
\begin{eqnarray}
\mathcal{A}_\mathrm{vertex} (p + \pi^0 \to p + \pi^0) & = & 0 \nonumber \\
& = & \mathcal{A}_\mathrm{vertex} (n + \pi^0 \to n + \pi^0),
\end{eqnarray}
\begin{eqnarray}
\mathcal{A}_\mathrm{vertex} (p + \pi^- \to n + \pi^0) & = & \frac{1}{2\sqrt{2}} \frac{k^0}{f^2} \nonumber \\
& = & \mathcal{A}_\mathrm{vertex} (n + \pi^+ \to p + \pi^0), \label{eq:nppppzvtx}
\end{eqnarray}
from which the following amplitudes can be obtained,
\begin{equation}
\mathcal{A}_\mathrm{vertex}^{(3/2)} = \frac14 \frac{k^0}{f^2}, \label{eq:aa32}
\end{equation}
and
\begin{equation}
\mathcal{A}_\mathrm{vertex}^{(1/2)} = -\frac12 \frac{k^0}{f^2}. \label{eq:aa12}
\end{equation}

\subsubsection{Isospin relations}

In a close analogy the the previous case, the isospin relations between these scattering amplitudes are
\begin{equation}
\mathcal{A}_\mathrm{vertex} (p + \pi^- \to p + \pi^-) - \mathcal{A}_\mathrm{vertex} (p + \pi^0 \to p + \pi^0) + \frac{1}{\sqrt{2}} \mathcal{A}_\mathrm{vertex} (p + \pi^- \to n + \pi^0) = 0,
\end{equation}

\begin{equation}
\mathcal{A}_\mathrm{vertex} (p + \pi^+ \to p + \pi^+) - \mathcal{A}_\mathrm{vertex} (p + \pi^- \to p + \pi^-) - \sqrt{2} \mathcal{A}_\mathrm{vertex} (p + \pi^- \to n + \pi^0) = 0,
\end{equation}

\begin{equation}
\mathcal{A}_\mathrm{vertex} (p + \pi^+ \to p + \pi^+) + \mathcal{A}_\mathrm{vertex} (p + \pi^- \to p + \pi^-) - 2 \mathcal{A}_\mathrm{vertex} (p + \pi^0 \to n + \pi^0) = 0,
\end{equation}

\begin{equation}
\mathcal{A}_\mathrm{vertex} (n + \pi^- \to n + \pi^-) - \mathcal{A}_\mathrm{vertex} (n + \pi^0 \to n + \pi^0) - \frac{1}{\sqrt{2}} \mathcal{A}_\mathrm{vertex} (n + \pi^{+ } \to p + \pi^0) = 0,
\end{equation}

\begin{equation}
\mathcal{A}_\mathrm{vertex} (n + \pi^+ \to n + \pi^+) - \mathcal{A}_\mathrm{vertex} (n + \pi^- \to n + \pi^-) + \sqrt{2} \mathcal{A}_\mathrm{vertex} (n + \pi^+ \to p + \pi^0) = 0,
\end{equation}

\begin{equation}
\mathcal{A}_\mathrm{vertex} (n + \pi^+ \to n + \pi^+) + \mathcal{A}_\mathrm{vertex} (n + \pi^- \to n + \pi^-) - 2 \mathcal{A}_\mathrm{vertex} (n + \pi^0 \to n + \pi^0) = 0.
\end{equation}

\section{\label{sec:str}Processes with strangeness: Two case studies}

To test the applicability of the approach, two processes including strangeness have been selected with no particular criteria. They are simply two case studies. They are $\Lambda + K^+ \to p + \pi^0$ and $\Xi^0 + K^0 \to \Lambda + \eta$. Without further ado, the respective scattering amplitudes from Fig.~\ref{fig:sp}(a,b) read
\begin{eqnarray}
& & 4 \sqrt{3} f^2 k^0 \mathcal{A}_\mathrm{LO}(\Lambda + K^+ \to p + \pi^0) \nonumber \\
& & \mbox{\hglue0.2truecm} = \left[ -\frac{17}{12} a_1^2 - \frac12 a_1b_2 - \frac{17}{18} a_1b_3 - \frac{1}{12} b_2^2 - \frac16 b_2b_3 - \frac{17}{108} b_3^2 \right. \nonumber \\
& & \mbox{\hglue0.6truecm} + \left. \frac23 \left[ 1 + \frac{\Delta}{k^0} + \frac{\Delta^2}{{k^0}^2} \right] \left[ a_1^2 + a_1c_3 + \frac14 c_3^2 \right] \right] \mathbf{k}\cdot \mathbf{k}^\prime \nonumber \\
& & \mbox{\hglue0.6truecm} + \left[ - \frac{13}{12} a_1^2 - \frac56 a_1b_2 - \frac{13}{18} a_1b_3 - \frac{1}{12} b_2^2 -\frac{5}{18} b_2b_3 - \frac{13}{108} b_3^2 \right. \nonumber \\
& & \mbox{\hglue1.1truecm} + \left. \frac13 \left[ 1 + \frac{\Delta}{k^0} + \frac{\Delta^2}{{k^0}^2} \right] \left[ a_1^2 + a_1c_3 + \frac14 c_3^2 \right]\right] (\mathbf{k}\times \mathbf{k}^\prime)_3 + \mathcal{O}\left[\frac{\Delta}{k^0}\right]^3,
\end{eqnarray}
and
\begin{eqnarray}
& & 4 \sqrt{3} f^2 k^0 \mathcal{A}_\mathrm{LO}(\Xi^0 + K^0 \to \Lambda + \eta) \nonumber \\
& & \mbox{\hglue0.2truecm} = \left[ - \frac34 a_1^2 - \frac16 a_1b_2 - \frac12 a_1b_3 - \frac{1}{12} b_2^2 - \frac{1}{18} b_2b_3 - \frac{1}{12} b_3^2 \right. \nonumber \\
& & \mbox{\hglue0.6truecm} + \left. \frac43 \frac{\Delta}{k^0} \left[ a_1^2 + a_1c_3 + \frac14 c_3^2 \right] \right] \mathbf{k}\cdot \mathbf{k}^\prime \nonumber \\
& & \mbox{\hglue0.6truecm} + \left[ - \frac{11}{12} a_1^2 - \frac16 a_1b_2 - \frac{11}{18} a_1b_3 + \frac{1}{12} b_2^2 - \frac{1}{18} b_2b_3 - \frac{11}{108} b_3^2 \right. \nonumber \\
& & \mbox{\hglue1.1truecm} + \left. \frac23 \left[ 1 + \frac{\Delta^2}{{k^0}^2} \right] \left[ a_1^2 + a_1c_3 + \frac14 c_3^2 \right] \right] (\mathbf{k}\times \mathbf{k}^\prime)_3 + \mathcal{O}\left[\frac{\Delta}{k^0}\right]^3,
\end{eqnarray}
or equivalently,
\begin{eqnarray}
& & 4 \sqrt{3} f^2 k^0 \mathcal{A}_\mathrm{LO}(\Lambda + K^+ \to p + \pi^0) \nonumber \\
& & \mbox{\hglue0.2truecm} = \left[ - 3D^2 - 2DF - 3F^2 + \frac23 \left[ \frac{k^0}{k^0 - \Delta} \right] \mathcal{C}^2 \right] \mathbf{k}\cdot \mathbf{k}^\prime \nonumber \\
& & \mbox{\hglue0.6truecm} + \left[ D^2 - 6 DF - 3F^2 + \frac13 \left[ \frac{k^0}{k^0 - \Delta} \right] \mathcal{C}^2 \right] (\mathbf{k}\times \mathbf{k}^\prime)_3 + \mathcal{O}\left[\frac{\Delta}{k^0}\right]^3,
\end{eqnarray}
and
\begin{eqnarray}
& & 4 \sqrt{3} f^2 k^0 \mathcal{A}_\mathrm{LO}(\Xi^0 + K^0 \to \Lambda + \eta) \nonumber \\
& & \mbox{\hglue0.2truecm} = \left[ - 3 D^2 + 2DF - 3F^2 + \frac23 \left[ \frac{k^0}{k^0 - \Delta} - \frac{k^0}{k^0 + \Delta} \right] \mathcal{C}^2 \right] \mathbf{k}\cdot \mathbf{k}^\prime \nonumber \\
& & \mbox{\hglue0.6truecm} + \left[ - D^2 - 6DF + 3 F^2 + \frac13 \left[ \frac{k^0}{k^0 - \Delta} + \frac{k^0}{k^0 + \Delta} \right] \mathcal{C}^2 \right] (\mathbf{k}\times \mathbf{k}^\prime)_3 + \mathcal{O}\left[\frac{\Delta}{k^0}\right]^3.
\end{eqnarray}

The above expressions have been obtained in a complete parallelism to the nucleon-pion processes, so no additional details are necessary here.

\section{\label{sec:sb}First-order $SU(3)$ symmetry breaking in the scattering amplitude}

$SU(3)$ flavor symmetry is not an exact symmetry and it is actually broken. Flavor symmetry breaking (SB) and strong isospin breaking (IB) refer to how the strong force deviates from the ideal symmetric limit where all quark flavors are treated on an equal footing (flavor symmetry) and where the up and down quarks are considered identical (isospin symmetry).

Two major sources of $SU(3)$ symmetry breaking are identified. The first one is due to the light quark masses and the perturbation transforms as the adjoint (octet) irreducible representation of $SU(3)$,
\begin{equation}
\epsilon \mathcal{H}^8 + \epsilon^\prime \mathcal{H}^3. \label{eq:sb}
\end{equation}
The first term in Eq.~(\ref{eq:sb}) is regarded as the dominant $SU(3)$ breaking and transforms as the eighth component of a flavor octet, where $\epsilon \sim m_s/{\Lambda_{\mathrm{QCD}}}$ is a (dimensionless) measure of SB; $\epsilon \sim 0.3$, which is comparable to a $1/N_c$ effect. The second term represents the leading QCD isospin breaking effect, i.e., the one associated with the difference of the up and down quark masses and transforms as the third component of a flavor octet, where $\epsilon^\prime \sim (m_d-m_u)/\Lambda_{\mathrm{QCD}}$, so $\epsilon^\prime \ll \epsilon$. This isospin breaking mechanism is referred to as strong isospin breaking.

The second source of symmetry breaking is induced by electromagnetic interactions. Electromagnetic mass splittings are second order in the quark charge matrix so they get a suppression factor of $\epsilon^{\prime\prime} \sim \alpha_{\mathrm{em}}/4\pi$. To a good approximation
\begin{equation}
\frac{m_d - m_u}{\Lambda_{\mathrm{QCD}}} \sim \frac{\alpha_{\mathrm{em}}}{4\pi}.
\end{equation}

In this section, effects due to first-order SB and IB to the scattering amplitude are discussed by extending the projection operator technique previously discussed, applied to the diagrams displayed in Fig.~\ref{fig:sp}(a,b) and \ref{fig:sp}(c) separately as they involve different operator structures. These effects will be added to the lowest-order results $\mathcal{A}_\mathrm{LO}$ to have more accurate expressions. Loop graphs that complement the analysis will be attempted elsewhere in the framework of large-$N_c$ chiral perturbation theory.

\subsection{\label{sec:po}Flavor projection operators for the product of 3 adjoints}

First-order flavor symmetry breaking contributions to the scattering amplitude are computed from the tensor product of the scattering amplitude itself, which transforms under the spin-flavor symmetry $SU(2) \times SU(3)$ as $(2,8\otimes 8)$, and the perturbation, which transforms as $(0,8)$. The tensor product of three adjoint representations $8 \otimes 8 \otimes 8$ decomposes as
\begin{equation}
8 \otimes 8 \otimes 8 = 2(1) \oplus 8(8) \oplus 4(10\oplus \overline{10}) \oplus 6(27) \oplus 2(35\oplus \overline{35}) \oplus 64. \label{eq:888}
\end{equation}
Thus, effects of SB can be evaluated by constructing the $1/N_c$ expansions of the pieces of the scattering amplitude transforming as $(2,1)$, $(2,8)$, $(2,10\oplus \overline{10})$, $(2,27)$, $(2,35\oplus \overline{35})$ and $(2,64)$ under $SU(2) \times SU(3)$. These $1/N_c$ expansions need be expressed in terms of a complete basis of linearly independent operators $\{R_k^{(ij)(a_1a_2a_3)}\}$, where a generic operator $R_k^{(ij)(a_1a_2a_3)}$ thus represents a spin-2 object with three adjoint indices. For $N_c=3$, up to 3-body operators should be retained in the series. Accordingly, first-order SB can be accounted for by setting one of the flavor indices to 8, {\it v.gr.}, $a_3=8$, whereas first-order strong IB can be accounted for by setting one of the flavor indices to 3, {\it v.gr.}, $a_3=3$. For completeness, the set of up to 3-body operators used as a basis is listed in the Online Resource. The set contains 170 linearly independent operators, where $R^{(ij)(a_1a_28)}$ and $R^{(ij)(a_1a_23)}$ denote operators with $I=0$ and $I=1$, respectively. Naively, isospin breaking induced by electromagnetism should appear from operators with $I=2$ and $I=3$, which emerge from the tensor product of 4 and 5 adjoint presentations, respectively. These tensor products will not be treated here.

The task of constructing the operators that yield SB effects facilitates considerably with the implementation of the projection operator technique presented in Ref.~\cite{banda}, extended to the decomposition (\ref{eq:888}). The projection operators can be constructed as
\begin{eqnarray}
& & \left[\pr{m}\right]^{c_1c_2c_3b_1b_2b_3} \nonumber \\
& &\mbox{\hglue0.2truecm} = \left[ \left( \frac{C-c^{n_1} {\mathcal{I}}}{c^m-c^{n_1}} \right) \left( \frac{C-c^{n_2} {\mathcal{I}}}{c^m-c^{n_2}} \right) \left( \frac{C-c^{n_3} {\mathcal{I}}}{c^m-c^{n_3}} \right) \left(\frac{C-c^{n_4} {\mathcal{I}}}{c^m-c^{n_4}} \right)\left( \frac{C-c^{n_5} {\mathcal{I}}}{c^m-c^{n_5}} \right) \right]^{c_1c_2c_3b_1b_2b_3},			
\end{eqnarray}
where $m$ labels the flavor representation of each projector and $n_i$ label flavor representations other than $m$. The quadratic Casimir operator, in turn, reads
\begin{eqnarray}
[C]^{c_1c_2c_3b_1b_2b_3} & = & 6 \delta^{c_1b_1} \delta^{c_2b_2} \delta^{c_3b_3} - 2 \delta^{c_1b_1} f^{ac_2b_2} f^{ac_3b_3} - 2 \delta^{c_2b_2} f^{ac_1b_1} f^{ac_3b_3} \nonumber \\
& & \mbox{} - 2 \delta^{c_3b_3} f^{ac_1b_1} f^{ac_2b_2},	
\end{eqnarray}
and
\begin{equation}
c^1 = 0, \hspace{0.5cm} c^8 = 3, \hspace{0.5cm} c^{10+\overline{10}} = 6, \hspace{0.5cm} c^{27} = 8, \hspace{0.5cm} c^{35+\overline{35}} = 12, \hspace{0.5cm} c^{64} = 15,
\end{equation}
are its corresponding eigenvalues.

Therefore, the product $\left[\pr{\mathrm{dim}} R_k^{(ij)}\right]^{c_1c_2c_3}$ will effectively provide the component of the operator $R_k^{(ij)(c_1c_2c_3)}$ transforming in the irreducible representation of dimension $\mathrm{dim}$, according to decomposition (\ref{eq:888}).

The explicit analytic construction of $\left[\pr{\mathrm{dim}}\right]^{c_1c_2c_3b_1b_2b_3}$, however, faces several algebraic challenges. The most evident one is how to deal with the products of up to ten $f$ symbols contained in the $C^5$ operator, which, in general, can not be reduced in terms containing fewer $f$ or $d$ symbols. Thus, the algebraic forms of $\left[\pr{\mathrm{dim}}\right]^{c_1c_2c_3b_1b_2b_3}$ contain hundreds of terms, which, in practice, become unmanageable. A more pragmatic approach should be adopted to solve the problem; it turns out that a matrix method is the most suitable one for this purpose.

To start with, notice that each projection operator (or quadratic Casimir operator) is an object with six adjoint indices, each one with eight possible values, so all these objects have $8^6$ elements. However, Casimir operators have all or half of their indices contracted and the projectors are applied on $3$-body operators with three adjoint indices, so half of the projector indices will be always contracted. Therefore, it is possible to collect the first three indices ($c_1$,$c_2$,$c_3$) and the last three indices ($b_1$,$b_2$,$b_3$) of both the Casimir and projectors in only two indices, one for each set. These new indices have $8^3=512$ values. In this way, a matrix representation for the projectors can be constructed, which is constituted by $512 \times512$ matrices. Similarly, the $3$-body operators with three adjoint indices can be represented as vectors with $512$ entries. Therefore, instead of performing the index contractions $\left[\pr{\mathrm{dim}} R_k^{(ij)}\right]^{c_1c_2c_3}$, the problem reduces to ordinary matrix multiplications. The whole procedure is very reliable and effectively simplifies the analysis.

Let ${\sf P}^{(\mathrm{dim})}$ represent the matrix corresponding to the projection operator $\left[\pr{\mathrm{dim}}\right]^{c_1c_2c_3b_1b_2b_3}$. With the method implemented, a series of consistency checks have been performed, namely,
\begin{equation}
{\sf P}^{(m)} {\sf P}^{(m)} = {\sf P}^{(m)}, \qquad \qquad {\sf P}^{(m)} {\sf P}^{(n)} = 0, \quad n\neq m,
\end{equation}
along with
\begin{equation}
{\sf P}^{(1)} + {\sf P}^{(8)} + {\sf P}^{(10+\overline{10})} + {\sf P}^{(27)} + {\sf P}^{(35+\overline{35})} + {\sf P}^{(64)} = {\sf I}_{512},
\end{equation}
where ${\sf I}_{512}$ stands for the identity matrix of order $512$. The above relations are the usual properties that projection operators must satisfy. No further details on the method are needed here.

In passing, it should be mentioned that the matrix method to construct projection operators can be extended to the tensor products of 4 and 5 adjoint representations. In the first case,
\begin{eqnarray}
8 \otimes 8 \otimes 8 \otimes 8 & = & 8 (1) \oplus 32 (8) \oplus 33 (27) \oplus 12 (64) \oplus 125 \nonumber \\
& & \mbox{} \oplus 20 (10 \oplus \overline{10}) \oplus 2 (28 \oplus \overline{28}) \oplus 15 (35 \oplus \overline{35}) \oplus 3 (81 \oplus \overline{81}). \label{eq:8888}
\end{eqnarray}
Decomposition (\ref{eq:8888}) contains operators with four flavor indices, two of which can be fixed to \{8,8\}, \{3,8\}, and \{3,3\}, which will identify operators with $I=0$, $I=1$, and $I=2$, respectively. Numerically, the procedure to construct projections operators would be rather involved, requiring a considerable amount of computing time. The procedure is nevertheless doable.

\subsubsection{Flavor SB effects to the scattering amplitude from Fig.~\ref{fig:sp}(a,b)}

The way flavor projection operators work can be better seen through a few examples. The operator $\{T^a,\{T^b,T^c\}\}$, for instance, contributes to the scattering amplitude of the process $n+\pi^+ \to n+\pi^+$ through the components with flavor indices $a=(1-i2)/\sqrt{2}$, $b=(1-i2)/\sqrt{2}$, and $c=8$. Using the matrix method, the \{1,1,8\} component of the flavor $8$ piece becomes,
\begin{eqnarray}
& & [{\pr{8}}]^{118cde}\{T^c,\{T^d,T^e\}\} \nonumber \\
& & \mbox{\hglue0.2truecm} = \frac{1}{15} T^1T^1T^8 + \frac{1}{30\sqrt{3}} T^1T^4T^6 + \frac{1}{30\sqrt{3}} T^1T^5T^7 + \frac{1}{30\sqrt{3}} T^1T^6T^4 + \frac{1}{30\sqrt{3}} T^1T^7T^5 \nonumber \\
& & \mbox{\hglue0.6truecm} + \frac{4}{15} T^1T^8T^1 + \frac{1}{15} T^2T^2T^8 - \frac{1}{30\sqrt{3}} T^2T^4T^7 + \frac{1}{30\sqrt{3}} T^2T^5T^6 + \frac{1}{30\sqrt{3}} T^2T^6T^5 \nonumber \\
& & \mbox{\hglue0.6truecm} - \frac{1}{30\sqrt{3}} T^2T^7T^4 + \frac{4}{15} T^2T^8T^2 + \frac{1}{15} T^3T^3T^8 + \frac{1}{30\sqrt{3}} T^3T^4T^4 + \frac{1}{30\sqrt{3}} T^3T^5T^5 \nonumber \\
& & \mbox{\hglue0.6truecm} - \frac{1}{30\sqrt{3}} T^3T^6T^6 - \frac{1}{30\sqrt{3}} T^3T^7T^7 + \frac{4}{15} T^3T^8T^3 - \frac{1}{15\sqrt{3}} T^4T^1T^6 + \frac{1}{15\sqrt{3}} T^4T^2T^7 \nonumber \\
& & \mbox{\hglue0.6truecm} - \frac{1}{15\sqrt{3}} T^4T^3T^4 + \frac{1}{30\sqrt{3}} T^4T^4T^3 + \frac{1}{10} T^4T^4T^8 + \frac{1}{30\sqrt{3}} T^4T^6T^1 - \frac{1}{30\sqrt{3}} T^4T^7T^2 \nonumber \\
& & \mbox{\hglue0.6truecm} + \frac15 T^4T^8T^4 - \frac{1}{15\sqrt{3}} T^5T^1T^7 - \frac{1}{15\sqrt{3}} T^5T^2T^6 - \frac{1}{15\sqrt{3}} T^5T^3T^5 + \frac{1}{30\sqrt{3}} T^5T^5T^3 \nonumber \\
& & \mbox{\hglue0.6truecm} + \frac{1}{10} T^5T^5T^8 + \frac{1}{30\sqrt{3}} T^5T^6T^2 + \frac{1}{30\sqrt{3}} T^5T^7T^1 + \frac15 T^5T^8T^5 - \frac{1}{15\sqrt{3}} T^6T^1T^4 \nonumber \\
& & \mbox{\hglue0.6truecm} - \frac{1}{15\sqrt{3}} T^6T^2T^5 + \frac{1}{15\sqrt{3}} T^6T^3T^6 + \frac{1}{30\sqrt{3}} T^6T^4T^1 + \frac{1}{30\sqrt{3}} T^6T^5T^2 \nonumber \\
& & \mbox{\hglue0.6truecm} - \frac{1}{30\sqrt{3}} T^6T^6T^3 + \frac{1}{10} T^6T^6T^8 + \frac15 T^6T^8T^6 - \frac{1}{15\sqrt{3}} T^7T^1T^5 + \frac{1}{15\sqrt{3}} T^7T^2T^4 \nonumber \\
& & \mbox{\hglue0.6truecm} + \frac{1}{15\sqrt{3}} T^7T^3T^7 - \frac{1}{30\sqrt{3}} T^7T^4T^2 + \frac{1}{30\sqrt{3}} T^7T^5T^1 - \frac{1}{30\sqrt{3}} T^7T^7T^3 + \frac{1}{10} T^7T^7T^8 \nonumber \\
& & \mbox{\hglue0.6truecm} + \frac15 T^7T^8T^7 + \frac{1}{15} T^8T^1T^1 + \frac{1}{15} T^8T^2T^2 + \frac{1}{15} T^8T^3T^3 + \frac{1}{10} T^8T^4T^4 \nonumber \\
& & \mbox{\hglue0.6truecm} + \frac{1}{10} T^8T^5T^5 + \frac{1}{10} T^8T^6T^6 + \frac{1}{10} T^8T^7T^7 + \frac25 T^8T^8T^8. \label{eq:118}
\end{eqnarray}
Similar expressions to Eq.~(\ref{eq:118}) can be found for the \{2,2,8\}, \{1,2,8\}, and \{2,1,8\} components required in the example so it can be shown that
\begin{equation}
[\pr{1} + \pr{8} + \pr{10+\overline{10}} + \pr{27} + \pr{35+\overline{35}} + \pr{64}]^{118cde} \{T^c,\{T^d,T^e\}\} = \{T^1,\{T^1,T^8\}\},
\end{equation}
which is the expected result. Computing the matrix elements of the operator (\ref{eq:118}) is straightforward; therefore,
\begin{equation}
\langle\pi^+n|[{\pr{8}}]^{118cde}\{T^c,\{T^d,T^e\}\}|\pi^+n\rangle = \frac12 \sqrt{3},
\end{equation}
and
\begin{equation}
\langle\pi^+n|[{\pr{r}}]^{118cde}\{T^c,\{T^d,T^e\}\}|\pi^+n\rangle = 0,
\end{equation}
for $r\neq 8$.

The procedure can be repeated for each flavor combination so the different contributions of the operator $\{T^a,\{T^b,T^c\}\}$ to the scattering amplitude of the process $n + \pi^+ \to n + \pi^+$ can be available. For the canonical example, the final expression can be summarized as
\begin{equation}
\frac{1}{\sqrt{2}} \frac{1}{\sqrt{2}} k^i{k^\prime}^j \delta^{ij} \langle\pi^+n|[{\pr{8}}]^{(1-i2)(1-i2)8cde}\{T^c,\{T^d,T^e\}\}|\pi^+n\rangle = \frac12 \sqrt{3} \mathbf{k} \cdot \mathbf{k}^\prime,
\end{equation}
and
\begin{equation}
\frac{1}{\sqrt{2}} \frac{1}{\sqrt{2}} k^i{k^\prime}^j \delta^{ij} \left[ \langle\pi^+n|[{P^{(r)}}]^{(1-i2)(1-i2)8cde} \right] \{T^c,\{T^d,T^e\}\}|\pi^+n\rangle = 0,
\end{equation}
for $r\neq 8$.

Gathering together partial results, first-order SB to the scattering amplitude $\mathcal{A}_\mathrm{LO}$ Eq.~(\ref{eq:sam1}), denoted hereafter by $\delta \mathcal{A}_\mathrm{SB}$ and for which $I=0$, can be organized as
\begin{eqnarray}
& & f^2 k^0 \delta \mathcal{A}_\mathrm{SB}(B + \pi^a \to B^\prime + \pi^b) = \nonumber \\
& & \mbox{\hglue0.2truecm} \sum_\mathrm{dim} \Big[ N_c g_1^{(\mathrm{dim})} k^i{k^\prime}^j\langle \pi^b B^\prime|[\pr{\mathrm{dim}} R_1^{(ij)}]^{(ab8)}|\pi^a B\rangle \nonumber \\
& & \mbox{\hglue0.6truecm} + N_c g_2^{(\mathrm{dim})} k^i{k^\prime}^j\langle \pi^b B^\prime|[\pr{\mathrm{dim}}R_2^{(ij)}]^{(ab8)}|\pi^a B \rangle \nonumber \\
& & \mbox{\hglue0.6truecm} + \sum_{r=3}^{16} g_r^{(\mathrm{dim})} k^i{k^\prime}^j\langle \pi^b B^\prime|[\pr{\mathrm{dim}}R_r^{(ij)}]^{(ab8)}|\pi^a B\rangle \nonumber \\
& & \mbox{\hglue0.6truecm} + \frac{1}{N_c} \sum_{r=17}^{71} g_r^{(\mathrm{dim})} k^i{k^\prime}^j\langle \pi^b B^\prime|[\pr{\mathrm{dim}}R_r^{(ij)}]^{(ab8)}|\pi^a B\rangle \nonumber \\
& & \mbox{\hglue0.6truecm} \nonumber \\
& & \mbox{\hglue0.6truecm} + \frac{1}{N_c^2} \sum_{r=72}^{170} g_r^{(\mathrm{dim})} k^i{k^\prime}^j\langle \pi^b B^\prime|[\pr{\mathrm{dim}}R_r^{(ij)}]^{(ab8)}|\pi^a B\rangle \Big], \label{eq:dA}
\end{eqnarray}
where $g_r^{(\mathrm{dim})}$, $r=1,\ldots,170$, are undetermined coefficients, which are expected to be of order one, the sum over $\mathrm{dim}$ covers all six irreducible representations indicated in relation (\ref{eq:888}) and the sums over $i$ and $j$ are implicit.

For instance, the flavor $1$ piece of $\delta \mathcal{A}_\mathrm{SB}(n+\pi^+ \to n+\pi^+)$, using the corresponding matrix elements of the operators listed in the Online Resource, becomes
\begin{eqnarray}
& & 2\sqrt{3} f^2 k^0 \delta \mathcal{A}_\mathrm{SB}(n+\pi^+ \to n+\pi^+) \nonumber \\
& & \mbox{\hglue0.2truecm} = \Big[ 6 g_{2}^{(1)} + \frac13 {g_{18}^{(1)} + \frac12 g_{20}^{(1)}} + \frac12 g_{52}^{(1)} + \frac12 g_{53}^{(1)} + \frac12 g_{54}^{(1)} +\frac{1}{18} g_{95}^{(1)} + \frac{1}{18} g_{96}^{(1)} + \frac{1}{18} g_{97}^{(1)} \nonumber \\
& & \mbox{\hglue0.6truecm} + \frac{1}{18} g_{98}^{(1)} + \frac{1}{18} g_{99}^{(1)} + \frac{1}{18} g_{100}^{(1)} + \frac13 g_{110}^{(1)} + \frac13 g_{111}^{(1)} + \frac13 g_{112}^{(1)} + \frac19 g_{116}^{(1)} + \frac19 g_{117}^{(1)} + \frac19 g_{118}^{(1)} \nonumber \\
& & \mbox{\hglue0.6truecm} + \frac19 g_{119}^{(1)} + \frac19 g_{120}^{(1)} + \frac19 g_{121}^{(1)} + \frac16 g_{134}^{(1)} + \frac16 g_{135}^{(1)} + \frac16 g_{136}^{(1)} + \frac16 g_{137}^{(1)} + \frac16 g_{138}^{(1)} + \frac16 g_{139}^{(1)} \nonumber \\
& & \mbox{\hglue0.6truecm} + \frac16 g_{140}^{(1)} + \frac16 g_{141}^{(1)} + \frac16 g_{142}^{(1)} + \frac16 g_{143}^{(1)} + \frac16 g_{144}^{(1)} + \frac16 g_{145}^{(1)} - \frac{1}{18} g_{146}^{(1)} -\frac{1}{18} g_{147}^{(1)} - \frac{1}{18} g_{148}^{(1)} \nonumber \\
& & \mbox{\hglue0.6truecm} - \frac{1}{18} g_{149}^{(1)} \Big] \mathbf{k} \cdot \mathbf{k}^\prime + \Big[ g_{4}^{(1)} + \frac16 g_{63}^{(1)} + \frac16 g_{64}^{(1)} + \frac16 g_{65}^{(1)} + \frac16 g_{66}^{(1)} + \frac16 g_{67}^{(1)} + \frac16 g_{68}^{(1)} \nonumber \\
& & \mbox{\hglue0.6truecm} + \frac16 g_{73}^{(1)} + \frac16 g_{80}^{(1)} + \frac16 g_{81}^{(1)} + \frac16 g_{82}^{(1)} + \frac{1}{12} g_{122}^{(1)} + \frac{1}{12} g_{123}^{(1)} + \frac{1}{12} g_{124}^{(1)} + \frac18 g_{125}^{(1)} + \frac18 g_{126}^{(1)} \nonumber \\
& & \mbox{\hglue0.6truecm} + \frac18 g_{127}^{(1)} - \frac{1}{24} g_{128}^{(1)} - \frac{1}{24} g_{129}^{(1)} - \frac{1}{24} g_{130}^{(1)} - \frac{1}{16} g_{150}^{(1)} - \frac{1}{16} g_{151}^{(1)} - \frac{1}{16} g_{152}^{(1)} - \frac{1}{16} g_{153}^{(1)} \nonumber \\
& & \mbox{\hglue0.6truecm} - \frac{1}{16} g_{154}^{(1)} - \frac{1}{16} g_{155}^{(1)} - \frac{1}{16} g_{156}^{(1)} - \frac{1}{16} g_{157}^{(1)} - \frac{1}{16} g_{158}^{(1)} - \frac{1}{16} g_{159}^{(1)} - \frac{1}{16} g_{160}^{(1)} - \frac{1}{16} g_{161}^{(1)} \nonumber \\
& & \mbox{\hglue0.6truecm} - \frac{1}{16} g_{162}^{(1)} - \frac{1}{16} g_{163}^{(1)} - \frac{1}{16} g_{164}^{(1)} + \frac{1}{16} g_{165}^{(1)} + \frac{1}{16} g_{166}^{(1)} + \frac{1}{16} g_{167}^{(1)} + \frac{1}{16} g_{168}^{(1)} + \frac{1}{16} g_{169}^{(1)} \nonumber \\
& & \mbox{\hglue0.6truecm} + \frac{1}{16} g_{170}^{(1)} \Big] i (\mathbf{k} \times \mathbf{k}^\prime)_3. \label{eq:de11}
\end{eqnarray}
The applicability of expressions such as (\ref{eq:de11}) is, however, hindered by several disadvantages. The obvious one is the impossibility of determining all free parameters. For the $N\pi \to N\pi$ process, simpler expressions are obtained by defining effective coefficients expressed in terms of linear combinations of the $g_r^{(\mathrm{dim})}$ ones. In view of this, Eq.~(\ref{eq:de11}) can be simply written as
\begin{equation}
f^2 k^0 \delta \mathcal{A}_\mathrm{SB}(n+\pi^+ \to n+\pi^+) = d_1^{(1)} \mathbf{k} \cdot \mathbf{k}^\prime + e_1^{(1)} i (\mathbf{k} \times \mathbf{k}^\prime)_3. \label{eq:de11d}
\end{equation}
where the $d_1^{(1)}$ and $e_1^{(1)}$ coefficients are easily read off Eq.~(\ref{eq:de11}).

Thus, the final expressions obtained for first-order SB effects to the scattering amplitudes for the $N+\pi\to N+\pi$ process are given by
\begin{eqnarray}
f^2 k^0 \delta \mathcal{A}_\mathrm{SB} (p + \pi^+ \to p + \pi^+) & = & (d_{1}^{(1)} + d_{1}^{(8)} + d_{1}^{(10+\overline{10})} + d_{1}^{(27)}) \mathbf{k} \cdot \mathbf{k}^\prime \nonumber \\
& & \mbox{} + (e_{1}^{(1)} + e_{1}^{(8)} + e_{1}^{(10+\overline{10})} + e_{1}^{(27)}) i (\mathbf{k} \times \mathbf{k}^\prime)_3 \nonumber \\
& = & f^2 k^0 \delta \mathcal{A}_\mathrm{SB} (n + \pi^- \to n + \pi^-), \label{eq:sb1}
\end{eqnarray}
\begin{eqnarray}
f^2 k^0 \delta \mathcal{A}_\mathrm{SB} (p + \pi^- \to p + \pi^-) & = & (d_{1}^{(1)} + d_{1}^{(8)} - d_{1}^{(10+\overline{10})} - d_{1}^{(27)} + d_{2}^{(8)} + d_{2}^{(27)}) \mathbf{k} \cdot \mathbf{k}^\prime \nonumber \\
& & \mbox{} + (e_{1}^{(1)} + e_{1}^{(8)} - e_{1}^{(10+\overline{10})} + e_{2}^{(8)}) i (\mathbf{k} \times \mathbf{k}^\prime)_3 \nonumber \\
& = & f^2 k^0 \delta \mathcal{A}_\mathrm{SB} (n + \pi^+ \to n + \pi^+),
\end{eqnarray}
\begin{eqnarray}
f^2 k^0 \delta \mathcal{A}_\mathrm{SB} (p + \pi^0 \to p + \pi^0) & = & \frac12 (2 d_{1}^{(1)} + 2 d_{1}^{(8)} + d_{2}^{(8)} + d_{2}^{(27)}) \mathbf{k} \cdot \mathbf{k}^\prime \nonumber \\
& & \mbox{} + \frac12 (2 e_{1}^{(1)} + 2 e_{1}^{(8)} + e_{1}^{(27)}+ e_{2}^{(8)}) i (\mathbf{k} \times \mathbf{k}^\prime)_3 \nonumber \\
& = & f^2 k^0 \delta \mathcal{A}_\mathrm{SB} (n + \pi^0 \to n + \pi^0),
\end{eqnarray}
\begin{eqnarray}
\sqrt{2} f^2 k^0 \delta \mathcal{A}_\mathrm{SB} (p + \pi^- \to n + \pi^0) & = & (2 d_{1}^{(10+\overline{10})} + 2 d_{1}^{(27)} - d_{2}^{(8)} - d_{2}^{(27)}) \mathbf{k} \cdot \mathbf{k}^\prime \nonumber \\
& & \mbox{} + (2 e_{1}^{(10+\overline{10})} + e_{1}^{(27)} - e_{2}^{(8)}) i (\mathbf{k} \times \mathbf{k}^\prime)_3 \nonumber \\
& = & \sqrt{2} f^2 k^0 \delta \mathcal{A}_\mathrm{SB} (n + \pi^+ \to p + \pi^0). \label{eq:sb8}
\end{eqnarray}
Expressions (\ref{eq:sb1})-(\ref{eq:sb8}) are written in terms of 11 unknown parameters, which contain implicit suppression factors in $N_c$, so they are expected to be $\mathcal{O}(N_c^0)$, $\mathcal{O}(N_c^{-1})$, and $\mathcal{O}(N_c^{-2})$ for coefficients coming from $1$, $8$ and $10+\overline{10}$, and $27$ representations, respectively. It should be highlighted that neither flavor $35+\overline{35}$ nor flavor $64$ representation participates in the final expressions.

The isospin relations (\ref{eq:is1})-(\ref{eq:is6}) are satisfied by corrections to scattering amplitudes (\ref{eq:sb1})-(\ref{eq:sb8}), which is a completely expected result.

Furthermore,
\begin{eqnarray}
f^2 k^0 \delta \mathcal{A}_\mathrm{SB}^{(3/2)} & = & (d_{1}^{(1)} + d_{1}^{(8)} + d_{1}^{(10+\overline{10})} + d_{1}^{(27)}) \mathbf{k} \cdot \mathbf{k}^\prime \nonumber \\
& & \mbox{} + (e_{1}^{(1)} + e_{1}^{(8)} + e_{1}^{(10+\overline{10})} + e_{1}^{(27)}) i (\mathbf{k} \times \mathbf{k}^\prime)_3, \label{eq:asb32}
\end{eqnarray}
and
\begin{eqnarray}
f^2 k^0 \delta \mathcal{A}_\mathrm{SB}^{(1/2)} & = & \left[ d_{1}^{(1)} + d_{1}^{(8)} - 2 d_{1}^{(10+\overline{10})} - 2 d_{1}^{(27)} + \frac32 d_{2}^{(8)} + \frac32 d_{2}^{(27)} \right] \mathbf{k} \cdot \mathbf{k}^\prime \nonumber \\
& & \mbox{} + \left[ e_{1}^{(1)} + e_{1}^{(8)} - 2 e_{1}^{(10+\overline{10})} - \frac12 e_{1}^{(27)} + \frac32 d_{2}^{(8)} \right] i (\mathbf{k} \times \mathbf{k}^\prime)_3. \label{eq:asb12}
\end{eqnarray}

\subsubsection{SB effects to the scattering amplitude from Fig.~\ref{fig:sp}(c)}

The SB effects to the scattering amplitude from Fig.~\ref{fig:sp}(c) are obtained following the lines of the previous section. In this case, $A_{\mathrm{vertex}}^{ab}$ Eq.~(\ref{eq:vtx}) is a spin-zero object and contains two adjoint indices. A straightforward way to obtain the spin-0 operators with three adjoint indices to account for SB is by forming tensor products of $R_k^{(ij)(abcd)}$ listed in the Online Resource with $\delta^{ij}$ to saturate spin indices. With this procedure, out of the 170 original operators, only 59 remain. The corresponding operator basis $\{V^{abc}\}$ is also listed in the Online Resource. However, after repeating the computation of the action of the flavor projectors on these 59 operators, computing matrix elements and gathering together partial results, only one unknown parameter is required to parametrize SB effects from Fig.~\ref{fig:sp}(c). The final forms of the amplitudes read
\begin{eqnarray}
\delta \mathcal{A}_\mathrm{vertex} (p + \pi^ + \to p + \pi^+) & = & - \frac14 \frac{k^0}{f^2} h_1 \nonumber \\
& = & \delta \mathcal{A}_\mathrm{vertex} (n + \pi^- \to n + \pi^-), \label{eq:ppppppvtxSB}
\end{eqnarray}
\begin{eqnarray}
\delta \mathcal{A}_\mathrm{vertex} (p + \pi^- \to p + \pi^-) & = & - \frac14 \frac{k^0}{f^2} h_1 \nonumber \\
& = & \delta \mathcal{A}_\mathrm{vertex} (n + \pi^+ \to n + \pi^+),
\end{eqnarray}
\begin{eqnarray}
\delta \mathcal{A}_\mathrm{vertex} (p + \pi^0 \to p + \pi^0) & = & - \frac14 \frac{k^0}{f^2} h_1 \nonumber \\
& = & \delta \mathcal{A}_\mathrm{vertex} (n + \pi^0 \to n + \pi^0),
\end{eqnarray}
\begin{eqnarray}
\delta \mathcal{A}_\mathrm{vertex} (p + \pi^- \to n + \pi^0) & = & 0 \nonumber \\
& = & \delta \mathcal{A}_\mathrm{vertex} (n + \pi^+ \to p + \pi^0), \label{eq:nppppzvtxSB}
\end{eqnarray}
where $h_1$ is a new unknown parameter, which is a linear combination of $1$, $8$ and $27$ operator coefficients only. Notice that $\mathcal{A}_\mathrm{vertex} (p + \pi^0 \to p + \pi^0)$ and $\mathcal{A}_\mathrm{vertex} (n + \pi^0 \to n + \pi^0)$ are no longer vanishing due to SB, whereas $\mathcal{A}_\mathrm{vertex} (p + \pi^- \to n + \pi^0)$ and $\mathcal{A}_\mathrm{vertex} (n + \pi^+ \to p + \pi^0)$ are unchanged. Also notice that the isospin relations (\ref{eq:ppppppvtx})-(\ref{eq:nppppzvtx}) are unaffected by SB effects, as expected.

Similarly,
\begin{equation}
\delta \mathcal{A}_\mathrm{SB,vertex}^{(3/2)} = \frac14 \frac{k^0}{f^2} h_1, \label{eq:aav32}
\end{equation}
and
\begin{equation}
\delta \mathcal{A}_\mathrm{SB,vertex}^{(1/2)} = - \frac12 \frac{k^0}{f^2} h_1. \label{eq:aav12}
\end{equation}

\subsubsection{Strong isospin breaking to the scattering amplitude from Fig.~\ref{fig:sp}(a,b)}

The evaluation of IB corrections to the scattering amplitudes, hereafter denoted by $\delta \mathcal{A}_\mathrm{IB}$, can be performed in a similar fashion to flavor SB described in the previous sections, except that the free flavor index is now fixed to 3. The corresponding $1/N_c$ expansion, for which $I=1$, reads,
\begin{eqnarray}
& & f^2 k^0 \delta \mathcal{A}_\mathrm{IB}(B + \pi^a \to B^\prime + \pi^b) = \nonumber \\
& & \mbox{\hglue0.2truecm} \sum_\mathrm{dim} \Big[ N_c s_1^{(\mathrm{dim})} k^i{k^\prime}^j\langle \pi^b B^\prime|[\pr{\mathrm{dim}} R_1^{(ij)}]^{(ab3)}|\pi^a B\rangle \nonumber \\
& & \mbox{\hglue0.6truecm} + N_c s_2^{(\mathrm{dim})} k^i{k^\prime}^j\langle \pi^b B^\prime|[\pr{\mathrm{dim}}R_2^{(ij)}]^{(ab3)}|\pi^a B\rangle \nonumber \\
& & \mbox{\hglue0.6truecm} + \sum_{r=3}^{16} s_r^{(\mathrm{dim})} k^i{k^\prime}^j\langle \pi^b B^\prime|[\pr{\mathrm{dim}}R_r^{(ij)}]^{(ab3)}|\pi^a B\rangle \nonumber \\
& & \mbox{\hglue0.6truecm} + \frac{1}{N_c} \sum_{r=17}^{71} s_r^{(\mathrm{dim})} k^i{k^\prime}^j\langle \pi^b B^\prime|[\pr{\mathrm{dim}}R_r^{(ij)}]^{(ab3)}|\pi^a B\rangle \nonumber \\
& & \mbox{\hglue0.6truecm} \nonumber \\
& & \mbox{\hglue0.6truecm} + \frac{1}{N_c^2} \sum_{r=72}^{170} s_r^{(\mathrm{dim})} k^i{k^\prime}^j\langle \pi^b B^\prime|[\pr{\mathrm{dim}}R_r^{(ij)}]^{(ab3)}|\pi^a B\rangle \Big], \label{eq:stri}
\end{eqnarray}
where $s_r^{(\mathrm{dim})}$, $r=1,\ldots,170$, are undetermined coefficients, which are expected to be of order one, the sum over $\mathrm{dim}$ covers all six irreducible representations indicated in relation (\ref{eq:888}) and the sums over $i$ and $j$ are implicit.

Matrix elements of expression (\ref{eq:stri}) can be straightforwardly obtained following the lines of the previous sections. This allows one to obtain violations to isospin relations (\ref{eq:is1})-(\ref{eq:is6}) as
\begin{eqnarray}
& & f^2k^0 \Big[ \delta \mathcal{A}_\mathrm{IB}(p + \pi^- \to p + \pi^-) - \delta \mathcal{A}_\mathrm{IB}(p + \pi^0 \to p + \pi^0) + \frac{1}{\sqrt{2}}\delta \mathcal{A}_\mathrm{IB}(p + \pi^- \to n + \pi^0) \Big] \nonumber \\
& & \mbox{\hglue0.2truecm} = \Big[ \Big[ - N_c w_{1}^{(1)} - \frac{1}{N_c} w_{3}^{(1)} - \frac{1}{N_c} w_{4}^{(1)} \Big] + \Big[ \frac34 w_{1}^{(8)} - w_{2}^{(8)} + w_{3}^{(8)} + \frac{3}{4N_c} w_{19}^{(8)} - \frac{1}{N_c} w_{20}^{(8)} \nonumber \\
& & \mbox{\hglue0.9truecm} + \frac{1}{N_c} w_{21}^{(8)} + \frac{1}{2N_c} w_{22}^{(8)} + \frac{1}{2N_c} w_{23}^{(8)} + \frac{3}{4N_c} w_{24}^{(8)} - \frac{1}{N_c} w_{25}^{(8)} + \frac{3}{4N_c} w_{26}^{(8)} - \frac{1}{N_c} w_{27}^{(8)} \nonumber \\
& & \mbox{\hglue0.9truecm} + \frac{1}{N_c} w_{28}^{(8)} - \frac{1}{N_c} w_{29}^{(8)} + \frac{3}{4N_c} w_{30}^{(8)} - \frac{1}{N_c} w_{31}^{(8)} + \frac{1}{N_c} w_{32}^{(8)} \Big] + \Big[ - \frac{1}{N_c} w_{1}^{(27)} \nonumber \\
& & \mbox{\hglue0.9truecm} + \frac{3}{4N_c} w_{2}^{(27)} - \frac{1}{N_c} w_{3}^{(27)} + \frac{1}{N_c} w_{4}^{(27)} \Big] \Big] \mathbf{k}\cdot \mathbf{k}^\prime + \Big[ \Big[ - w_{2}^{(1)} - \frac{1}{N_c} w_{4}^{(1)} \Big] \nonumber \\
& & \mbox{\hglue0.9truecm} + \Big[ \frac34 w_{4}^{(8)} - w_{5}^{(8)} + w_{6}^{(8)} + \frac{1}{2} w_{7}^{(8)} + \frac{1}{2} w_{8}^{(8)} + \frac34 w_{9}^{(8)} - w_{10}^{(8)} + \frac{3}{4N_c} w_{11}^{(8)} - \frac{1}{N_c} w_{12}^{(8)} \nonumber \\
& & \mbox{\hglue0.9truecm} + \frac{1}{N_c} w_{13}^{(8)} + \frac{1}{2N_c} w_{14}^{(8)} + \frac{1}{2N_c} w_{15}^{(8)} + \frac{3}{4N_c} w_{16}^{(8)} - \frac{1}{N_c} w_{17}^{(8)} - \frac{1}{N_c} w_{18}^{(8)} - \frac{1}{N_c} w_{33}^{(8)} \nonumber \\
& & \mbox{\hglue0.9truecm} + \frac{3}{4N_c} w_{34}^{(8)} + \frac{1}{N_c} w_{35}^{(8)} - \frac{1}{N_c} w_{36}^{(8)} \Big] + \Big[ \frac{3}{4N_c} w_{1}^{(10+\overline{10})} - \frac{1}{N_c} w_{2}^{(10+\overline{10})} + \frac{1}{N_c} w_{3}^{(10+\overline{10})} \Big] \nonumber \\
& & \mbox{\hglue0.9truecm} + \Big[ - \frac{1}{N_c} w_{5}^{(27)} + \frac{3}{4N_c} w_{6}^{(27)} + \frac{1}{N_c} w_{7}^{(27)} - \frac{1}{N_c} w_{8}^{(27)} + \frac{3}{4N_c} w_{9}^{(27)} + \frac{1}{N_c} w_{10}^{(27)} \Big] \Big] i (\mathbf{k}\times \mathbf{k}^\prime)_3, \nonumber \\
& & \mbox{\hglue0.9truecm} + \mathcal{O}\left[ \frac{1}{N_c^2} \right] \label{eq:isb1}
\end{eqnarray}

\begin{eqnarray}
& & f^2k^0 \Big[ \delta \mathcal{A}_\mathrm{IB}(p + \pi^+ \to p + \pi^+) - \delta \mathcal{A}_\mathrm{IB}(p + \pi^- \to p + \pi^-) - \sqrt{2}\delta \mathcal{A}_\mathrm{IB}(p + \pi^- \to n + \pi^0) \Big] \nonumber \\
& & \mbox{\hglue0.2truecm} = \Big[ \Big[ 2 N_c w_{1}^{(1)} + \frac{2}{N_c} w_{3}^{(1)} + \frac{2}{N_c} w_{4}^{(1)} \Big] + \Big[ - 4 w_{3}^{(8)} - \frac{4}{N_c} w_{21}^{(8)} - \frac{1}{N_c} w_{22}^{(8)} + \frac{1}{N_c} w_{23}^{(8)} \nonumber \\
& & \mbox{\hglue0.9truecm} + \frac{2}{N_c} w_{25}^{(8)} - \frac{4}{N_c} w_{28}^{(8)} + \frac{2}{N_c} w_{29}^{(8)} - \frac{4}{N_c} w_{32}^{(8)} \Big] + \Big[ \frac{2}{N_c} w_{1}^{(27)} - \frac{4}{N_c} w_{4}^{(27)} \Big] \Big] \mathbf{k}\cdot \mathbf{k}^\prime \nonumber \\
& & \mbox{\hglue0.9truecm} + \Big[ \Big[ 2 w_{2}^{(1)} + \frac{2}{N_c} w_{4}^{(1)} \Big] + \Big[ - 4 w_{6}^{(8)} - w_{7}^{(8)} + w_{8}^{(8)} + 2 w_{10}^{(8)} - \frac{4}{N_c} w_{13}^{(8)} - \frac{1}{N_c} w_{14}^{(8)} \nonumber \\
& & \mbox{\hglue0.9truecm} + \frac{1}{N_c} w_{15}^{(8)} + \frac{2}{N_c} w_{17}^{(8)} + \frac{2}{N_c} w_{18}^{(8)} + \frac{2}{N_c} w_{33}^{(8)} + \frac{4}{N_c} w_{36}^{(8)} \Big] + \Big[ - \frac{4}{N_c} w_{3}^{(10+\overline{10})} \Big] \nonumber \\
& & \mbox{\hglue0.9truecm} + \Big[ \frac{2}{N_c} w_{5}^{(27)} + \frac{4}{N_c} w_{8}^{(27)} \Big] \Big]
 i (\mathbf{k}\times \mathbf{k}^\prime)_3 + \mathcal{O}\left[ \frac{1}{N_c^2} \right], \label{eq:isb2}
\end{eqnarray}

\begin{eqnarray}
& & f^2k^0 \Big[ \delta \mathcal{A}_\mathrm{IB}(p + \pi^+ \to p + \pi^+) + \delta \mathcal{A}_\mathrm{IB}(p + \pi^- \to p + \pi^-) - 2 \delta \mathcal{A}_\mathrm{IB}(p + \pi^0 \to p + \pi^0) \Big] \nonumber \\
& & \mbox{\hglue0.2truecm} = \Big[ \Big[ \frac32 w_{1}^{(8)} - 2 w_{2}^{(8)} - 2 w_{3}^{(8)} + \frac{3}{2N_c} w_{19}^{(8)} - \frac{2}{N_c} w_{20}^{(8)} - \frac{2}{N_c} w_{21}^{(8)} + \frac{2}{N_c} w_{23}^{(8)} + \frac{3}{2N_c} w_{24}^{(8)} \nonumber \\
& & \mbox{\hglue0.9truecm} + \frac{3}{2N_c} w_{26}^{(8)} - \frac{2}{N_c} w_{27}^{(8)} - \frac{2}{N_c} w_{28}^{(8)} + \frac{3}{2N_c} w_{30}^{(8)} - \frac{2}{N_c} w_{31}^{(8)} - \frac{2}{N_c} w_{32}^{(8)} \Big] + \Big[ \frac{3}{2N_c} w_{2}^{(27)} \nonumber \\
& & \mbox{\hglue0.9truecm} - \frac{2}{N_c} w_{3}^{(27)} - \frac{2}{N_c} w_{4}^{(27)} \Big] \Big] \mathbf{k}\cdot \mathbf{k}^\prime + \Big[ \Big[ \frac32 w_{4}^{(8)} - 2 w_{5}^{(8)} - 2 w_{6}^{(8)} + 2 w_{8}^{(8)} + \frac32 w_{9}^{(8)} \nonumber \\
& & \mbox{\hglue0.9truecm} + \frac{3}{2N_c} w_{11}^{(8)} - \frac{2}{N_c} w_{12}^{(8)} - \frac{2}{N_c} w_{13}^{(8)} + \frac{2}{N_c} w_{15}^{(8)} + \frac{3}{2N_c} w_{16}^{(8)} + \frac{3}{2N_c} w_{34}^{(8)} + \frac{2}{N_c} w_{35}^{(8)} \nonumber \\
& & \mbox{\hglue0.9truecm} + \frac{2}{N_c} w_{36}^{(8)} \Big] + \Big[ \frac{3}{2N_c} w_{1}^{(10+\overline{10})} - \frac{2}{N_c} w_{2}^{(10+\overline{10})} - \frac{2}{N_c} w_{3}^{(10+\overline{10})} \Big] + \Big[ \frac{3}{2N_c} w_{6}^{(27)} \nonumber \\
& & \mbox{\hglue0.9truecm} + \frac{2}{N_c} w_{7}^{(27)} + \frac{2}{N_c} w_{8}^{(27)} + \frac{3}{2N_c} w_{9}^{(27)} + \frac{2}{N_c}
 w_{10}^{(27)} \Big] \Big] i (\mathbf{k}\times \mathbf{k}^\prime)_3 + \mathcal{O}\left[ \frac{1}{N_c^2} \right], \label{eq:isb3}
\end{eqnarray}

\begin{eqnarray}
& & f^2k^0 \Big[ \delta \mathcal{A}_\mathrm{IB}(n + \pi^- \to n + \pi^-) - \delta \mathcal{A}_\mathrm{IB}(n + \pi^0 \to n + \pi^0) - \frac{1}{\sqrt{2}}\delta \mathcal{A}_\mathrm{IB}(n + \pi^+ \to p + \pi^0) \Big] \nonumber \\
& & \mbox{\hglue0.2truecm} = \Big[ \Big[ - N_c w_{1}^{(1)} - \frac{1}{N_c} w_{3}^{(1)} - \frac{1}{N_c} w_{4}^{(1)} \Big] + \Big[ - \frac34 w_{1}^{(8)} + w_{2}^{(8)} - w_{3}^{(8)} - \frac{3}{4N_c} w_{19}^{(8)} \nonumber \\
& & \mbox{\hglue0.9truecm} + \frac{1}{N_c} w_{20}^{(8)} - \frac{1}{N_c} w_{21}^{(8)} - \frac{1}{2N_c} w_{22}^{(8)} - \frac{1}{2N_c} w_{23}^{(8)} - \frac{3}{4N_c} w_{24}^{(8)} - \frac{1}{N_c} w_{25}^{(8)} - \frac{3}{4N_c} w_{26}^{(8)} \nonumber \\
& & \mbox{\hglue0.9truecm} + \frac{1}{N_c} w_{27}^{(8)} - \frac{1}{N_c} w_{28}^{(8)} - \frac{1}{N_c} w_{29}^{(8)} - \frac{3}{4N_c} w_{30}^{(8)} + \frac{1}{N_c} w_{31}^{(8)} - \frac{1}{N_c} w_{32}^{(8)} \Big] \nonumber \\
& & \mbox{\hglue0.9truecm} + \Big[ - \frac{1}{N_c}w_{1}^{(27)} - \frac{3}{4N_c} w_{2}^{(27)} + \frac{1}{N_c} w_{3}^{(27)} - \frac{1}{N_c} w_{4}^{(27)} \Big] \Big] \mathbf{k}\cdot \mathbf{k}^\prime \nonumber \\
& & \mbox{\hglue0.9truecm} + \Big[ \Big[ - w_{2}^{(1)} - \frac{1}{N_c} w_{4}^{(1)} \Big] + \Big[ - \frac34 w_{4}^{(8)} + w_{5}^{(8)} - w_{6}^{(8)} - \frac{1}{2} w_{7}^{(8)} - \frac{1}{2} w_{8}^{(8)} - \frac34 w_{9}^{(8)} \nonumber \\
& & \mbox{\hglue0.9truecm} - w_{10}^{(8)} - \frac{3}{4N_c} w_{11}^{(8)} + \frac{1}{N_c} w_{12}^{(8)} - \frac{1}{N_c} w_{13}^{(8)} - \frac{1}{2N_c} w_{14}^{(8)} - \frac{1}{2N_c} w_{15}^{(8)} - \frac{3}{4N_c} w_{16}^{(8)} \nonumber \\
& & \mbox{\hglue0.9truecm} - \frac{1}{N_c} w_{17}^{(8)} - \frac{1}{N_c} w_{18}^{(8)} - \frac{1}{N_c} w_{33}^{(8)} - \frac{3}{4N_c} w_{34}^{(8)} - \frac{1}{N_c} w_{35}^{(8)} + \frac{1}{N_c} w_{36}^{(8)} \Big] \nonumber \\
& & \mbox{\hglue0.9truecm} + \Big[ - \frac{3}{4N_c} w_{1}^{(10+\overline{10})} + \frac{1}{N_c} w_{2}^{(10+\overline{10})} - \frac{1}{N_c} w_{3}^{(10+\overline{10})} \Big] + \Big[ - \frac{1}{N_c} w_{5}^{(27)} - \frac{3}{4N_c} w_{6}^{(27)} \nonumber \\
& & \mbox{\hglue0.9truecm} - \frac{1}{N_c} w_{7}^{(27)} + \frac{1}{N_c} w_{8}^{(27)} - \frac{3}{4N_c} w_{9}^{(27)} - \frac{1}{N_c} w_{10}^{(27)} \Big] \Big] i (\mathbf{k}\times \mathbf{k}^\prime)_3 + \mathcal{O}\left[ \frac{1}{N_c^2} \right], \label{eq:isb4}
\end{eqnarray}

\begin{eqnarray}
& & f^2k^0 \Big[ \delta \mathcal{A}_\mathrm{IB}(n + \pi^+ \to n + \pi^+) - \delta \mathcal{A}_\mathrm{IB}(n + \pi^- \to n + \pi^-) + \sqrt{2} \delta \mathcal{A}_\mathrm{IB}(n + \pi^+ \to p + \pi^0) \Big] \nonumber \\
& & \mbox{\hglue0.2truecm} = \Big[ \Big[ 2 N_c w_{1}^{(1)} + \frac{2}{N_c} w_{3}^{(1)} + \frac{2}{N_c} w_{4}^{(1)} \Big] + \Big[ 4 w_{3}^{(8)} + \frac{4}{N_c} w_{21}^{(8)} + \frac{1}{N_c} w_{22}^{(8)} - \frac{1}{N_c} w_{23}^{(8)} \nonumber \\
& & \mbox{\hglue0.9truecm} + \frac{2}{N_c} w_{25}^{(8)} + \frac{4}{N_c} w_{28}^{(8)} + \frac{2}{N_c} w_{29}^{(8)} + \frac{4}{N_c} w_{32}^{(8)} \Big] + \Big[ \frac{2}{N_c} w_{1}^{(27)} + \frac{4}{N_c} w_{4}^{(27)} \Big] \Big] \mathbf{k}\cdot \mathbf{k}^\prime \nonumber \\
& & \mbox{\hglue0.9truecm} + \Big[ \Big[ 2 w_{2}^{(1)} + \frac{2}{N_c} w_{4}^{(1)} \Big] + \Big[ 4 w_{6}^{(8)} + w_{7}^{(8)} - w_{8}^{(8)} + 2 w_{10}^{(8)} + \frac{4}{N_c} w_{13}^{(8)} + \frac{1}{N_c} w_{14}^{(8)} \nonumber \\
& & \mbox{\hglue0.9truecm} - \frac{1}{N_c} w_{15}^{(8)} + \frac{2}{N_c} w_{17}^{(8)} + \frac{2}{N_c} w_{18}^{(8)} + \frac{2}{N_c} w_{33}^{(8)} - \frac{4}{N_c} w_{36}^{(8)} \Big] + \Big[ \frac{4}{N_c} w_{3}^{(10+\overline{10})} \Big] \nonumber \\
& & \mbox{\hglue0.9truecm} + \Big[ \frac{2}{N_c} w_{5}^{(27)} - \frac{4}{N_c} w_{8}^{(27)} \Big] \Big] i (\mathbf{k}\times \mathbf{k}^\prime)_3 + \mathcal{O}\left[ \frac{1}{N_c^2} \right], \label{eq:isb5}
\end{eqnarray}

\begin{eqnarray}
& & f^2k^0 \Big[ \delta \mathcal{A}_\mathrm{IB}(n + \pi^+ \to n + \pi^+) + \delta \mathcal{A}_\mathrm{IB}(n + \pi^- \to n + \pi^-) - 2 \delta \mathcal{A}_\mathrm{IB}(n + \pi^0n \to \pi^0) \nonumber \\
& & \mbox{\hglue0.2truecm} = \Big[ \Big[ - \frac32 w_{1}^{(8)} + 2 w_{2}^{(8)} + 2 w_{3}^{(8)} - \frac{3}{2N_c} w_{19}^{(8)} + \frac{2}{N_c} w_{20}^{(8)} + \frac{2}{N_c} w_{21}^{(8)} - \frac{2}{N_c} w_{23}^{(8)} \nonumber \\
& & \mbox{\hglue0.9truecm} - \frac{3}{2N_c} w_{24}^{(8)} - \frac{3}{2N_c} w_{26}^{(8)} + \frac{2}{N_c} w_{27}^{(8)} + \frac{2}{N_c} w_{28}^{(8)} - \frac{3}{2N_c} w_{30}^{(8)} + \frac{2}{N_c} w_{31}^{(8)} + \frac{2}{N_c} w_{32}^{(8)} \Big] \nonumber \\
& & \mbox{\hglue0.9truecm} + \Big[ - \frac{3}{2N_c} w_{2}^{(27)} + \frac{2}{N_c} w_{3}^{(27)} + \frac{2}{N_c} w_{4}^{(27)} \Big] \Big] \mathbf{k}\cdot \mathbf{k}^\prime + \Big[ \Big[ - \frac32 w_{4}^{(8)} + 2 w_{5}^{(8)} + 2 w_{6}^{(8)} \nonumber \\
& & \mbox{\hglue0.9truecm} - 2 w_{8}^{(8)} - \frac32 w_{9}^{(8)} - \frac{3}{2N_c} w_{11}^{(8)} + \frac{2}{N_c} w_{12}^{(8)} + \frac{2}{N_c} w_{13}^{(8)} - \frac{2}{N_c} w_{15}^{(8)} - \frac{3}{2N_c} w_{16}^{(8)} \nonumber \\
& & \mbox{\hglue0.9truecm} - \frac{3}{2N_c} w_{34}^{(8)} - \frac{2}{N_c} w_{35}^{(8)} - \frac{2}{N_c} w_{36}^{(8)} \Big] + \Big[ - \frac{3}{2N_c} w_{1}^{(10+\overline{10})} + \frac{2}{N_c} w_{2}^{(10+\overline{10})} \nonumber \\
& & \mbox{\hglue0.9truecm} + \frac{2}{N_c} w_{3}^{(10+\overline{10})} \Big] + \Big[ - \frac{3}{2N_c} w_{6}^{(27)} - \frac{2}{N_c} w_{7}^{(27)} - \frac{2}{N_c} w_{8}^{(27)} - \frac{3}{2N_c} w_{9}^{(27)} \nonumber \\
& & \mbox{\hglue0.9truecm} - \frac{2}{N_c} w_{10}^{(27)} \Big] \Big] i (\mathbf{k}\times \mathbf{k}^\prime)_3 + \mathcal{O}\left[ \frac{1}{N_c^2} \right]. \label{eq:isb6}
\end{eqnarray}

The effective coefficients $w_m^{(\mathrm{dim})}$ can be written in terms of the original ones as
\begin{eqnarray}
& & w_{1}^{(1)} = -2 s_{1}^{(1)}, \\
& & w_{2}^{(1)} = s_{3}^{(1)}, \\
& & w_{3}^{(1)} = -s_{17}^{(1)} - \frac32 s_{19}^{(1)}, \\
& & w_{4}^{(1)} = -\frac32 s_{51}^{(1)}, \\
& & w_{1}^{(8)} = s_{5}^{(8)}, \\
& & w_{2}^{(8)} = \frac14 s_{6}^{(8)}, \\
& & w_{3}^{(8)} = \frac14 s_{7}^{(8)}, \\
& & w_{4}^{(8)} = \frac56 s_{8}^{(8)}, \\
& & w_{5}^{(8)} = \frac{5}{24} s_{9}^{(8)}, \\
& & w_{6}^{(8)} = \frac{5}{24} s_{10}^{(8)}, \\
& & w_{7}^{(8)} = \frac56 s_{11}^{(8)}, \\
& & w_{8}^{(8)} = -\frac56 s_{12}^{(8)}, \\
& & w_{9}^{(8)} = \frac{5}{18} s_{13}^{(8)}, \\
& & w_{10}^{(8)} = \frac16 (s_{14}^{(8)} - s_{15}^{(8)}), \\
& & w_{11}^{(8)} = \frac12 s_{21}^{(8)}, \\
& & w_{12}^{(8)} = \frac18 s_{22}^{(8)}, \\
& & w_{13}^{(8)} = \frac18 s_{23}^{(8)}, \\
& & w_{14}^{(8)} = \frac12 s_{24}^{(8)}, \\
& & w_{15}^{(8)} = -\frac12 s_{25}^{(8)}, \\
& & w_{16}^{(8)} = \frac16 s_{26}^{(8)}, \\
& & w_{17}^{(8)} = \frac12 (s_{27}^{(8)} - s_{28}^{(8)}), \\
& & w_{18}^{(8)} = \frac12 s_{29}^{(8)}, \\
& & w_{19}^{(8)} = \frac56 (s_{30}^{(8)} + s_{33}^{(8)}), \\
& & w_{20}^{(8)} = \frac{5}{24} (s_{31}^{(8)} + s_{34}^{(8)}), \\
& & w_{21}^{(8)} = \frac{5}{24} (s_{32}^{(8)} + s_{35}^{(8)}), \\
& & w_{22}^{(8)} = \frac56 (s_{36}^{(8)} + s_{38}^{(8)}),
\end{eqnarray}
\begin{eqnarray}
& & w_{23}^{(8)} = -\frac56 (s_{37}^{(8)} + s_{39}^{(8)}), \\
& & w_{24}^{(8)} = \frac{5}{18} (s_{40}^{(8)} + s_{41}^{(8)}), \\
& & w_{25}^{(8)} = \frac16 (-s_{42}^{(8)} + s_{43}^{(8)} - s_{44}^{(8)} - s_{45}^{(8)} - s_{46}^{(8)} - s_{47}^{(8)}), \\
& & w_{26}^{(8)} = 3 s_{48}^{(8)}, \\
& & w_{27}^{(8)} = \frac34 s_{49}^{(8)}, \\
& & w_{28}^{(8)} = \frac34 s_{50}^{(8)}, \\
& & w_{29}^{(8)} = \frac35 s_{51}^{(8)}, \\
& & w_{30}^{(8)} = \frac35 s_{52}^{(8)}, \\
& & w_{31}^{(8)} = \frac{3}{20} s_{53}^{(8)}, \\
& & w_{32}^{(8)} = \frac{3}{20} s_{54}^{(8)}, \\
& & w_{33}^{(8)} = -\frac{3}{10} (s_{57}^{(8)} - s_{58}^{(8)} + s_{59}^{(8)} + s_{60}^{(8)} - s_{61}^{(8)} + s_{62}^{(8)}), \\
& & w_{34}^{(8)} = \frac{11}{30} (s_{63}^{(8)} + s_{66}^{(8)}), \\
& & w_{35}^{(8)} = -\frac{11}{120} (s_{64}^{(8)} + s_{67}^{(8)}), \\
& & w_{36}^{(8)} = -\frac{11}{120} (s_{65}^{(8)} + s_{68}^{(8)}), \\
& & w_{1}^{(10+\overline{10})} = -\frac13 s_{66}^{(10+\overline{10})}, \\
& & w_{2}^{(10+\overline{10})} = -\frac{1}{12} s_{67}^{(10+\overline{10})}, \\
& & w_{3}^{(10+\overline{10})} = -\frac{1}{12} (s_{65}^{(10+\overline{10})} - s_{68}^{(10+\overline{10})}), \\
& & w_{1}^{(27)} = -\frac{1}{10} s_{51}^{(27)}, \\
& & w_{2}^{(27)} = \frac25 s_{52}^{(27)}, \\
& & w_{3}^{(27)} = \frac{1}{10} s_{53}^{(27)}, \\
& & w_{4}^{(27)} = \frac{1}{10} s_{54}^{(27)}, \\
& & w_{5}^{(27)} = \frac{1}{30} (-s_{57}^{(27)} + s_{58}^{(27)} - s_{59}^{(27)} - s_{60}^{(27)} + s_{61}^{(27)} - s_{62}^{(27)}), \\
& & w_{6}^{(27)} = \frac{2}{15} s_{63}^{(27)}, \\
& & w_{7}^{(27)} = -\frac{1}{30} s_{64}^{(27)}, \\
& & w_{8}^{(27)} = -\frac{1}{30} (s_{65}^{(27)} + s_{68}^{(27)}), \\
& & w_{9}^{(27)} = \frac{2}{15} s_{66}^{(27)}, \\
& & w_{10}^{(27)} = -\frac{1}{30} s_{67}^{(27)}.
\end{eqnarray}

Unfortunately, unlike flavor SB corrections, strong IB corrections cannot be further simplified in terms of fewer effective operator coefficients. The reason is quite simple. The operator basis $\{R^{(ij)(abc)}\}$ is constituted by 170 linearly independent operators so they all contribute to the $1/N_c$ expansion (\ref{eq:stri}) alike. There is not a single rule to eliminate some of them. Notice that, in Eqs.~(\ref{eq:isb1})-(\ref{eq:isb6}), the explicit dependence on $N_c$ that comes along the operators involved in the $1/N_c$ expansion have been kept. Because those expressions are evaluated at $N_c=3$, this is a useful artifact to identify leading and subleading terms in them. Just recall that $f^2 \sim\mathcal{O}(N_c)$, so unitarity of the scattering amplitudes is not compromised.

Therefore, the usefulness of relations (\ref{eq:isb1})-(\ref{eq:isb6}) can be better appreciated by retaining leading and subleading terms in $N_c$. Specifically, Eqs.~(\ref{eq:isb3}) and (\ref{eq:isb6}) get leading corrections from the 8 representation whereas $10+\overline{10}$ and $27$ representations are $1/N_c$ suppressed. The relevant operators for the symmetric part are $\delta^{ij}\delta^{ab}T^3$, $\delta^{ij}\delta^{a3}T^b$, and $\delta^{ij}\delta^{b3}T^a$, whereas for the antisymmetric part the relevant operators are
$i \epsilon^{ijm} \delta^{ab} G^{m3}$, $i \epsilon^{ijm} \delta^{a3} G^{mb}$, $i \epsilon^{ijm} \delta^{b3} G^{ma}$, $i \epsilon^{ijm} f^{a3e} f^{beg} G^{mg}$, and $i \epsilon^{ijm} d^{abe} d^{3eg} G^{mg}$.

Relations (\ref{eq:isb1}), (\ref{eq:isb2}), (\ref{eq:isb4}), and (\ref{eq:isb5}) get important corrections from the singlet and octet representations, whereas $10+\overline{10}$ and $27$ representations get $1/N_c$-suppressed factors.

Additionally,
\begin{eqnarray}
& & f^2k^0 \big[ \delta \mathcal{A}_\mathrm{IB} (p + \pi^0 \to p + \pi^0) - \delta \mathcal{A}_\mathrm{IB} (n + \pi^0 \to n + \pi^0) \big] \nonumber \\
& & \mbox{\hglue0.2truecm} = \Big[ \Big[ \frac12 w_{1}^{(8)} + 2 w_{2}^{(8)} + 2 w_{3}^{(8)} + \frac{1}{2N_c} w_{19}^{(8)} + \frac{2}{N_c} w_{20}^{(8)} + \frac{2}{N_c} w_{21}^{(8)} + \frac{1}{2N_c} w_{24}^{(8)} \nonumber \\
& & \mbox{\hglue0.9truecm} + \frac{1}{2N_c} w_{26}^{(8)} + \frac{2}{N_c} w_{27}^{(8)} + \frac{2}{N_c} w_{28}^{(8)} + \frac{1}{2N_c} w_{30}^{(8)} + \frac{2}{N_c} w_{31}^{(8)} + \frac{2}{N_c} w_{32}^{(8)} \Big] \nonumber \\
& & \mbox{\hglue0.9truecm} + \Big[ \frac{1}{2N_c} w_{2}^{(27)} + \frac{2}{N_c} w_{3}^{(27)} + \frac{2}{N_c} w_{4}^{(27)} \Big] \Big] \mathbf{k} \cdot \mathbf{k}^\prime + \Big[ \Big[ \frac12 w_{4}^{(8)} + 2 w_{5}^{(8)} + 2 w_{6}^{(8)} \nonumber \\
& & \mbox{\hglue0.9truecm} + \frac12 w_{9}^{(8)} + \frac{1}{2N_c} w_{11}^{(8)} + \frac{2}{N_c} w_{12}^{(8)} + \frac{2}{N_c} w_{13}^{(8)} + \frac{1}{2N_c} w_{16}^{(8)} + \frac{1}{2N_c} w_{34}^{(8)} - \frac{2}{N_c} w_{35}^{(8)} \nonumber \\
& & \mbox{\hglue0.9truecm} - \frac{2}{N_c} w_{36}^{(8)} \Big] + \Big[ \frac{1}{2N_c} w_{1}^{(10+\overline{10})} + \frac{2}{N_c} w_{2}^{(10+\overline{10})} + \frac{2}{N_c} w_{3}^{(10+\overline{10})} \Big] \nonumber \\
& & \mbox{\hglue0.9truecm} + \Big[ \frac{1}{2N_c} w_{6}^{(27)} - \frac{2}{N_c} w_{7}^{(27)} - \frac{2}{N_c} w_{8}^{(27)} + \frac{1}{2N_c} w_{9}^{(27)} - \frac{2}{N_c} w_{10}^{(27)} \Big] \Big] i (\mathbf{k} \times \mathbf{k}^\prime)_3 \nonumber \\
& & \mbox{\hglue0.9truecm} + \mathcal{O} \left[ \frac{1}{N_c^2} \right]. \nonumber \\
\end{eqnarray}
In this case, there is a kind of octet dominance because the $10+\overline{10}$ representation actually starts contributing at order $\mathcal{O}(1/N_c^2)$ and the 27 representation is at least one factor of $1/N_c$ suppressed relative to the octet representation.

\subsubsection{Strong isospin breaking to the scattering amplitude from Fig.~\ref{fig:sp}(c)}

Strong IB corrections emerging from Fig.~\ref{fig:sp}(c) can be cast into
\begin{eqnarray}
& & f^2k^0 [\delta \mathrm{A}_\mathrm{IB} (p + \pi^- \to p + \pi^-) - \delta \mathrm{A}_\mathrm{IB} (p + \pi^0 \to p + \pi^0) + \frac{1}{\sqrt{2}} \delta \mathrm{A}_\mathrm{IB} (p + \pi^- \to n + \pi^0)] \nonumber \\
& & \mbox{\hglue0.2truecm} = \Big[ 2 N_c v_{1}^{(1)} + \frac{3}{2N_c} v_{6}^{(1)} \Big] + \Big[ \frac34 v_{3}^{(8)} - \frac14 v_{4}^{(8)} + \frac14 v_{5}^{(8)} + \frac{15}{8N_c} v_{8}^{(8)} - \frac{5}{8N_c} v_{9}^{(8)} \nonumber \\
& & \mbox{\hglue0.9truecm} + \frac{5}{8N_c} v_{10}^{(8)} + \frac{5}{4N_c} v_{11}^{(8)} - \frac{5}{4N_c} v_{12}^{(8)} + \frac{5}{8N_c} v_{13}^{(8)} + \frac{1}{2N_c} v_{14}^{(8)} - \frac{1}{2N_c} v_{15}^{(8)} + \frac{1}{2N_c} v_{16}^{(8)} \Big] \nonumber \\
& & \mbox{\hglue0.9truecm} + \mathcal{O} \left[ \frac{1}{N_c^2} \right], \label{eq:ib1}
\end{eqnarray}

\begin{eqnarray}
& & f^2k^0 [\delta \mathrm{A}_\mathrm{IB} (p + \pi^+ \to p + \pi^+) - \delta \mathrm{A}_\mathrm{IB} (p + \pi^- \to p + \pi^-) - \sqrt{2} \delta \mathrm{A}_\mathrm{IB} (p + \pi^- \to n + \pi^0)] \nonumber \\
& & \mbox{\hglue0.2truecm} = \Big[ - 4 N_c v_{1}^{(1)} - \frac{3}{N_c} v_{6}^{(1)} \Big] + \Big[ - v_{5}^{(8)} - \frac{5}{2N_c} v_{10}^{(8)} - \frac{5}{2N_c} v_{11}^{(8)} - \frac{5}{2N_c} v_{12}^{(8)} - \frac{1}{N_c} v_{14}^{(8)} \nonumber \\
& & \mbox{\hglue0.9truecm} + \frac{1}{N_c} v_{15}^{(8)} - \frac{1}{N_c} v_{16}^{(8)} \Big] + \mathcal{O} \left[ \frac{1}{N_c^2} \right], \label{eq:ib2}
\end{eqnarray}

\begin{eqnarray}
& & f^2k^0 [\delta \mathrm{A}_\mathrm{IB} (p + \pi^+ \to p + \pi^+) + \delta \mathrm{A}_\mathrm{IB} (p + \pi^- \to p + \pi^-) - 2 \delta \mathrm{A}_\mathrm{IB} (p + \pi^0 \to p + \pi^0)] \nonumber \\
& & \mbox{\hglue0.2truecm} = \frac32 v_{3}^{(8)} - \frac12 v_{4}^{(8)} - \frac12 v_{5}^{(8)} + \frac{15}{4N_c} v_{8}^{(8)} - \frac{5}{4N_c} v_{9}^{(8)} - \frac{5}{4N_c} v_{10}^{(8)} - \frac{5}{N_c} v_{12}^{(8)} + \frac{5}{4N_c} v_{13}^{(8)} \nonumber \\
& & \mbox{\hglue0.9truecm} + \mathcal{O} \left[ \frac{1}{N_c^2} \right], \label{eq:ib3}
\end{eqnarray}

\begin{eqnarray}
& & f^2k^0 [\delta \mathrm{A}_\mathrm{IB} (n + \pi^- \to n + \pi^-) - \delta \mathrm{A}_\mathrm{IB} (n + \pi^0 \to n + \pi^0) - \frac{1}{\sqrt{2}} \delta \mathrm{A}_\mathrm{IB} (n + \pi^+ \to p + \pi^0)] \nonumber \\
& & \mbox{\hglue0.2truecm} = \Big[ 2 N_c v_{1}^{(1)} + \frac{3}{2N_c} v_{6}^{(1)} \Big] + \Big[ \frac54 v_{3}^{(8)} + \frac14 v_{4}^{(8)} - \frac14 v_{5}^{(8)} + \frac{25}{8N_c} v_{8}^{(8)} + \frac{5}{8N_c} v_{9}^{(8)} \nonumber \\
& & \mbox{\hglue0.9truecm} - \frac{5}{8N_c} v_{10}^{(8)} - \frac{5}{4N_c} v_{11}^{(8)} - \frac{15}{4N_c} v_{12}^{(8)} + \frac{25}{24N_c} v_{13}^{(8)} + \frac{1}{2N_c} v_{14}^{(8)} - \frac{1}{2N_c} v_{15}^{(8)} + \frac{1}{2N_c} v_{16}^{(8)} \Big] \nonumber \\
& & \mbox{\hglue0.9truecm} + \mathcal{O} \left[ \frac{1}{N_c^2} \right], \label{eq:ib4}
\end{eqnarray}

\begin{eqnarray}
& & f^2k^0 [\delta \mathrm{A}_\mathrm{IB} (n + \pi^+ \to n + \pi^+) - \delta \mathrm{A}_\mathrm{IB} (n + \pi^- \to n + \pi^-) + \sqrt{2} \delta \mathrm{A}_\mathrm{IB} (n + \pi^+ \to p + \pi^0)] \nonumber \\
& & \mbox{\hglue0.2truecm} = \Big[ - 4 N_c v_{1}^{(1)} - \frac{3}{N_c} v_{6}^{(1)} \Big] + \Big[ - 2 v_{3}^{(8)} + v_{5}^{(8)} - \frac{5}{N_c} v_{8}^{(8)} + \frac{5}{2N_c} v_{10}^{(8)} + \frac{5}{2N_c} v_{11}^{(8)} \nonumber \\
& & \mbox{\hglue0.9truecm} + \frac{15}{2N_c} v_{12}^{(8)} - \frac{5}{3N_c} v_{13}^{(8)} - \frac{1}{N_c} v_{14}^{(8)} + \frac{1}{N_c} v_{15}^{(8)} - \frac{1}{N_c} v_{16}^{(8)} \Big] + \mathcal{O} \left[ \frac{1}{N_c^2} \right], \label{eq:ib5}
\end{eqnarray}

\begin{eqnarray}
& & f^2k^0 [\delta \mathrm{A}_\mathrm{IB} (n + \pi^+ \to n + \pi^+) + \delta \mathrm{A}_\mathrm{IB} (n + \pi^- \to n + \pi^-) - 2 \delta \mathrm{A}_\mathrm{IB} (n + \pi^0 \to n + \pi^0)] \nonumber \\
& & \mbox{\hglue0.2truecm} = \frac12 v_{3}^{(8)} + \frac12 v_{4}^{(8)} + \frac12 v_{5}^{(8)} + \frac{5}{4N_c} v_{8}^{(8)} + \frac{5}{4N_c} v_{9}^{(8)} + \frac{5}{4N_c} v_{10}^{(8)} + \frac{5}{12N_c} v_{13}^{(8)} + \mathcal{O} \left[ \frac{1}{N_c^2} \right] \nonumber \\
& & \mbox{\hglue0.9truecm}, \label{eq:ib6}
\end{eqnarray}

Relations (\ref{eq:ib1})-(\ref{eq:ib6}) cannot be reduced further in terms of effective operator coefficients. In a similar fashion to the previous section, notice that relations (\ref{eq:ib3}) and (\ref{eq:ib6}) are also dominated by corrections from the octet representations and numerically they should be at least a factor of $1/N_c$ smaller than relations (\ref{eq:ib1}), (\ref{eq:ib2}), (\ref{eq:ib4}), and (\ref{eq:ib5}), which are dominated by the singlet representation.

Additionally, for the relation
\begin{eqnarray}
& & f^2k^0 [\delta \mathrm{A}_\mathrm{IB} (p + \pi^0 \to p + \pi^0) - \delta \mathrm{A}_\mathrm{IB} (n + \pi^0 \to n + \pi^0)] \nonumber \\
& & \mbox{\hglue0.2truecm} = \frac12 v_{3}^{(8)} + \frac12 v_{4}^{(8)} + \frac12 v_{5}^{(8)} + \frac{5}{4N_c} v_{8}^{(8)} + \frac{5}{4N_c} v_{9}^{(8)} + \frac{5}{4N_c} v_{10}^{(8)} + \frac{5}{12N_c} v_{13}^{(8)} + \mathcal{O} \left[ \frac{1}{N_c^2} \right] \nonumber \\
& & \mbox{\hglue0.9truecm}. \label{eq:ib7}
\end{eqnarray}
a kind of octet dominance is also found in the sense that flavor $10+\overline{10}$ and $27$ representations actually start contributing at relative order $\mathcal{O}(1/N_c^2)$ so they can be safely ignored.

\subsection{\label{sec:compa}Some remarks about a comparison with HBChPT expressions}

The scattering amplitudes for the $N\pi$ system obtained here through the use of $SU(3)$ flavor projection operators
can be (partially) compared with HBChPT theory results at tree-level order. At this point, the urge of performing the full evaluation of the three terms retained in Eq.~(\ref{eq:atree3}) for $N_c=3$ becomes manifest. The success of $SU(2)$ HBChPT to investigate the low-energy processes of pions and nucleons is undeniable. However, the inclusion of particles with strangeness requires the use of $SU(3)$ HBChPT. For instance, $s$-wave pseudoscalar meson octet-baryon scattering lengths to the third chiral order in that framework have been studied with only baryon octet contributions \cite{liu06} and both baryon octet and decuplet contributions \cite{liu07}. Specifically, in the latter reference decuplet contributions to the threshold $T$-matrices are found to vanish, in complete opposition to the present analysis where non-vanishing decuplet baryon contributions proportional to $\mathcal{C}$ are obtained, even in the degeneracy limit $\Delta\to 0$. In a more recent work \cite{huang}, the $T$-matrices of pseudoscalar meson octet-baryon scattering to one-loop order are computed in HBChPT. For elastic meson-baryon scattering, the leading order $\mathcal{O}(q)$ amplitudes resulting from tree diagrams for $\pi N$ scattering contributing at first chiral order are given in Eq.~(10) and (11) of that reference, which can be compared to Eqs.~(\ref{eq:a32}) and (\ref{eq:aa32}) and (\ref{eq:a12}) and (\ref{eq:aa12}) of this work in the limit $\Delta\to 0$ and excluding decuplet baryon contributions; aside from kinematic factors relating the rest system of the initial baryon and the center of mass system, which can be linked through a Lorentz transformation, the Clebsch-Gordan structures coincide, up to a global minus sign which might be traced back to the different conventions used. Other scattering processes such as $\pi\Sigma$, $\pi\Xi$, $KN$, and so on discussed in Ref.~\cite{huang} can also be evaluated in the present formalism. A recent analysis with the inclusion of decuplet effects \cite{huang2} reveals some interesting facts in the comparison with the present analysis. Again, except for some kinematic factors, the comparison is achieved for $\Delta$ replaced by $-\Delta$ in Eqs.~(13)-(16) of that reference.
 
At next-to-leading order, the explicit chiral symmetry breaking part of the meson-baryon effective chiral Lagrangian $\mathcal{L}^{(2,\mathrm{ct})}$, with no inclusion of decuplet baryon effects, is presented in Eq.~(8) of Ref.~\cite{huang}. It yields the amplitudes $T^{(I)}_{\pi N}$ in terms of 11 LECs. For $\pi N$ system they are given in Eqs.~(64) and (65) of that reference. In principle, these LECs should (partially) correspond to the 12 operator coefficients contained in expressions (\ref{eq:asb32}) and (\ref{eq:aav32}) and (\ref{eq:asb12}) and (\ref{eq:aav12}), respectively. Although the relations among them should be linear, it is hard to identify them, except for the vertex diagram for which $C_3=-k_0 h_1/8$ for fixed incident meson energy. A full identification requires the inclusion of the decuplet baryons in the framework of that reference and the computation of additional amplitudes in the framework discussed here. The latter will be attempted elsewhere.

\section{\label{sec:sa}$S$-wave scattering lengths}

The $N\pi$ forward scattering amplitude for a nucleon at rest can be readily obtained from Eqs.~(\ref{eq:a12}) and (\ref{eq:a32}) at threshold. Following the lines of Ref.~\cite{bernard}, here the $s$-wave scattering lengths including the baryon mass splitting and first-order SB are found to be
\begin{eqnarray}
a^{(1/2)} & = & \frac{1}{4\pi} \frac{m_\pi}{f^2} \left[1+\frac{m_\pi}{M_N}\right]^{-1} \Bigg[ (D + F)^2 - \frac49 \left[1 + \frac{\Delta}{m_\pi} + \frac{\Delta^2}{m_\pi^2} \right] \mathcal{C}^2 \nonumber \\
& & \mbox{} + d_{1}^{(1)} + d_{1}^{(8)} + d_{1}^{(10+\overline{10})} + d_{1}^{(27)} \Bigg] \label{eq:a12f} \nonumber \\
& = & a^+ + 2 a^-,
\end{eqnarray}
and
\begin{eqnarray}
a^{(3/2)} & = & \frac{1}{4\pi} \frac{m_\pi}{f^2} \left[1+\frac{m_\pi}{M_N}\right]^{-1} \left[ - \frac12 (D + F)^2 + \frac29 \left[ 1 - \frac{2\Delta}{m_\pi} + \frac{\Delta^2}{m_\pi^2}\right] \mathcal{C}^2 \right. \nonumber \\
& & \mbox{} + \left. d_{1}^{(1)} + d_{1}^{(8)} - 2 d_{1}^{(10+\overline{10})} - 2 d_{1}^{(27)} + \frac32 d_{2}^{(8)} + \frac32 d_{2}^{(27)} \right] \label{eq:a32f} \nonumber \\
& = & a^+ - a^-,
\end{eqnarray}
which are valid to order $\mathcal{O}(\Delta^3/m_\pi^3)$.

Notice that in the limit $\Delta\to 0$ and removing SB effects,
\begin{equation}
a^{(1/2)} + 2 a^{(3/2)} = 0, \label{eq:curr}
\end{equation}
which is a well-known result obtained in the context of current algebra \cite{wein,tomo}. It is important to remark that relation (\ref{eq:curr}) is fulfilled even in the presence of the $\mathcal{C}^2$ term, which accounts for the contribution of decuplet baryons. Thus, violations to expression (\ref{eq:curr}) arise not only from SB but also from a linear term in $\Delta$.

The usefulness of Eqs.~(\ref{eq:a12f}) and (\ref{eq:a32f}) relies entirely on the precise determination of the $SU(3)$ invariants $D$, $F$, and $\mathcal{C}$ and the six parameters $d_{k}^{(\mathrm{dim})}$ involved in those equations. The invariants can be extracted from baryon semileptonic decays, for instance. The latter set can be obtained by comparing the theoretical expressions with the available experimental data \cite{part} via a least-squares fit. A detailed analysis requires additional theoretical expressions for which data are available and would involve processes including strangeness.

Isospin IB effects as obtained here can also be incorporated into relations (\ref{eq:a12f}) and (\ref{eq:a32f}) is a straightforward manner.

\section{\label{sec:cr}Concluding remarks}

The material discussed in this work represents an enterprising program to understand the baryon-meson scattering processes in the context of the $1/N_c$ expansion. It presents new ideas, perspectives, or analytical frameworks that contribute to a more comprehensive understanding of the subject matter. Specifically, the scattering amplitude for the process $B\pi \to B^\prime \pi$, including the decuplet-octet baryon mass splitting and flavor symmetry breaking, has been computed, specialized to the process $N\pi \to N\pi$. Evidently, processes such as $\Delta \pi\to N\pi$, $\Delta \pi \to \Delta \pi$ or processes including strangeness can straightforwardly be evaluated because the formalism is general enough to cover the cases when $B$ and $B^\prime$ are any baryon states and $\pi^a$ and $\pi^b$ are any pseudo scalar mesons, provided that the Gell-Mann--Nishijima scheme is fulfilled. The expressions for $N\pi \to N\pi$ scattering amplitudes obtained here get simple forms [Eqs.~(\ref{eq:ppppppch})-(\ref{eq:nppppzch}) and Eqs.~(\ref{eq:ppppppvtx})-(\ref{eq:nppppzvtx})] once all the ingredients are put together, regardless of the breathtaking original expressions, such as (\ref{eq:sam1}). However, the inclusion of strong isospin breaking introduces a rather large number of operators coefficients so the series seems to have minimal utility, unless stringent suppressions in $1/N_c$ are done to keep only leading contributions. Violations to strong isospin breaking, uncovered by relations (\ref{eq:isb1})-(\ref{eq:isb6}), reveal which $SU(3)$ flavor representations dominate over the others.

One important result extracted from the present analysis is worth mentioning: Once more, it is evident that the spin-1/2 and spin-3/2 baryons are present from the outset because together form an irreducible representation of the spin-flavor symmetry.

As it was mentioned in the introductory section, previous analyses about scattering amplitudes in the context of the $1/N_c$ expansion \cite{c1,kwee,c2} have focused their goals on some specific aspects of the theory. The analysis presented here, with the extensive use of projection operators to classify operator structures, contributes to the subject from a different perspective; the approaches complement among them.

On the other hand, a comparison of the results obtained here with HBChPT results performed at tree-level order can be made. The idea of rewriting scattering amplitudes in terms of the $SU(3)$ invariant baryon-meson couplings $D$, $F$, $\mathcal{C}$, and $\mathcal{H}$, Eqs.~(\ref{eq:ppppppch})-(\ref{eq:nppppzch}), allows a comparison with the tree-level values [in the $SU(3)$ exact limit] from HBChPT, mostly by dropping the mass difference $\Delta$ and possibly the $\mathcal{C}^2$ terms, {\it i.e.}, under the degeneracy limit and with the decuplet baryon degrees of freedom integrated out, which is usually the common procedure advocated in the literature. A full comparison will require the computation of loops in the combined formalism in $1/N_c$ and chiral corrections. This requires formidable effort that will be attempted elsewhere.

\section{Supplementary information}

This paper is complemented by some supplementary material where explicit reductions of baryon operators are presented, as well as their corresponding matrix elements in the form of tables. The pdf file can be obtained from authors by request.

\section*{Acknowledgement}
The authors are grateful to Consejo Nacional de Humanidades, Ciencias y Tecnolog{\'\i}as (Mexico) for support through the {\it Ciencia de Frontera} project CF-2023-I-162.

\appendix

\section{\label{sec:appa}Baryon operator basis used in baryon-meson scattering}

The operators $S_m^{(ij)(ab)}$ that constitute the basis used in baryon-meson scattering at lowest order, comprising up to $7$-body operators, read
\begin{align}
S_{1}^{(ij)(ab)} & = i \delta^{ij} f^{abe} T^e, \nonumber &
S_{2}^{(ij)(ab)} & = i \epsilon^{ijr} \delta^{ab} J^r, \nonumber \\
S_{3}^{(ij)(ab)} & = i \epsilon^{ijr} d^{abe} G^{re}, \nonumber &
S_{4}^{(ij)(ab)} & = \delta^{ab} \{J^i,J^j\}, \nonumber \\
S_{5}^{(ij)(ab)} & = \delta^{ij} \delta^{ab} J^2, \nonumber &
S_{6}^{(ij)(ab)} & = \{G^{ia},G^{jb}\}, \nonumber \\
S_{7}^{(ij)(ab)} & = \{G^{ib},G^{ja}\}, \nonumber &
S_{8}^{(ij)(ab)} & = \delta^{ij} \{G^{ra},G^{rb}\}, \nonumber \\
S_{9}^{(ij)(ab)} & = i \epsilon^{ijr} \{G^{ra},T^b\}, \nonumber &
S_{10}^{(ij)(ab)} & = i \epsilon^{ijr} \{G^{rb},T^a\}, \nonumber \\
S_{11}^{(ij)(ab)} & = d^{abe} \{J^j,G^{ie}\}, \nonumber &
S_{12}^{(ij)(ab)} & = i f^{abe} \{J^i,G^{je}\}, \nonumber \\
S_{13}^{(ij)(ab)} & = i f^{abe} \{J^j,G^{ie}\}, \nonumber &
S_{14}^{(ij)(ab)} & = \delta^{ij} d^{abe} \{J^r,G^{re}\}, \nonumber \\
S_{15}^{(ij)(ab)} & = \epsilon^{ijr} f^{abe} \mathcal{D}_2^{re}, \nonumber &
S_{16}^{(ij)(ab)} & = i \epsilon^{ijr} d^{abe} \mathcal{D}_3^{re}, \nonumber \\
S_{17}^{(ij)(ab)} & = i \epsilon^{ijr} d^{abe} \mathcal{O}_3^{re}, \nonumber &
S_{18}^{(ij)(ab)} & = i \epsilon^{ijr} \{J^r,\{T^a,T^b\}\}, \nonumber \\
S_{19}^{(ij)(ab)} & = i \epsilon^{ijm} \{J^m,\{G^{ra},G^{rb}\}\}, \nonumber &
S_{20}^{(ij)(ab)} & = i f^{abe} \{T^e,\{J^i,J^j\}\}, \nonumber \\
S_{21}^{(ij)(ab)} & = i \delta^{ij} f^{abe} \{J^2,T^e\}, \nonumber &
S_{22}^{(ij)(ab)} & = i \epsilon^{ijr} \delta^{ab} \{J^2,J^r\}, \nonumber \\
S_{23}^{(ij)(ab)} & = i \epsilon^{imr} \{G^{ja},\{J^m,G^{rb}\}\}, \nonumber &
S_{24}^{(ij)(ab)} & = i \epsilon^{jmr} \{G^{ia},\{J^m,G^{rb}\}\}, \nonumber \\
S_{25}^{(ij)(ab)} & = i \epsilon^{imr} \{G^{jb},\{J^m,G^{ra}\}\}, \nonumber &
S_{26}^{(ij)(ab)} & = i \epsilon^{jmr} \{G^{ib},\{J^m,G^{ra}\}\}, \nonumber \\
S_{27}^{(ij)(ab)} & = i \epsilon^{ijm} \{G^{ma},\{J^r,G^{rb}\}\}, \nonumber &
S_{28}^{(ij)(ab)} & = i \epsilon^{ijm} \{G^{mb},\{J^r,G^{ra}\}\}, \nonumber \\
S_{29}^{(ij)(ab)} & = i f^{aeg} d^{beh} \{T^h,\{J^i,G^{jg}\}\}, \nonumber &
S_{30}^{(ij)(ab)} & = i d^{aeg} f^{beh} \{T^g,\{J^j,G^{ih}\}\}, \nonumber \\
S_{31}^{(ij)(ab)} & = d^{abe} [J^2,\{J^i,G^{je}\}], \nonumber &
S_{32}^{(ij)(ab)} & = d^{abe} [J^2,\{J^j,G^{ie}\}], \nonumber \\
S_{33}^{(ij)(ab)} & = i f^{abe} [J^2,\{J^i,G^{je}\}], \nonumber &
S_{34}^{(ij)(ab)} & = i f^{abe} [J^2,\{J^j,G^{ie}\}], \nonumber \\
S_{35}^{(ij)(ab)} & = i \epsilon^{ijr} [J^2,\{G^{ra},T^b\}], \nonumber &
S_{36}^{(ij)(ab)} & = i \epsilon^{ijr} [J^2,\{G^{rb},T^a\}], \nonumber \\
S_{37}^{(ij)(ab)} & = \epsilon^{ijr} f^{abe} \mathcal{D}_4^{re}, \nonumber &
S_{38}^{(ij)(ab)} & = i f^{abe} \{\{J^i,J^j\},\{J^r,G^{re}\}\}, \nonumber \\
S_{39}^{(ij)(ab)} & = i \epsilon^{ijm} \{\mathcal{D}_2^{mb},\{J^r,G^{ra}\}\}, \nonumber &
S_{40}^{(ij)(ab)} & = i \epsilon^{ijm} \{\mathcal{D}_2^{ma},\{J^r,G^{rb}\}\}, \nonumber \\
S_{41}^{(ij)(ab)} & = i \epsilon^{ijr} \{J^2,\{G^{ra},T^b\}\}, \nonumber &
S_{42}^{(ij)(ab)} & = i \epsilon^{ijr} \{J^2,\{G^{rb},T^a\}\}, \nonumber \\
S_{43}^{(ij)(ab)} & = i f^{abe} \{J^2,\{J^i,G^{je}\}\}, \nonumber &
S_{44}^{(ij)(ab)} & = i f^{abe} \{J^2,\{J^j,G^{ie}\}\}, \nonumber \\
S_{45}^{(ij)(ab)} & = \{J^2,\{G^{ia},G^{jb}\}\}, \nonumber &
S_{46}^{(ij)(ab)} & = \{J^2,\{G^{ib},G^{ja}\}\}, \nonumber \\
S_{47}^{(ij)(ab)} & = d^{abe} \{\{J^i,J^j\},\{J^r,G^{re}\}\}, \nonumber &
S_{48}^{(ij)(ab)} & = \delta^{ab} \{J^2,\{J^i,J^j\}\}, \nonumber \\
S_{49}^{(ij)(ab)} & = \epsilon^{ijk} \epsilon^{rml} \{J^k,\{G^{ra},\{J^m,G^{lb}\}\}\}, \nonumber &
S_{50}^{(ij)(ab)} & = i \epsilon^{iml} [\{J^j,\{J^m,G^{la}\}\},\{J^r,G^{rb}\}], \nonumber \\
S_{51}^{(ij)(ab)} & = i \epsilon^{jml} [\{J^i,\{J^m,G^{la}\}\},\{J^r,G^{rb}\}], \nonumber &
S_{52}^{(ij)(ab)} & = i \epsilon^{jml} [\{J^i,\{J^m,G^{lb}\}\},\{J^r,G^{ra}\}], \nonumber \\
S_{53}^{(ij)(ab)} & = i \epsilon^{iml} [\{J^j,\{J^m,G^{lb}\}\},\{J^r,G^{ra}\}], \nonumber &
S_{54}^{(ij)(ab)} & = \{G^{ia},\mathcal{O}_3^{jb}\}, \nonumber \\
S_{55}^{(ij)(ab)} & = i \epsilon^{ijm} [J^2,\{G^{mb},\{J^r,G^{ra}\}\}], \nonumber &
S_{56}^{(ij)(ab)} & = i \epsilon^{ijm} \{J^2,[G^{mb},\{J^r,G^{ra}\}]\}, \nonumber \\
S_{57}^{(ij)(ab)} & = \delta^{ij} \{J^2,\{G^{ra},G^{rb}\}\}, \nonumber &
S_{58}^{(ij)(ab)} & = \delta^{ij} d^{abe} \{J^2,\{J^r,G^{re}\}\}, \nonumber \\
\end{align}
\begin{align}
S_{59}^{(ij)(ab)} & = \delta^{ij} \delta^{ab} \{J^2,J^2\}, \nonumber &
S_{60}^{(ij)(ab)} & = \{[J^2,G^{ia}],[J^2,G^{jb}]\}, \nonumber \\
S_{61}^{(ij)(ab)} & = i \epsilon^{jmr} [J^2,\{G^{ib},\{J^m,G^{ra}\}\}], \nonumber &
S_{62}^{(ij)(ab)} & = i \epsilon^{ijr} d^{abe} \mathcal{D}_5^{re}, \nonumber \\
S_{63}^{(ij)(ab)} & = \epsilon^{ijr} f^{abe} \mathcal{O}_5^{re}, \nonumber &
S_{64}^{(ij)(ab)} & = i \epsilon^{ijr} d^{abe} \mathcal{O}_5^{re}, \nonumber \\
S_{65}^{(ij)(ab)} & = \{\mathcal{O}_3^{ia},\mathcal{D}_2^{jb}\}, \nonumber &
S_{66}^{(ij)(ab)} & = \{\mathcal{D}_2^{ia},\mathcal{O}_3^{jb}\}, \nonumber \\
S_{67}^{(ij)(ab)} & = \{J^2,\{T^a,\{J^j,G^{ib}\}\}\}, \nonumber &
S_{68}^{(ij)(ab)} & = \{J^2,\{T^b,\{J^i,G^{ja}\}\}\}, \nonumber \\
S_{69}^{(ij)(ab)} & = i f^{abe} \{J^2,\{T^e,\{J^i,J^j\}\}\}, \nonumber &
S_{70}^{(ij)(ab)} & = i \delta^{ij} f^{abe} \{J^2,\{J^2,T^e\}\}, \nonumber \\
S_{71}^{(ij)(ab)} & = i \epsilon^{ijr} \delta^{ab} \{J^2,\{J^2,J^r\}\}, \nonumber &
S_{72}^{(ij)(ab)} & = i \epsilon^{ijm} \{J^2,\{J^m,\{G^{ra},G^{rb}\}\}\}, \nonumber \\
S_{73}^{(ij)(ab)} & = i \epsilon^{imr} \{J^2,\{G^{ja},\{J^m,G^{rb}\}\}\}, \nonumber &
S_{74}^{(ij)(ab)} & = i \epsilon^{jmr} \{J^2,\{G^{ia},\{J^m,G^{rb}\}\}\}, \nonumber \\
S_{75}^{(ij)(ab)} & = i \epsilon^{imr} \{J^2,\{G^{jb},\{J^m,G^{ra}\}\}\}, \nonumber &
S_{76}^{(ij)(ab)} & = i \epsilon^{jmr} \{J^2,\{G^{ib},\{J^m,G^{ra}\}\}\}, \nonumber \\
S_{77}^{(ij)(ab)} & = i \epsilon^{ijm} \{J^2,\{G^{ma},\{J^r,G^{rb}\}\}\}, \nonumber &
S_{78}^{(ij)(ab)} & = i \epsilon^{ijm} \{J^2,\{G^{mb},\{J^r,G^{ra}\}\}\}, \nonumber \\
S_{79}^{(ij)(ab)} & = i f^{aeg} d^{beh} \{J^2,\{T^h,\{J^i,G^{jg}\}\}\}, \nonumber &
S_{80}^{(ij)(ab)} & = i d^{aeg} f^{beh} \{J^2,\{T^g,\{J^j,G^{ih}\}\}\}, \nonumber \\
S_{81}^{(ij)(ab)} & = d^{abe} \{J^2,[J^2,\{J^i,G^{je}\}]\}, \nonumber &
S_{82}^{(ij)(ab)} & = d^{abe} \{J^2,[J^2,\{J^j,G^{ie}\}]\}, \nonumber \\
S_{83}^{(ij)(ab)} & = i \epsilon^{ijl} \{J^l,\{\{J^r,G^{ra}\},\{J^m,G^{mb}\}\}\}, \nonumber &
S_{84}^{(ij)(ab)} & = \{\{J^i,J^j\},\{T^a,\{J^r,G^{rb}\}\}\}, \nonumber \\
S_{85}^{(ij)(ab)} & = \{\{J^i,J^j\},\{T^b,\{J^r,G^{ra}\}\}\}, \nonumber &
S_{86}^{(ij)(ab)} & = i \epsilon^{mlr} \{\{J^i,J^j\},\{G^{mb},\{J^l,G^{ra}\}\}\}, \nonumber \\
S_{87}^{(ij)(ab)} & = i \epsilon^{jlm} \{\{J^i,\{J^l,G^{ma}\}\},\{J^r,G^{rb}\}\}, \nonumber &
S_{88}^{(ij)(ab)} & = i \epsilon^{ilm} \{\{J^j,\{J^l,G^{ma}\}\},\{J^r,G^{rb}\}\}, \nonumber \\
S_{89}^{(ij)(ab)} & = i \epsilon^{jlm} \{\{J^i,\{J^l,G^{mb}\}\},\{J^r,G^{ra}\}\}, \nonumber &
S_{90}^{(ij)(ab)} & = i \epsilon^{ilm} \{\{J^j,\{J^l,G^{mb}\}\},\{J^r,G^{ra}\}\}, \nonumber \\
S_{91}^{(ij)(ab)} & = i \epsilon^{imr} \{G^{jb},\{J^2,\{J^m,G^{ra}\}\}\}, \nonumber &
S_{92}^{(ij)(ab)} & = i \epsilon^{ijr} \{J^2,[J^2,\{G^{ra},T^b\}]\}, \nonumber \\
S_{93}^{(ij)(ab)} & = i \epsilon^{ijr} \{J^2,[J^2,\{G^{rb},T^a\}]\}, \nonumber &
S_{94}^{(ij)(ab)} & = i f^{abe} \{J^2,[J^2,\{J^i,G^{je}\}]\}, \nonumber \\
S_{95}^{(ij)(ab)} & = i f^{abe} \{J^2,[J^2,\{J^j,G^{ie}\}]\}, \nonumber &
S_{96}^{(ij)(ab)} & = \epsilon^{ijr} f^{abe} \mathcal{D}_6^{re}, \nonumber \\
S_{97}^{(ij)(ab)} & = i f^{abe} \{J^2,\{\{J^i,J^j\},\{J^r,G^{re}\}\}\}, \nonumber &
S_{98}^{(ij)(ab)} & = i \epsilon^{ijm} \{J^2,\{\mathcal{D}_2^{mb},\{J^r,G^{ra}\}\}\}, \nonumber \\
S_{99}^{(ij)(ab)} & = i \epsilon^{ijm} \{J^2,\{\mathcal{D}_2^{ma},\{J^r,G^{rb}\}\}\}, \nonumber &
S_{100}^{(ij)(ab)} & = i \epsilon^{ijr} \{J^2,\{J^2,\{G^{ra},T^b\}\}\}, \nonumber \\
S_{101}^{(ij)(ab)} & = i \epsilon^{ijr} \{J^2,\{J^2,\{G^{rb},T^a\}\}\}, \nonumber &
S_{102}^{(ij)(ab)} & = i f^{abe} \{J^2,\{J^2,\{J^i,G^{je}\}\}\}, \nonumber \\
S_{103}^{(ij)(ab)} & = i f^{abe} \{J^2,\{J^2,\{J^j,G^{ie}\}\}\}, \nonumber &
S_{104}^{(ij)(ab)} & = \{J^2,\{J^2,\{G^{ia},G^{jb}\}\}\}, \nonumber \\
S_{105}^{(ij)(ab)} & = \{J^2,\{J^2,\{G^{ib},G^{ja}\}\}\}, \nonumber &
S_{106}^{(ij)(ab)} & = d^{abe} \{J^2,\{\{J^i,J^j\},\{J^r,G^{re}\}\}\}, \nonumber \\
S_{107}^{(ij)(ab)} & = \delta^{ab} \{J^2,\{J^2,\{J^i,J^j\}\}\}, \nonumber &
S_{108}^{(ij)(ab)} & = i \epsilon^{jml} \{J^2,[\{J^i,\{J^m,G^{la}\}\},\{J^r,G^{rb}\}]\}, \nonumber \\
S_{109}^{(ij)(ab)} & = i \epsilon^{jml} \{J^2,[\{J^i,\{J^m,G^{lb}\}\},\{J^r,G^{ra}\}]\}, \nonumber &
S_{110}^{(ij)(ab)} & = i \epsilon^{iml} \{J^2,[\{J^j,\{J^m,G^{lb}\}\},\{J^r,G^{ra}\}]\}, \nonumber \\
S_{111}^{(ij)(ab)} & = \{J^2,\{G^{ia},\mathcal{O}_3^{jb}\}\}, \nonumber &
S_{112}^{(ij)(ab)} & = i \epsilon^{ijm} \{J^2,[J^2,\{G^{mb},\{J^r,G^{ra}\}\}]\}, \nonumber \\
S_{113}^{(ij)(ab)} & = i \epsilon^{ijm} \{J^2,\{J^2,[G^{mb},\{J^r,G^{ra}\}]\}\}, \nonumber &
S_{114}^{(ij)(ab)} & = \delta^{ij} \{J^2,\{J^2,\{G^{ra},G^{rb}\}\}\}, \nonumber \\
S_{115}^{(ij)(ab)} & = \delta^{ij} d^{abe} \{J^2,\{J^2,\{J^r,G^{re}\}\}\}, \nonumber &
S_{116}^{(ij)(ab)} & = \delta^{ij} \delta^{ab} \{J^2,\{J^2,J^2\}\}, \nonumber \\
S_{117}^{(ij)(ab)} & = \{J^2,\{[J^2,G^{ia}],[J^2,G^{jb}]\}\}, \nonumber &
S_{118}^{(ij)(ab)} & = i \epsilon^{imr} \{J^2,[J^2,\{G^{jb},\{J^m,G^{ra}\}\}]\}, \nonumber \\
S_{119}^{(ij)(ab)} & = i \epsilon^{jmr} \{J^2,[J^2,\{G^{ib},\{J^m,G^{ra}\}\}]\}, \nonumber &
S_{120}^{(ij)(ab)} & = i \epsilon^{ijr} d^{abe} \mathcal{D}_7^{re}, \nonumber \\
S_{121}^{(ij)(ab)} & = i \epsilon^{ijr} d^{abe} \mathcal{O}_7^{re}, \nonumber &
S_{122}^{(ij)(ab)} & = i f^{abe} \{J^2,\{J^2,\{T^e,\{J^i,J^j\}\}\}\}, \nonumber \\
S_{123}^{(ij)(ab)} & = i \delta^{ij} f^{abe} \{J^2,\{J^2,\{J^2,T^e\}\}\}, \nonumber &
S_{124}^{(ij)(ab)} & = i \epsilon^{ijr} \delta^{ab} \{J^2,\{J^2,\{J^2,J^r\}\}\}, \nonumber \\
S_{125}^{(ij)(ab)} & = i \epsilon^{imr} \{J^2,\{J^2,\{G^{ja},\{J^m,G^{rb}\}\}\}\}, \nonumber &
S_{126}^{(ij)(ab)} & = i \epsilon^{jmr} \{J^2,\{J^2,\{G^{ia},\{J^m,G^{rb}\}\}\}\}, \nonumber \\
\end{align}
\begin{align}
S_{127}^{(ij)(ab)} & = i \epsilon^{imr} \{J^2,\{J^2,\{G^{jb},\{J^m,G^{ra}\}\}\}\}, \nonumber &
S_{128}^{(ij)(ab)} & = i \epsilon^{jmr} \{J^2,\{J^2,\{G^{ib},\{J^m,G^{ra}\}\}\}\}, \nonumber \\
S_{129}^{(ij)(ab)} & = i \epsilon^{ijm} \{J^2,\{J^2,\{G^{ma},\{J^r,G^{rb}\}\}\}\}, \nonumber &
S_{130}^{(ij)(ab)} & = i \epsilon^{ijm} \{J^2,\{J^2,\{G^{mb},\{J^r,G^{ra}\}\}\}\}, \nonumber \\
S_{131}^{(ij)(ab)} & = i f^{aeg} d^{beh} \{J^2,\{J^2,\{T^h,\{J^i,G^{jg}\}\}\}\}, \nonumber &
S_{132}^{(ij)(ab)} & = i d^{aeg} f^{beh} \{J^2,\{J^2,\{T^g,\{J^j,G^{ih}\}\}\}\}, \nonumber \\
S_{133}^{(ij)(ab)} & = d^{abe} \{J^2,\{J^2,[J^2,\{J^i,G^{je}\}]\}\}, \nonumber &
S_{134}^{(ij)(ab)} & = d^{abe} \{J^2,\{J^2,[J^2,\{J^j,G^{ie}\}]\}\}, \nonumber \\
S_{135}^{(ij)(ab)} & = i \epsilon^{ijl} \{J^2,\{J^l,\{\{J^r,G^{ra}\},\{J^m,G^{mb}\}\}\}\}, \nonumber &
S_{136}^{(ij)(ab)} & = i \epsilon^{mlr} \{J^2,\{\{J^i,J^j\},\{G^{mb},\{J^l,G^{ra}\}\}\}\}, \nonumber \\
S_{137}^{(ij)(ab)} & = i \epsilon^{jlm} \{J^2,\{\{J^i,\{J^l,G^{ma}\}\},\{J^r,G^{rb}\}\}\}, \nonumber &
S_{138}^{(ij)(ab)} & = i \epsilon^{jlm} \{J^2,\{\{J^i,\{J^l,G^{mb}\}\},\{J^r,G^{ra}\}\}\}, \nonumber \\
S_{139}^{(ij)(ab)} & = i \epsilon^{imr} \{J^2,\{G^{jb},\{J^2,\{J^m,G^{ra}\}\}\}\}. &
\label{opbasis}
\end{align}
For completeness, the operator coefficients $c^{(\mathrm{s})}_m$ and $c^{(\mathrm{a})}_m$ that accompany these operators are listed in the Online Resource to this paper.

\end{document}